\documentclass[two column]{aastex631}
\usepackage{amsmath}
\usepackage{xcolor}
\usepackage{footmisc}
\usepackage{rotating,longtable}
\newcommand{\kms}{\,km\,s$^{-1}$}

\usepackage{graphicx}
\usepackage{subfigure}
\usepackage{natbib} 
\usepackage{hyperref}
\usepackage{lipsum} 

\begin{document}





\title{Investigating the Star-forming Sites in the Outer Galactic Arm}

\author[0000-0002-6586-936X]{Aayushi Verma}
\affil{Aryabhatta Research Institute of Observational sciencES (ARIES),
Manora Peak, Nainital 263001, India}
\affil{M.J.P.Rohilkhand University, Bareilly, Uttar Pradesh-243006, India}

\author[0000-0001-5731-3057]{Saurabh Sharma}
\affil{Aryabhatta Research Institute of Observational sciencES (ARIES),
Manora Peak, Nainital 263001, India}

\author[0000-0001-6725-0483]{Lokesh K. Dewangan}
\affil{Physical Research Laboratory, Navrangpura, Ahmedabad - 380009, India}

\author[0000-0001-9312-3816]{Devendra K. Ojha} 
\affil{Tata Institute of Fundamental Research (TIFR), Homi Bhabha Road, Colaba, Mumbai - 400005, India}

\author[0000-0002-3873-6449]{Kshitiz Mallick}
\affil{National Astronomical Observatory of Japan, 2-21-1 Osawa, Mitaka, Tokyo 181-8588, Japan}

\author[0000-0002-6740-7425]{Ram Kesh Yadav}
\affil{National Astronomical Research Institute of Thailand (Public Organization), 260 Moo 4, T. Donkaew, A. Maerim, Chiangmai 50 180, Thailand}

\author{Harmeen Kaur}
\affil{Center of Advanced Study, Department of Physics DSB Campus, Kumaun University Nainital, 263001, India}

\author[0009-0008-8490-8601]{Tarak Chand}
\affil{Aryabhatta Research Institute of Observational sciencES (ARIES),
Manora Peak, Nainital 263001, India}
\affil{M.J.P.Rohilkhand University, Bareilly, Uttar Pradesh-243006, India}

\author{Mamta Agarwal}
\affil{Aryabhatta Research Institute of Observational sciencES (ARIES),
Manora Peak, Nainital 263001, India}

\author{Archana Gupta}
\affil{M.J.P.Rohilkhand University, Bareilly, Uttar Pradesh-243006, India}

\begin{abstract}
We aim to investigate the global star formation scenario in star-forming sites AFGL 5157, [FSR2007] 0807 (hereafter FSR0807), [HKS2019] E70 (hereafter E70), [KPS2012] MWSC 0620 (hereafter KPS0620), and IRAS 05331$+$3115 in the outer galactic arm.
The distribution of young stellar objects in these sites coincides with a higher extinction and H$_2$ column density, which agrees with the notion that star formation occurs inside the dense molecular cloud cores. 
We have found two molecular structures at different velocities in this direction; one contains AFGL 5157 and FSR0807, and the other contains E70, [KPS2012] MWSC 0620, and IRAS 05331$+$3115. All these clusters in our target region are in different evolutionary stages and might form stars through different mechanisms. The E70 cluster seems to be the oldest in our sample; AFGL 5157 and FSR0807 formed later, and KPS0620 and IRAS 05331$+$3115 are the youngest sites. AFGL 5157 and FSR0807 are physically connected and have cold filamentary structures and dense hub regions. Additionally, the near-infrared photometric analysis shows signatures of massive star formation in these sites. KPS0620 also seems to have cold filamentary structures with the central hub but lacks signatures of massive stars. Our analysis suggests molecular gas flow and the hub filamentary star formation scenario in these regions. IRAS 05331$+$3115 is a single clump of molecular gas favoring low-mass star formation. Our study suggests that the selected area is a menagerie of star-forming sites where the formation of the stars happens through different processes.
\end{abstract}


\section{Introduction} \label{sec:intro}

The formation of stars occurs in some ensemble of clusters and associations \citep{2003ARA&A..41...57L}, and it is assumed that their formation in isolation is rare. A significant fraction (70–90\%) of the stars formed in the Giant Molecular Clouds (GMCs) is found in Embedded young star clusters (YSCs; \citealt{2017ApJ...842...25G,2018MNRAS.481.1016G}). Star clusters are clustered to each other in large complexes and imprinted with the fractal structure of the GMCs from which they are born \citep{1978PAZh....4..125E,1996ApJ...471..816E,2010ApJ...720..541S}. Stars form in dense gas, so the initial distribution of this material in a cloud controls whether the cloud forms a distributed population of stars or compact clusters. \citep{2003ARA&A..41...57L}. 

Molecular clouds (MCs) in our Galaxy have complex geometries, such as filamentary structures, elongated networks, sheets, etc., and the distribution and physical properties of active star formation sites in them aid in understanding the role of these structures in star formation and evolution \citep{1972A&A....21..255A,1991IAUS..147..293E,1993prpl.conf...97E,2003ARA&A..41...57L,2008ApJ...688.1142K,2014ApJ...787..107K,2018MNRAS.477.1903S}. The present study examines such active star formation sites towards $l=176^{\circ}$ to $177^{\circ}$, $b=-0^{\circ}.44$ to $0.^{\circ}20$ in the outer galactic arm aiming to understand the global star formation scenario concerning MC morphology.

The paper's organization is as follows: Section \ref{selection} overviews the selected area. Section \ref{sec:archival_data} briefly summarises the data sets used in the present study. In Section \ref{sec:result}, we have presented the methodology to extract the YSCs' physical parameters and study the environment around them, including the morphology/kinematics of the associated MCs. In Section \ref{sec:discussion}, we have discussed the major findings of our analysis. Finally, we have concluded our study in Section \ref{sec:conclusion}.

\section{Source Selection}
\label{selection}

\begin{figure*}[!t]
    \centering
    \includegraphics[width=0.48\textwidth]{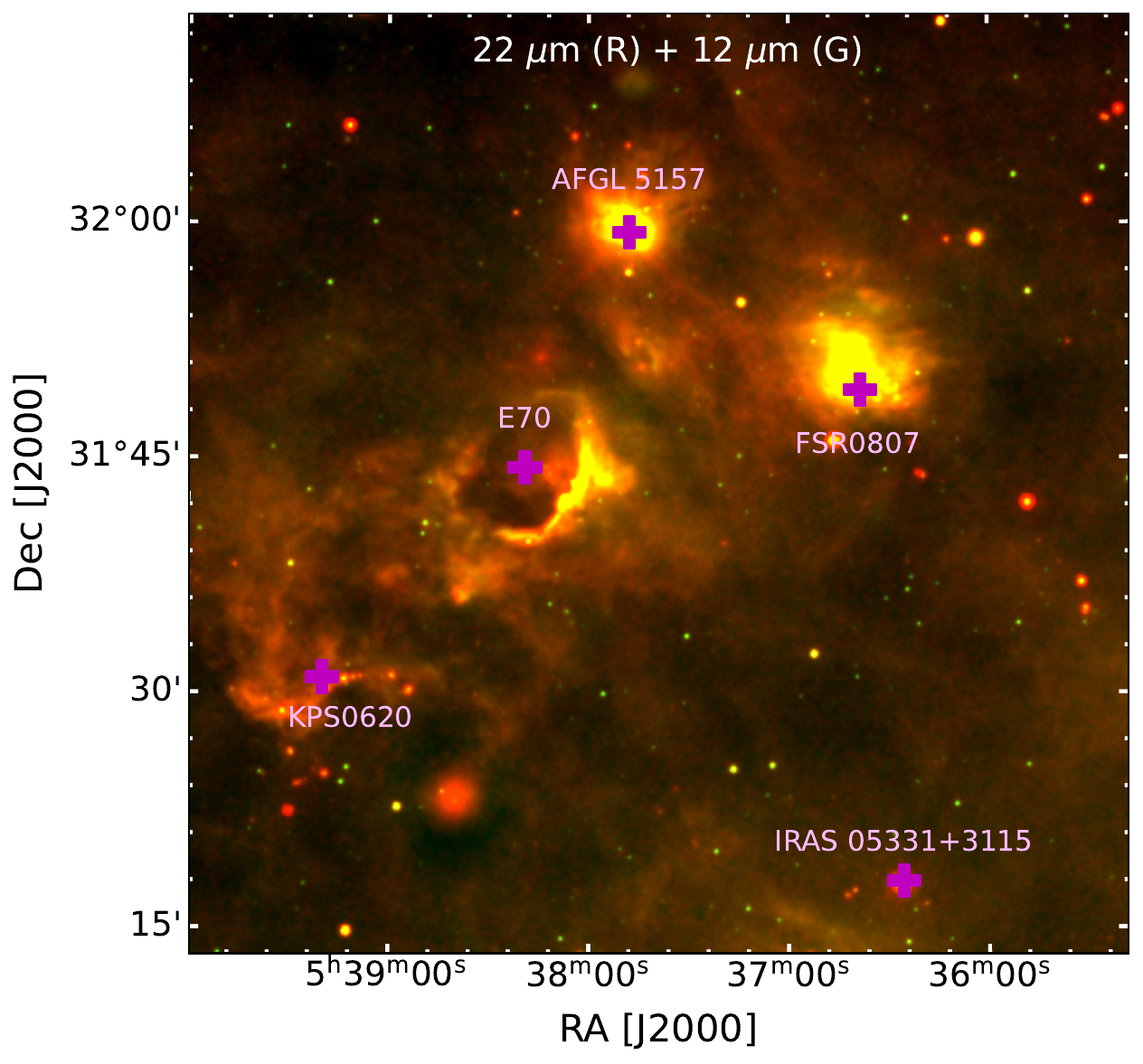}
    \includegraphics[width=0.48\textwidth]{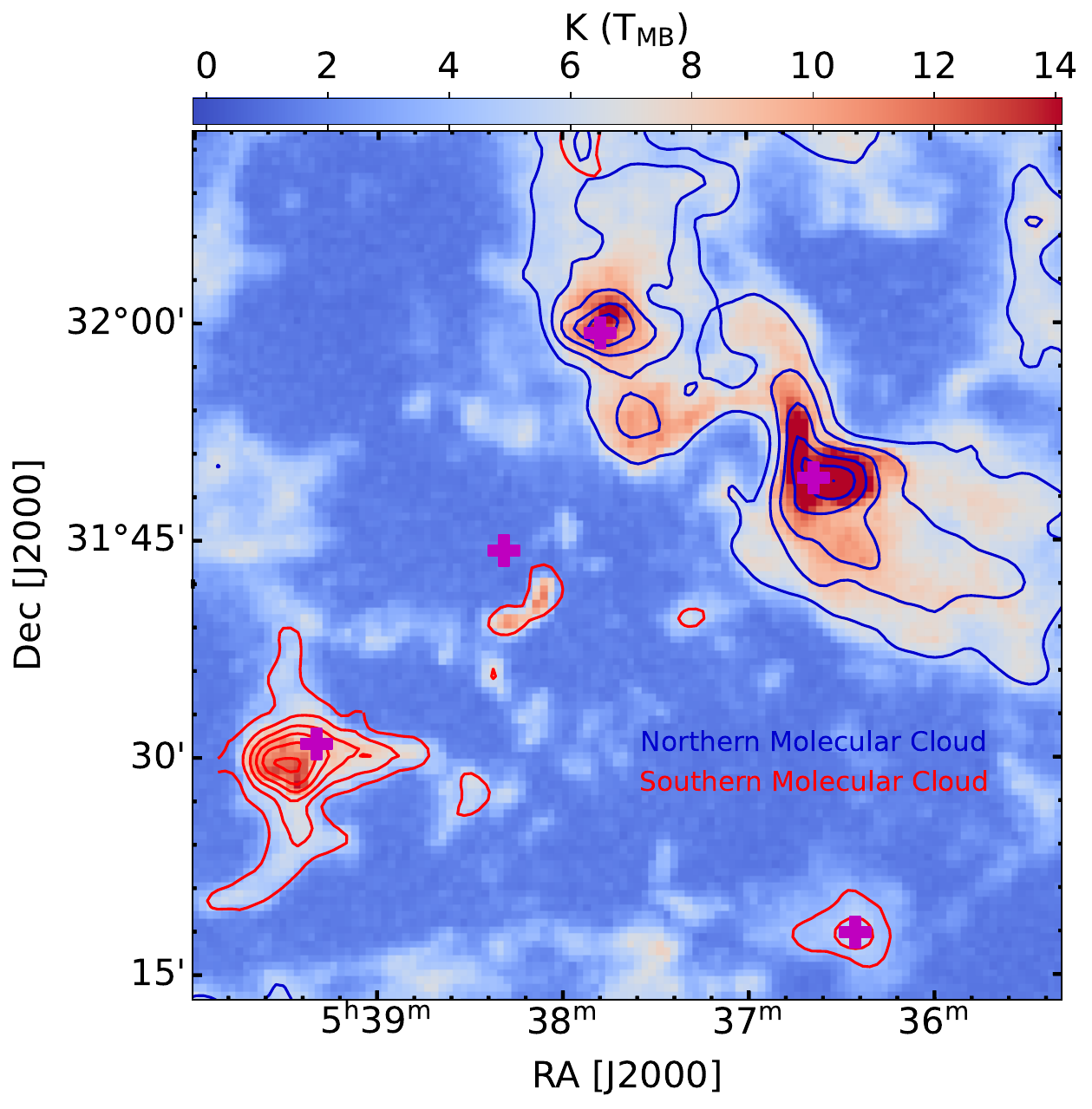}
    \caption{Left panel: Color-composite image (Red: WISE 22 $\mu$m; Green: WISE 12 $\mu$m) of present  $1^{\circ} \times 1^{\circ}$ target area. Right panel: $^{12}$CO peak intensity map of the selected region. 
    The blue and red contours represent the $^{12}$CO moment-0 contours for the northern and southern molecular structures/clouds, respectively (c.f. Section \ref{sec:channel_map}). The lowest level value for both the contours is the mean value with a step size of 1$\sigma$. The locations of five sites of active star formation are also marked with a magenta `$+$' marker in both panels.}   
    \label{fig:intro}
\end{figure*}

Active star-forming sites in the MC generally consist of YSCs, mid-infrared (MIR) bubbles, clouds/filaments, and young massive stars.
For the present study, we selected an interesting region of $1^{\circ} \times 1^{\circ}$ FOV consisting of five sites of active star formation, i.e., AFGL 5157, [FSR2007] 0807 (hereafter FSR0807), MIR bubble [HKS2019] E70 (hereafter E70), [KPS2012] MWSC 0620 (hereafter KPS0620) and an H\,{\sc ii} region IRAS 05331$+$3115. The coordinates of these regions are given in Table \ref{tab:ra_dec}. The left panel of Figure \ref{fig:intro} represents the color-composite image (Red: \emph{WISE} 22 $\mu$m; Green: \emph{WISE} 12 $\mu$m) of our selected target area marked with the locations above five regions. This figure depicts the distribution of warmed-up gas and dust and the embedded stars, suggesting recent star formation activities in the region. Brief details of each of these regions are given below:

\begin{table}[ht]
    \centering
    \caption{Locations of the five clusters}
    \begin{tabular}{c c c c c}
    \hline
    Clusters & $\alpha_{J2000}$ & $\delta_{J2000}$ & $l$ & $b$\\
     & (hh:mm:ss) & (dd:mm:ss) & (deg) & (deg)\\
    \hline
    AFGL 5157 & 05:37:48 & +31:59:24 & 176.52 & 0.18\\
    FSR0807 & 05:36:39 & +31:49:20 & 176.53 & $-$0.11\\
    E70 & 05:38:19 & +31:44:22 & 176.79 & 0.14\\
    KPS0620 & 05:39:20 & +31:30:58 & 177.09 & 0.20\\
    IRAS 05331$+$3115 & 05:36:26 & +31:17:43 & 176.95 & $-$0.44\\
    \hline\\
    \end{tabular}
   
    \label{tab:ra_dec}
\end{table}

\paragraph{AFGL 5157}{It is a reflection nebula hosting a cluster and is located at a distance of 1.8 kpc \citep{1992ApJ...384..528T}. An extended shell of H$_2$ surrounding the cluster of IR sources was detected in its direction \citep{2003A&A...405..655C,2008hsf1.book..869R,2017ApJ...844...38W}. It is reported as an evolved cluster having an age of $10^6$ yr \citep{1999AJ....117..446C,2003A&A...405..655C}. \citet{2019ApJ...884...84D} suggested that the triggered star formation of the young stars is occurring in this region by the collision of two filamentary MCs, which might have also initiated the formation of massive stars.}
\paragraph{FSR0807}{\citet{2013MNRAS.436.1465B} reported that FSR0807 is an open cluster located at a distance of 8.9 kpc. Later on, \citet{2018MNRAS.473..849D} reported that its age is 10 Myr and is located at a distance of 1.75 kpc from the sun with a core radius of 1.03 pc.}
\paragraph{E70}{MIR bubble E70, first cataloged by \citet{2019PASJ...71....6H}. \citet{2023ApJ...953..145V} reported that it is located at a distance of 3.26 kpc, having a small stellar clustering of radius $\sim$ 1.7 pc, and is hosting a massive O9V star possessing positive feedback and `collect and collapse scenario' might be a probable SF scenario at the rim of this bubble.}
\paragraph{KPS0620}{KPS0620 is a poorly studied star cluster which is located at a distance of 6.1 kpc \citet{2013MNRAS.436.1465B} whereas \citet{2013yCat..35580053K} reported it to be located at 2.2 kpc.}
\paragraph{IRAS 05331$+$3115}{IRAS 05331$+$3115 is an IRAS point source associated with an H\,{\sc ii} region \citep{beichman1988infrared,1996A&AS..115...81B,2002ApJS..141..157Y}. According to \citet{1989A&AS...80..149W}, this object is located at a kinematic galactocentric distance of 8.50 kpc. Additionally, it is 6.24 arcseconds away from its corresponding SIMBAD source (e.g., \citealt{2014PASJ...66...17T}).

\begin{table*}[!ht]
    \footnotesize
    \centering
    \caption{List of numerous surveys employed for the current study (NIR to Radio Wavelength)}
    \begin{tabular}{c c c c}
    \hline
    Survey & Wavelength/s & $\sim$ Resolution & Reference\\
    \hline
    Two Micron All Sky Survey\footnote{\citet{https://doi.org/10.26131/irsa2}} (2MASS) & 1.25, 1.65, and 2.17 $\mu$m & $2\arcsec.5$ & \citet{2006AJ....131.1163S}\\
    Gaia DR3\footnote{https://www.cosmos.esa.int/web/gaia/dr3} (magnitudes, parallax, and PM) & 330–1050 nm & 0.4 mas & \citet{2016gaia,2023gaia}\\
    Herschel Infrared Galactic Plane Survey\footnote{http://archives.esac.esa.int/hsa/whsa/} & 70, 160, 250, 350, 500 $\mu$m & $5\arcsec.8$, $12\arcsec$, $18\arcsec$, $25\arcsec$, $37\arcsec$ & \citet{2010PASP..122..314M}\\
    Milky Way Imaging Scroll Painting (MWISP) & CO (J =1$-$0), $^{13}$CO (J =1$-$0) &  $50\arcsec$ & \citet[]{Su_MWISP_2019ApJS} \\
    Spitzer GLIMPSE360 Survey\footnote{\citet{https://doi.org/10.26131/irsa214}} & 3.6, 4.5 $\mu$m & $2\arcsec$, $2\arcsec$ & \citet{2005ApJ...630L.149B}\\
    UKIRT InfraRed Deep Sky Survey\footnote{http://wsa.roe.ac.uk/} (UKIDSS) & 1.25, 1.65, and 2.22 $\mu$m & $0\arcsec.8$, $0\arcsec.8$, $0\arcsec.8$ & \citet{2008MNRAS.391..136L}\\
    Wide-field Infrared Survey Explorer\footnote{\citet{https://doi.org/10.26131/irsa1}} (\emph{WISE}) & 3.4, 4.6, 12, 22 $\mu$m & $6\arcsec.1$, $6\arcsec.4$, $6\arcsec.5$, $12\arcsec$ & \citet{2010AJ....140.1868W}\\
    \hline
    \end{tabular}
    \label{tab:archival_data}
\end{table*}

\section{Archival Data-sets used} \label{sec:archival_data}

\subsection{Molecular Line Data}
We obtained the $^{12}$CO$ (J=1-0)$, $^{13}$CO $(J=1-0)$, and C$^{18}$O$(J=1-0)$ molecular line data observed as a part of Milky Way Imaging Scroll Painting (MWISP) project by the Purple Mountain Observatory (PMO) 13.7\,m millimeter-wave radio telescope, China (\citealt{ 633694461037117445}; \citealt{Su_MWISP_2019ApJS}). All the Velocities were provided with respect to the local standard of rest (LSR). The $^{12}$CO spectral cube has a velocity resolution of $\sim$0.16 \kms, whereas the $^{13}$CO and C$^{18}$O have a velocity resolution of $\sim$0.17 \kms\, with a spatial resolution and a grid size of 50\arcsec and 30\arcsec, respectively. The rms noise for $^{12}$CO, $^{13}$CO, and C$^{18}$O was calculated to be $\sim$\,0.51\,K, $\sim$\,0.24\,K, and $\sim$\,0.25\,K, respectively. The difference in the rms noise is due to typical system temperatures ($\sim$250 K for $^{12}$CO at the upper sideband and $\sim$140 K for $^{13}$CO and C$^{18}$O at the lower sideband, respectively). The system temperature includes noises from the antenna (the optical system, the membrane, and the dome), the receiver, and the atmosphere \citep{Su_MWISP_2019ApJS}. A subset of this data comprising the E70 bubble has already been presented in \citet{2023ApJ...953..145V}.


\subsection{Other Archival Data Sets}

We used numerous archival data sets from near-infrared (NIR) to radio. Table \ref{tab:archival_data} briefly specifies these data sets. The Herschel column density and temperature maps (spatial resolution $\sim12\arcsec$) are taken from the open website\footnote{http://www.astro.cardiff.ac.uk/research/ViaLactea/}. These maps are acquired for EU-funded ViaLactea project \citep{2010PASP..122..314M} employing the Bayesian PPMAP technique \citep{2010A&A...518L.100M,2015MNRAS.454.4282M, 2017MNRAS.471.2730M}.

\begin{figure*}[!ht]
    \centering
    \includegraphics[width=\textwidth]{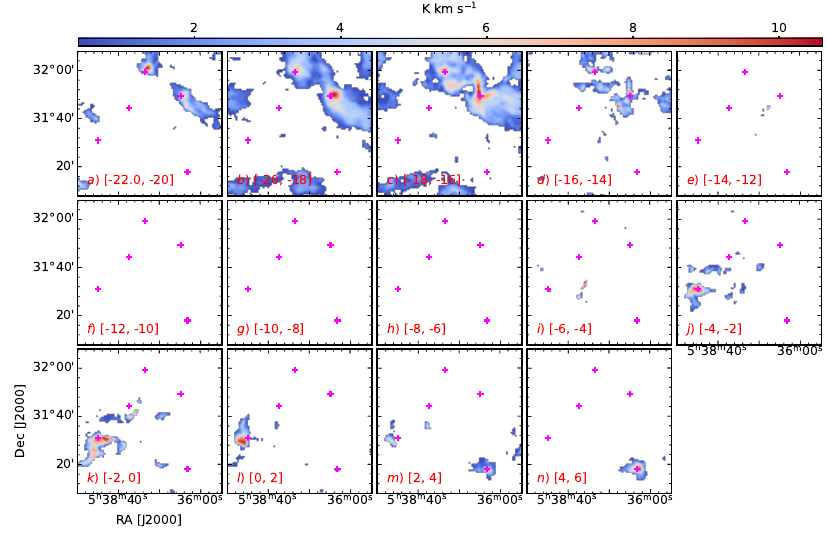}
    \caption{Channel Maps for $^{12}$CO$(J=1-0)$ emission integrated over marked velocity interval in each panel (in \kms). The emission is depicted above 5$\sigma$ value ($\sigma$ being the rms noise). The '+' magenta markers depict the clusters' locations.}
    \label{fig:channel_maps}
\end{figure*}

\section{Result and Analysis}\label{sec:result}

\subsection{Morphology and Dynamics of the Molecular Gas in the Selected Region}\label{sec:channel_map}

This section explores the molecular morphology and dynamics in our target area with the help of high-resolution CO data from the PMO survey.  Due to larger abundance (or larger optical thickness), $^{12}$CO is more relevant to uncover the spatial extent of the diffused extended gas having a density of $\sim$10$^2$ cm$^{-3}$ than $^{13}$CO, which traces $\sim$10$^{3-4}$ cm$^{-3}$ \citep{Su_MWISP_2019ApJS}. We confine our further analyses to the regions identified above the 5$\sigma$ threshold ($\sigma$ being the rms noise) to get the real physical features and remove any confusing artifacts. Since we could not find strong emission in C$^{18}$O transition, we do not include this in further analyses. 
The right panel of Figure \ref{fig:intro} depicts the selected region's $^{12}$CO peak intensity map. The locations of five star-forming sites in this region are also marked in both panels with a magenta `$+$' marker. All selected regions except E70 are associated with a peak in molecular emission. There is an arc of molecular emission surrounding the E70 bubble. 

Figure \ref{fig:channel_maps} depicts the channel maps for the target area in $^{12}$CO emission above 5$\sigma$ value ($\sigma$ being the rms noise). These channel maps show two different molecular structures with two prominent velocity components. The `northern molecular structure/cloud' has blue-shifted velocity components ranging from $-22$ to $-12$ \kms\, and it comprises AFGL 5157 and FSR0807. The `southern molecular structure/cloud' has a red-shifted velocity component ranging between $-4$ to 6 \kms\, and it comprises E70 bubble, KPS0620, IRAS 05331$+$3115. E70 bubble falls between these two structures
and is discussed in detail in \citet{2023ApJ...953..145V}. 
The blue and red contours in the right panel of Figure \ref{fig:intro} represent the $^{12}$CO moment-0 contours for the northern and southern molecular structures/clouds in the velocity ranges [$-$22, $-$12] \kms\, and [$-$4, +6] \kms, respectively. The lowest level value for both the contours is the mean value with a step size of 1$\sigma$.
We examine their morphology and dynamics in detail in the following sub-sections.

\subsubsection{The Northern Molecular Cloud: AFGL 5157 and FSR0807}\label{sec:afgl_f807}

Figure \ref{fig:moment_maps_-22_-12} depict the $^{12}$CO and $^{13}$CO  Integrated Intensity (moment-0 or m-0), Intensity-weighted velocity (moment-1 or m-1), and Intensity-weighted dispersion (linewidth) maps \citep{2023ApJ...944..228M,2023JApA...44...34M,2024MNRAS.527.9626P} for northern molecular structure consisting of AFGL 5157 and FSR0807 (velocity ranges [$-22$, $-12$] \kms\,). The inspection of $^{12}$CO and $^{13}$CO m-0 maps (cf. panels (a) and (d) of Figure \ref{fig:moment_maps_-22_-12}, respectively) in the velocity range [$-22$, $-12$] \kms\, affirm that AFGL 5157 and FSR0807 are connected with filamentary structures. On inspecting the m-1 maps (cf. panels (b) and (e) of Figure \ref{fig:moment_maps_-22_-12}), we find that the variation of velocity along AFGL 5157 to FSR0807 is almost uniform between $-19$ \kms\, and $-16$ \kms. The line-width map (dispersion of velocity; panel (c) and (f) of Figure \ref{fig:moment_maps_-22_-12}) displays that the central parts of AFGL 5157 and FSR0807 have a larger velocity dispersion than the filament connecting them.

\begin{figure*}[!ht]
    \centering
    \includegraphics[width=0.99\textwidth]{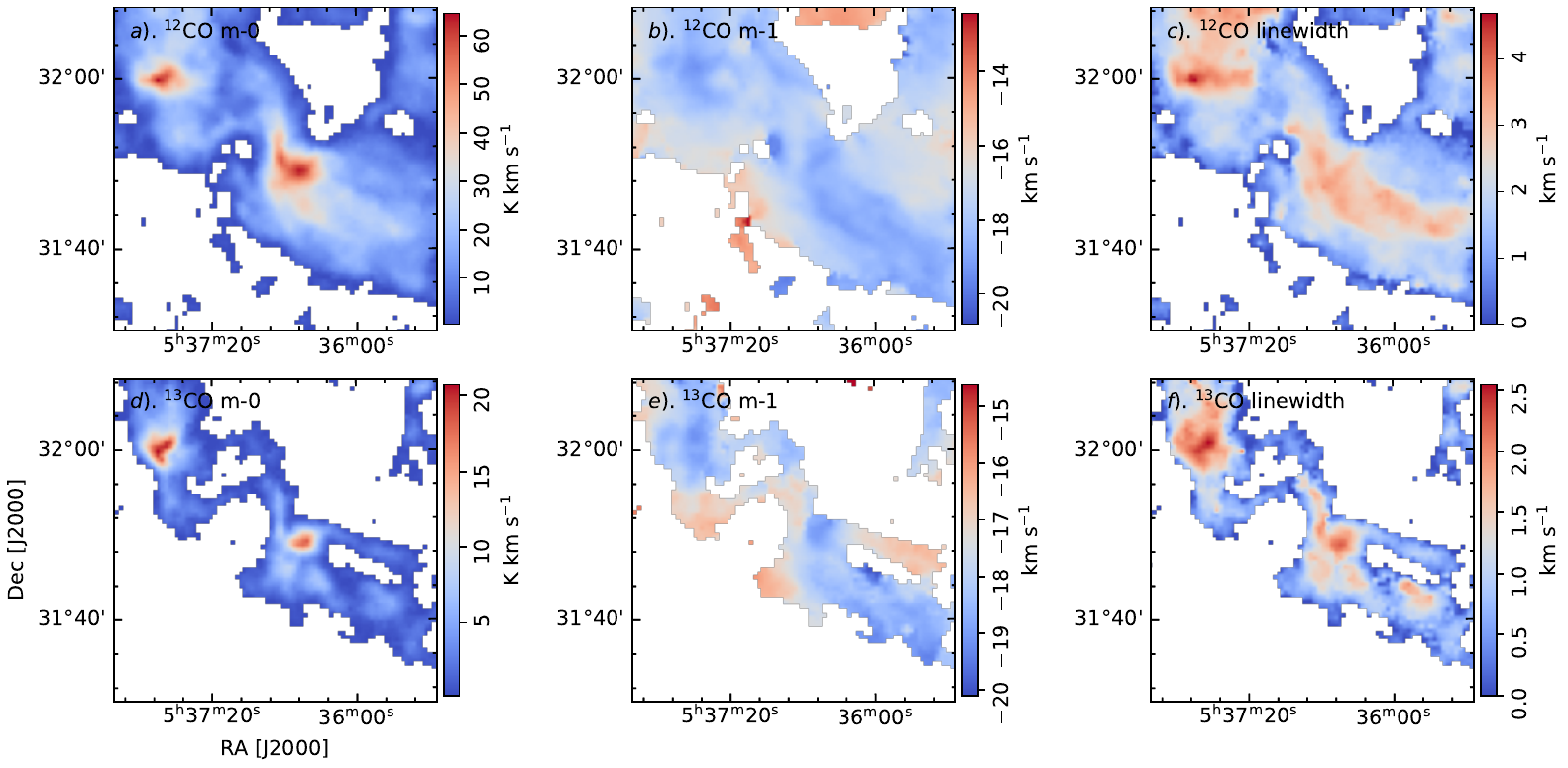}
    \caption{m-0, m-1 and linewidth maps (column-wise) for $^{12}$CO$(J=1-0)$ and $^{13}$CO$(J=1-0)$ emission, respectively in the velocity range [$-22$, $-12$] \kms\, for northern molecular cloud. The emission is depicted above 5$\sigma$ value ($\sigma$ being the rms noise for respective spectral cubes).}
    \label{fig:moment_maps_-22_-12}
\end{figure*}

To further examine the probable connection of AFGL 5157 and FSR0807 and velocity entanglements, we selected a set of square-shaped regions of size 30\arcsec\, along AFGL 5157 to FSR0807 (termed as N1 - N4 and marked in the upper and lower left panels of Figure \ref{fig:co_spectra_afgl_f807}) and determined the average spectra with Gaussian fitting using both $^{12/13}$CO spectral cubes. These spectra are shown in the right panel of Figure \ref{fig:co_spectra_afgl_f807}, and the derived Gaussian peaks are also mentioned in the figure. The $^{12}$CO spectra at location N1 (coinciding with the peak CO emission around AFGL 5157) reveal the presence of blue-shifted ($\sim -19.7$ \kms) and red-shifted ($\sim -17.0$ \kms) velocity peaks as compared to $^{13}$CO velocity peak at this location, which appear as a single velocity peak ($\sim -18.1$ \kms) in $^{13}$CO spectra. This single peak persists at other locations N2 - N4 in both $^{12/13}$CO spectra. We found a peculiar feature of optically thick $^{12}$CO spectra at N1, associated with AFGL 5157. It displays a red asymmetric profile (or `red profile', where the red-shifted peak of a double-peaked profile is more prominent in the case of an optically thick line), revealing outflow motion due to the absorption of colder blue-shifted gas in front of the hot core (e.g., \citealt{1998sao..rept.....M,2011ApJ...730..102L,2024MNRAS.530.1516P}).


\begin{figure*}[!ht]
\begin{minipage}{0.35\textwidth}
\includegraphics[width=\textwidth]{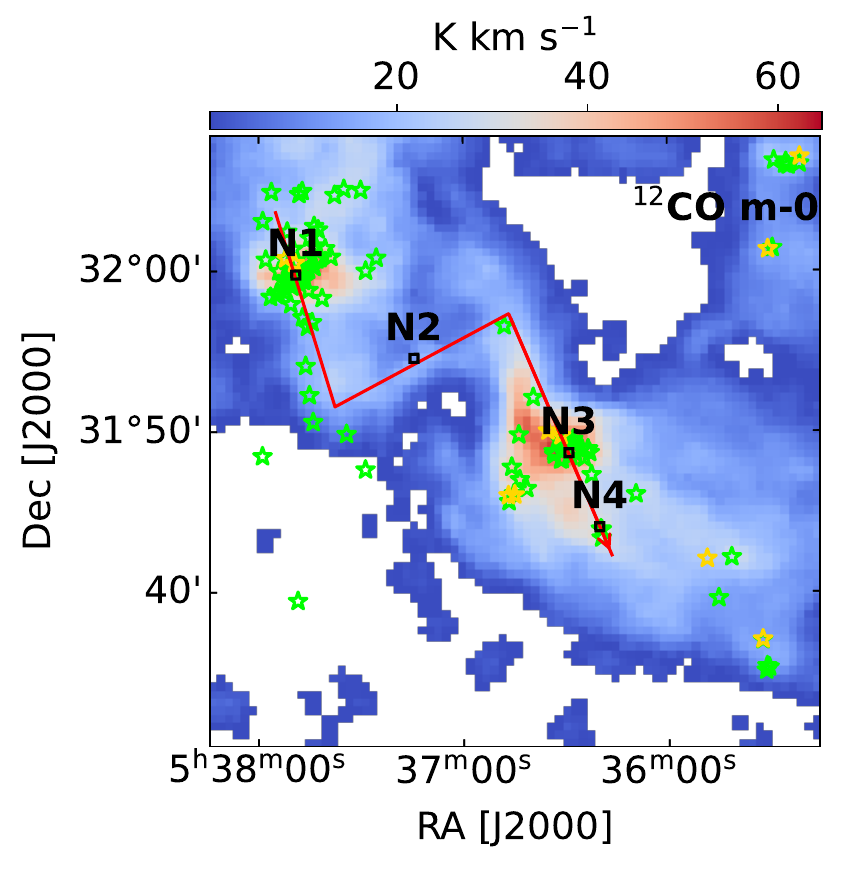}
\\[2mm]
\includegraphics[width=\textwidth]{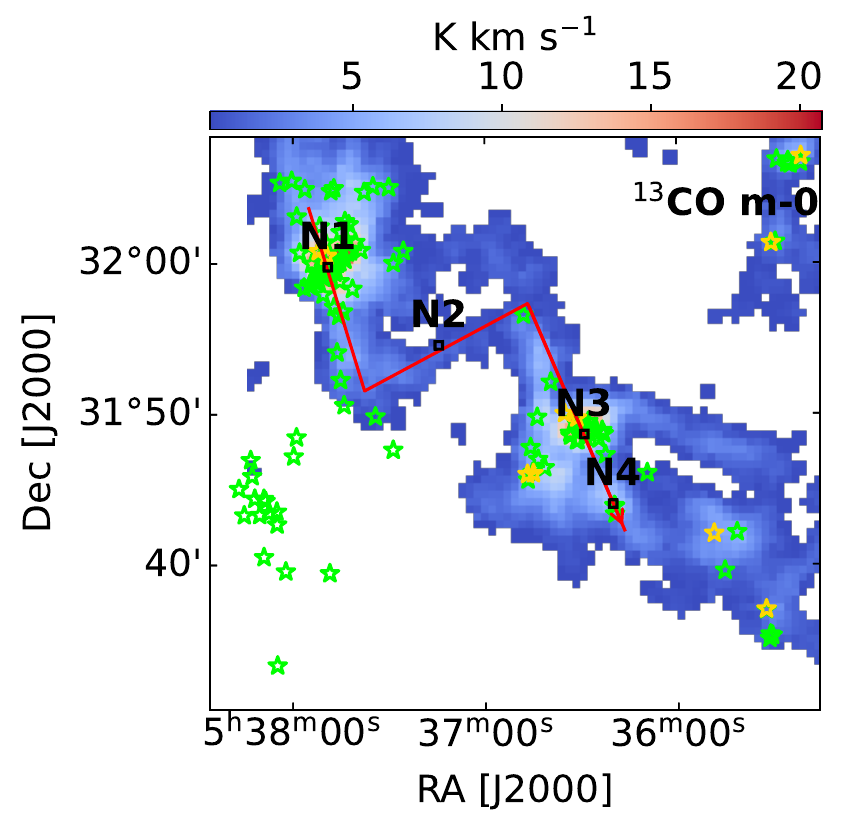}
\end{minipage}
\begin{minipage}{0.65\textwidth}
\includegraphics[width=\textwidth]{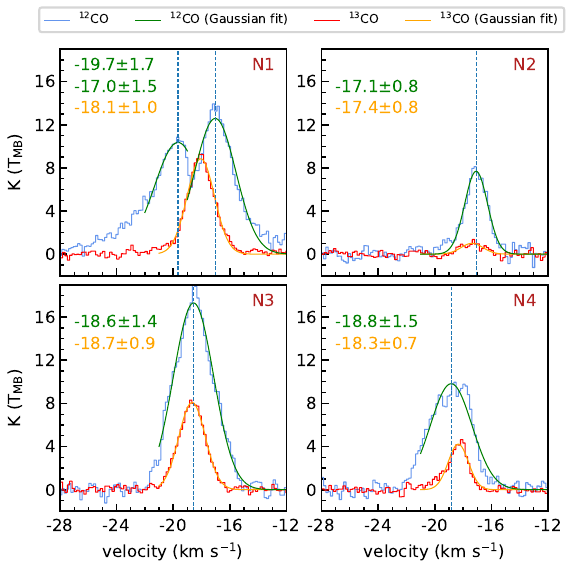}
\end{minipage}
\caption{ Upper left panel: $^{12}$CO$(J=1-0)$ m-0, and lower left Panel: $^{13}$CO$(J=1-0)$ m-0 maps overlaid with elected locations (N1 - N4) to extract spectra and red colored path to extract the PV diagram in the velocity range [$-22$, $-12$] \kms\, for northern molecular cloud along with YSOs (Class I (yellow asterisk) and Class II (green asterisk)). Right Panel: $^{12}$CO$(J=1-0)$ (blue color) and $^{13}$CO$(J=1-0)$ (red color) spectra at the selected locations marked in the left panel. The green and yellow curves represent the Gaussian fit/s to these spectra. The vertical dashed (blue) line represents the $^{12}$CO Gaussian fit/s peak.}
\label{fig:co_spectra_afgl_f807}
\end{figure*}

Probing the position-velocity (PV) diagrams is another observational aspect that can be used to examine the association of MCs in terms of velocity. We selected $^{13}$CO spectral data cube to generate PV diagrams since it depicts the filamentary nature more remarkably. Figure \ref{fig:pv_afgl_f807} represents the PV diagram along the path shown in the upper and lower left panels of Figure \ref{fig:co_spectra_afgl_f807} with a red color. These panels depict the $^{12/13}$CO m-0 maps in the velocity range [$-22$, $-12$] \kms. We chose this path to check the probable connection of AFGL 5157 and FSR0807. 
This PV diagram indicates a velocity gradient and a large spread in velocity at locations N1 and N3. N1 and N3 are connected by a structure, which, though not of uniform velocity, clearly shows a connection between N1 and N3 in velocity space. Apart from this, it indicates the existence of two clumps in FSR0807 that are also indicated in the MST analysis (cf. Section \ref{sec:mst}). A faint connecting feature is visible between AFGL 5157 and FSR0807. 

\begin{figure}[!ht]
    \includegraphics[width=0.44\textwidth]{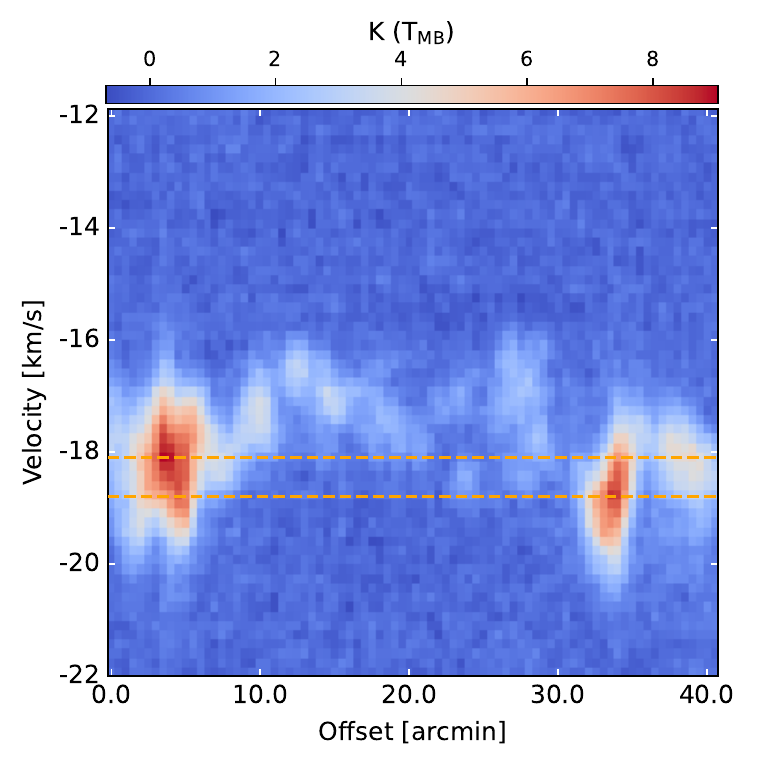}
    \caption{PV diagram generated using $^{13}$CO emission, along the path (red vectors) overlaid in Figure \ref{fig:co_spectra_afgl_f807} for the northern molecular structure. The orange straight lines represent the prominent peak $^{13}$CO emission of $-$18.1 \kms and $-$18.7 \kms in the direction of locations N1 and N3, respectively (see Figure \ref{fig:co_spectra_afgl_f807})}.
    \label{fig:pv_afgl_f807}
\end{figure}



The left, middle, and right panels of Figure \ref{fig:rgb_images_afgl_f807} depict the Herschel column density, temperature, and Spitzer ratio (4.5 $\mu$m/3.6 $\mu$m emission) maps for AFGL 5157 (top panels) and FSR0807 (bottom panels). The Herschel column and temperature density maps were procured for EU-funded ViaLactea project \citep{2010PASP..122..314M} adopting the Bayesian PPMAP technique on the Herschel data \citep{2010A&A...518L.100M} at 70, 160, 250, 350, and 500 $\mu$m wavelengths \citep{2015MNRAS.454.4282M, 2017MNRAS.471.2730M}. The Spitzer ratio map (4.5 $\mu$m$/$3.6 $\mu$m emission) is smoothened using a Gaussian function of two-pixel radius (for more details, please refer \citealt{2017ApJ...845...34D}). This map has some bright and dark regions. The 4.5 $\mu$m (brighter regions) is due to prominent Br-$\alpha$ emission at 4.05 $\mu$m and molecular hydrogen line at 4.693 $\mu$m and traces the outflow activities, whereas 3.6 $\mu$m (darker region) is due to the PAH emission at 3.3 $\mu$m, indicative of the presence of photo-dissociation region (PDR). 

The column density map reveals the presence of dominant central high column density regions at AFGL 5157 and FSR0807, along with several filamentary sub-structures. These central regions and structures coincide with the locations of YSOs. The central high column density region is also associated with warmer dust emission (i.e., $T_d \sim$ 17 - 18 K), whereas the filamentary sub-structures are associated with a bit colder dust emission. The central high column density region showing warmer dust emission is also surrounded by PDRs, suggesting the presence of feedback from the massive star$/$s or recent star formation activities in these regions.

\begin{figure*}[!ht]
    \centering
    \includegraphics[width=0.32\textwidth]{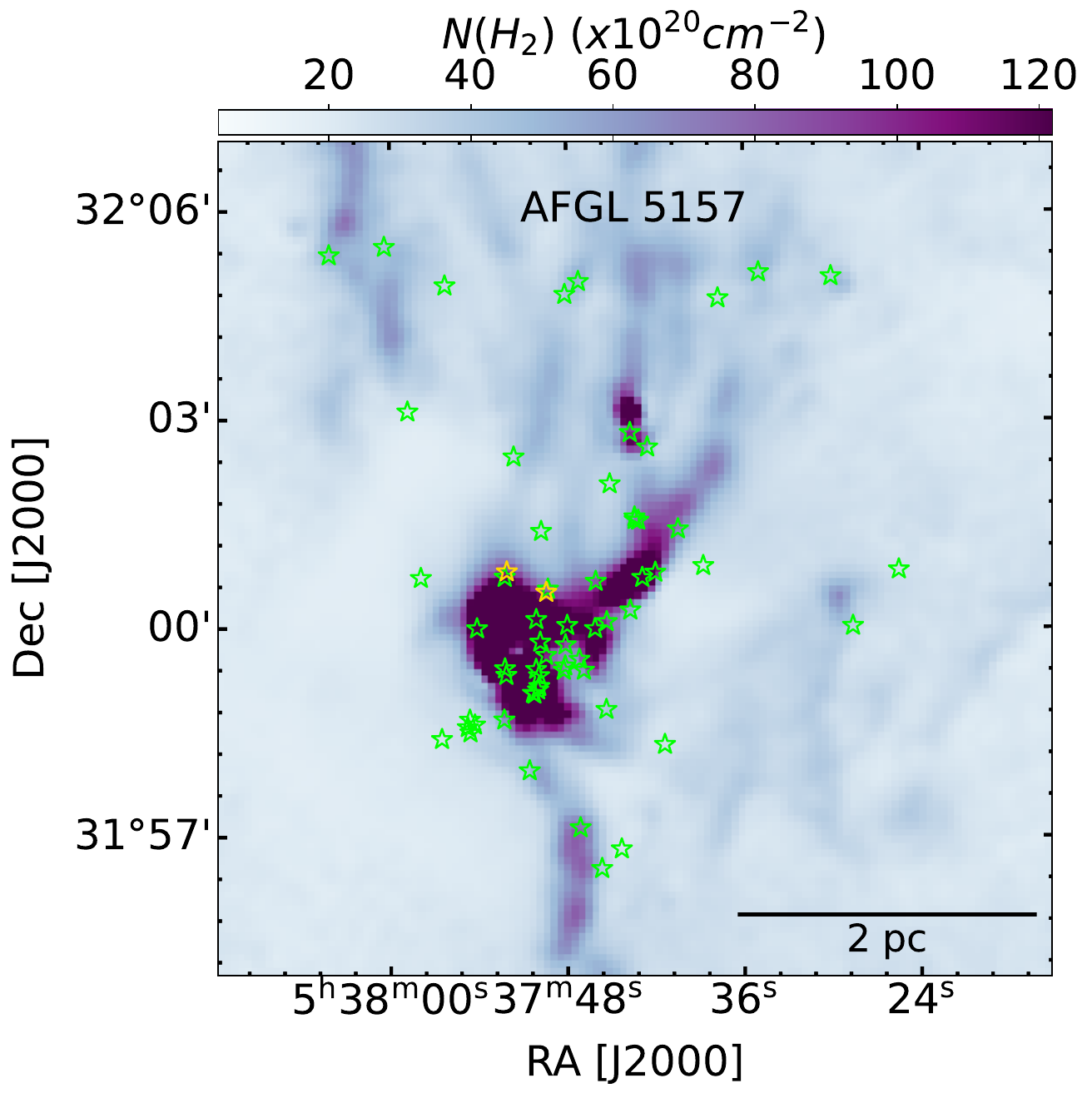}
    \includegraphics[width=0.32\textwidth]{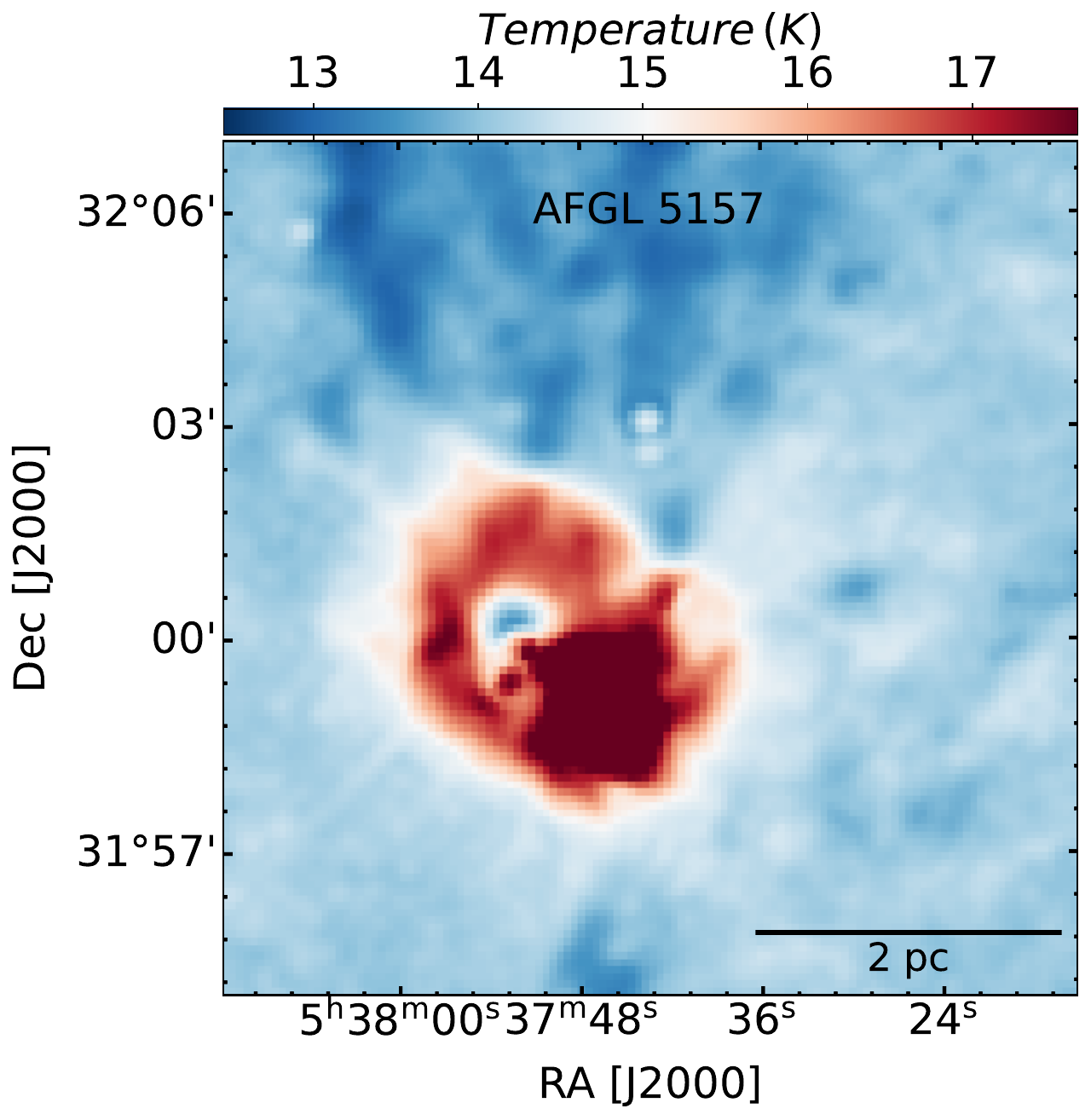}
    \includegraphics[width=0.32\textwidth]{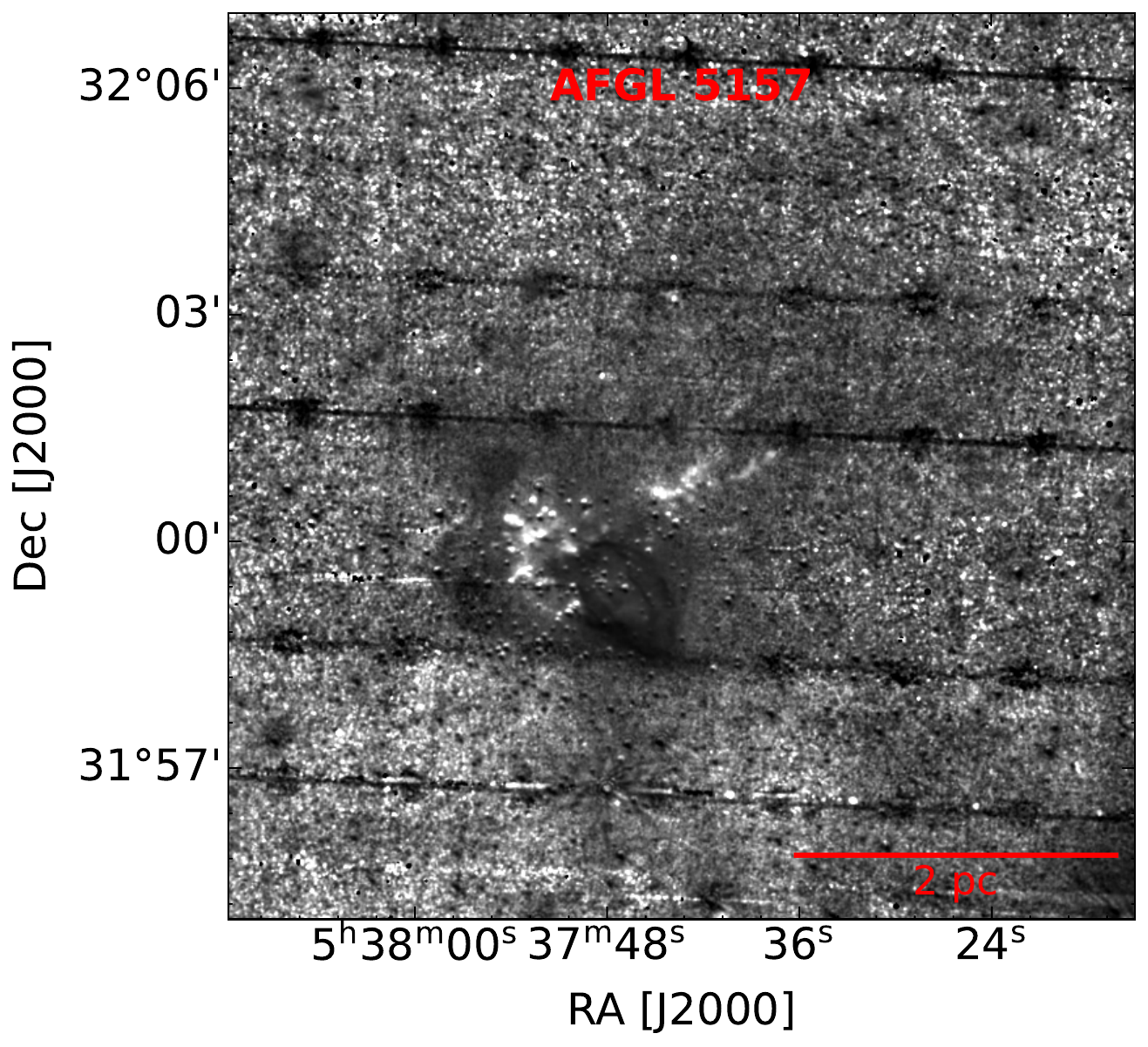}
    \includegraphics[width=0.32\textwidth]{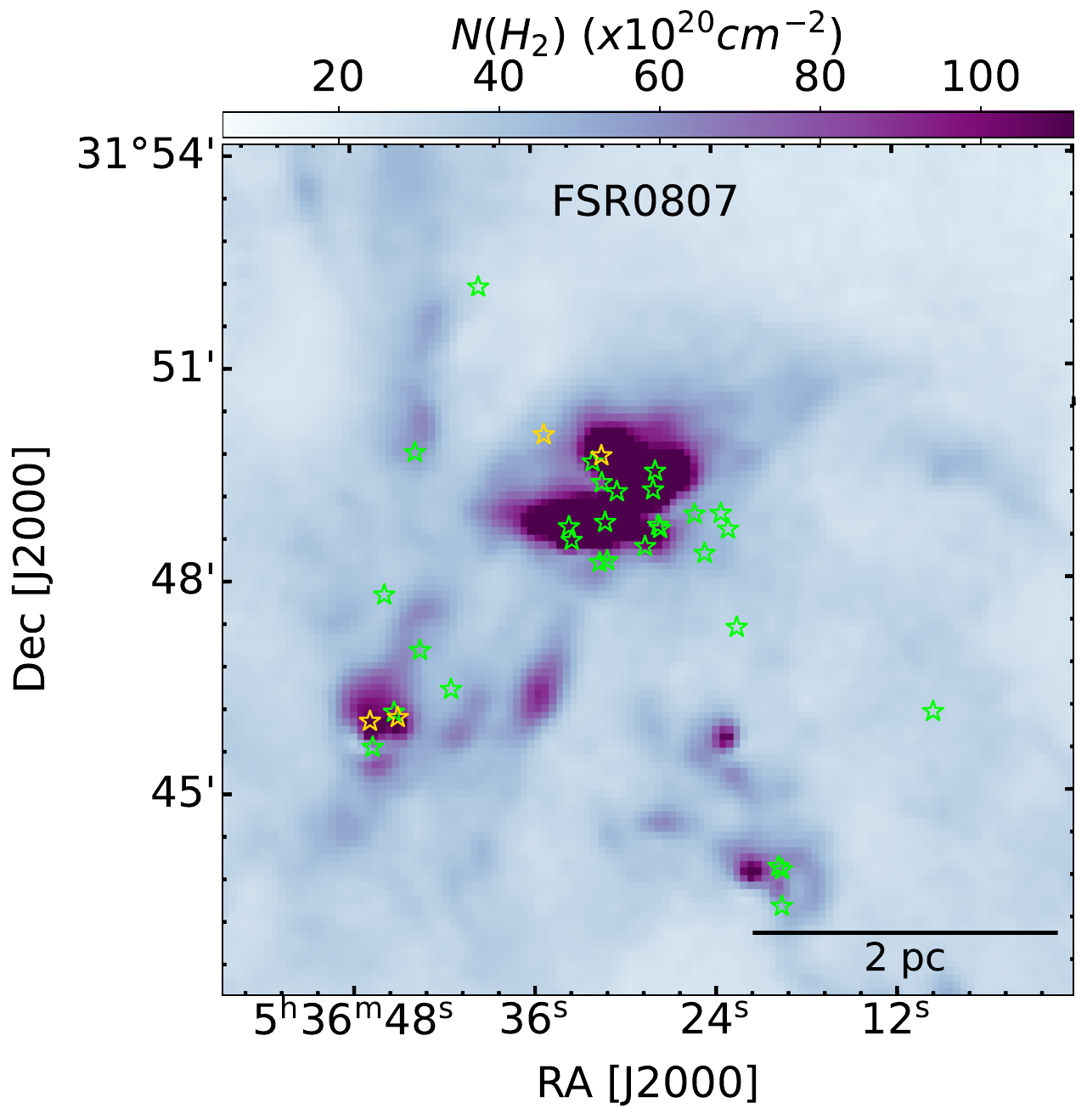}
    \includegraphics[width=0.32\textwidth]{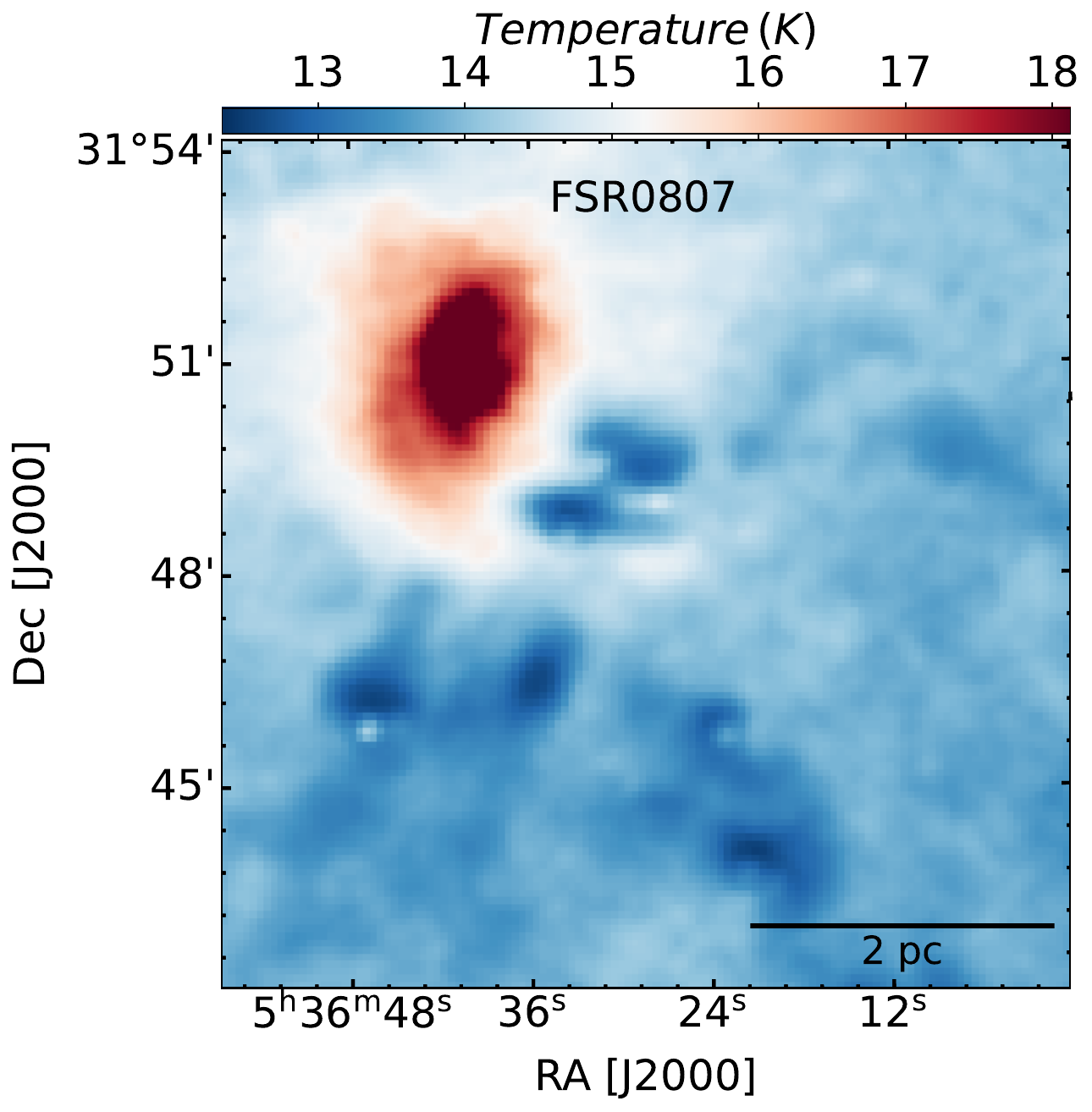}
    \includegraphics[width=0.32\textwidth]{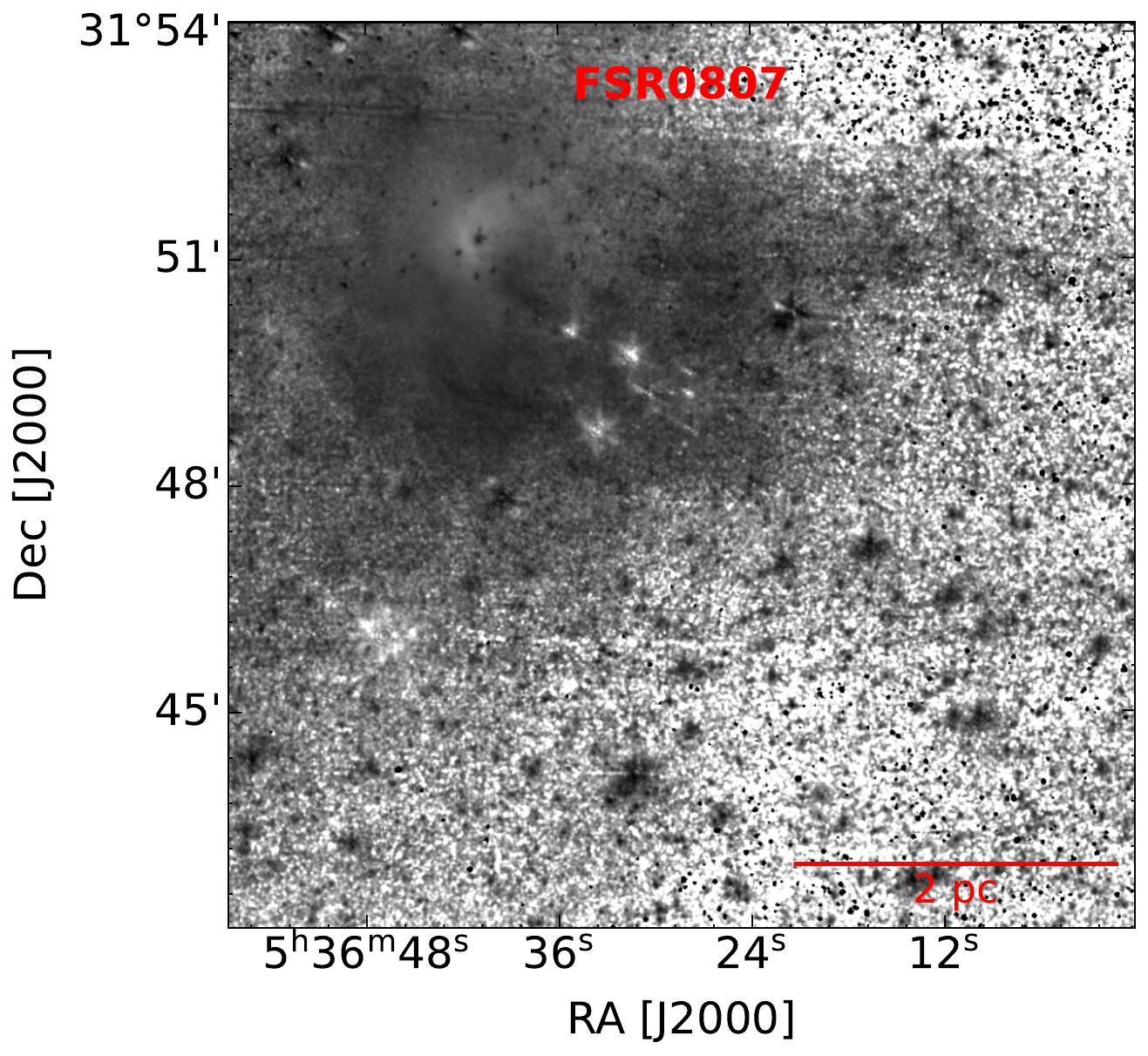}
    \caption{Left panels: Herschel column density map overlaid with the locations of Class I (yellow asterisk) and Class II (green asterisk) YSOs. Middle panels: Herschel temperature map. Right Panels: Spitzer ratio map (4.5 $\mu$m$/$3.6 $\mu$m emission) for AFGL 5157 (upper panels) and FSR0807 (bottom panels).}
    \label{fig:rgb_images_afgl_f807}
\end{figure*}

\subsubsection{The Southern Molecular Cloud: KPS0620 and IRAS 05331$+$3115}\label{sec:k620}

Figure \ref{fig:moment_maps_-4_6} depict the $^{12}$CO and $^{13}$CO m-0, m-1, and linewidth maps \citep{2023ApJ...944..228M,2023JApA...44...34M,2024MNRAS.527.9626P} for the KPS0620 in the velocity range [$-4$, 6] \kms. The inspection of $^{12}$CO and $^{13}$CO m-0 maps (cf. panels (a) and (d) of Figure \ref{fig:moment_maps_-4_6}, respectively) reveals the remarkable hub-filamentary system (HFS) of KPS0620. The presence of HFS is also supported by the dense molecular emission at the center of this system ($\sim$10$^{22}$\,cm$^{-2}$, upper-left panel of Figure \ref{fig:rgb_images_k620}). This kind of dense molecular emission is observed in the ``hubs'' of the HFS \citep{2009ApJ...700.1609M}. The filament in the northern direction is less prominent than that in the southern and western directions of the hub/junction, where an enhancement in the intensity is observed in both the $^{12}$CO and $^{13}$CO m-0 maps (cf. panels (a) and (d) of Figure \ref{fig:moment_maps_-4_6}, respectively). The $^{12}$CO and $^{13}$CO linewidth maps (cf. panels (c) and (f) of Figure \ref{fig:moment_maps_-4_6}, respectively) suggest that there is an existence of large velocity dispersion ($\sim$4 \kms) at the KPS0620 hub.

\begin{figure*}[!ht]
    \centering
    \includegraphics[width=0.99\textwidth]{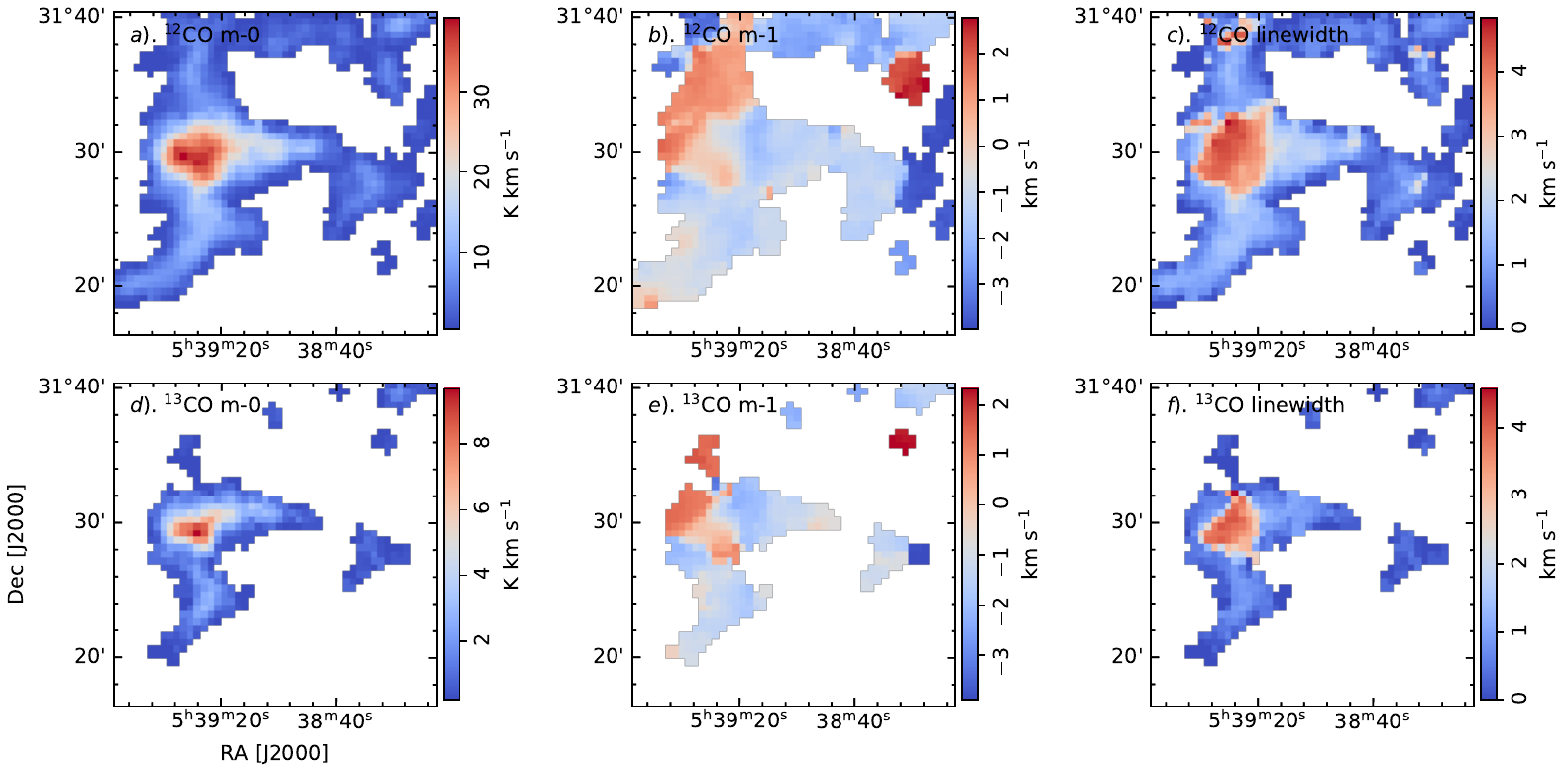}
    \caption{m-0, m-1 and linewidth maps (column-wise) for $^{12}$CO$(J=1-0)$ and $^{13}$CO$(J=1-0)$ emission, respectively in the velocity range [$-4$, 6] \kms\, for KPS0620. The emission is depicted above 5$\sigma$ value ($\sigma$ being the rms noise for respective spectral cubes).}
    \label{fig:moment_maps_-4_6}
\end{figure*}

Furthermore, we selected locations S1 - S4, marked in the upper and lower left panels of Figure \ref{fig:co_spectra_k620}, to examine the filamentary structure of KPS0620. We determined the average spectra with Gaussian fitting using the $^{12/13}$CO spectral cubes. These spectra are shown in the right panel of Figure \ref{fig:co_spectra_k620}, and the derived Gaussian peaks are also written in the figure. The locations S1 (associated with the hub of KPS0620) and S2 have two velocity peaks ($\sim -2$ \kms, blue-shifted and $\sim 1$ \kms, red-shifted). In contrast, locations S3 and S4 (along the filament to the south and west of the hub) have only a single velocity peak in both $^{12/13}$CO spectra. Similar to location N1 (refer Section \ref{sec:afgl_f807}), S2 also displays a large `red profile', both in optically thick $^{12}$CO and optically thin $^{13}$CO lines. So, though we could not find any physical feature in the optically thinner $^{18}$CO transition, we looked for it particularly at this location to check the outflow motion. We found a single-peaked profile, suggesting outflow motion at this location.


\begin{figure*}[!ht]
\begin{minipage}{0.35\textwidth}
\includegraphics[width=\textwidth]{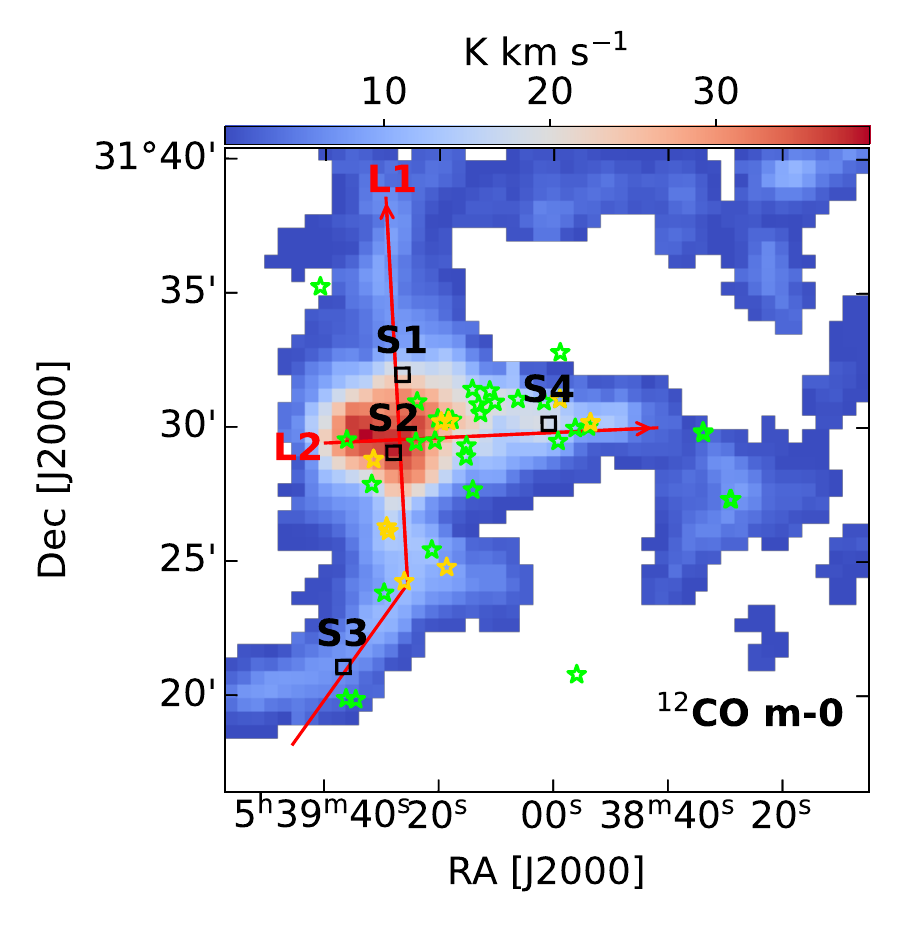}
\\[2mm]
\includegraphics[width=\textwidth]{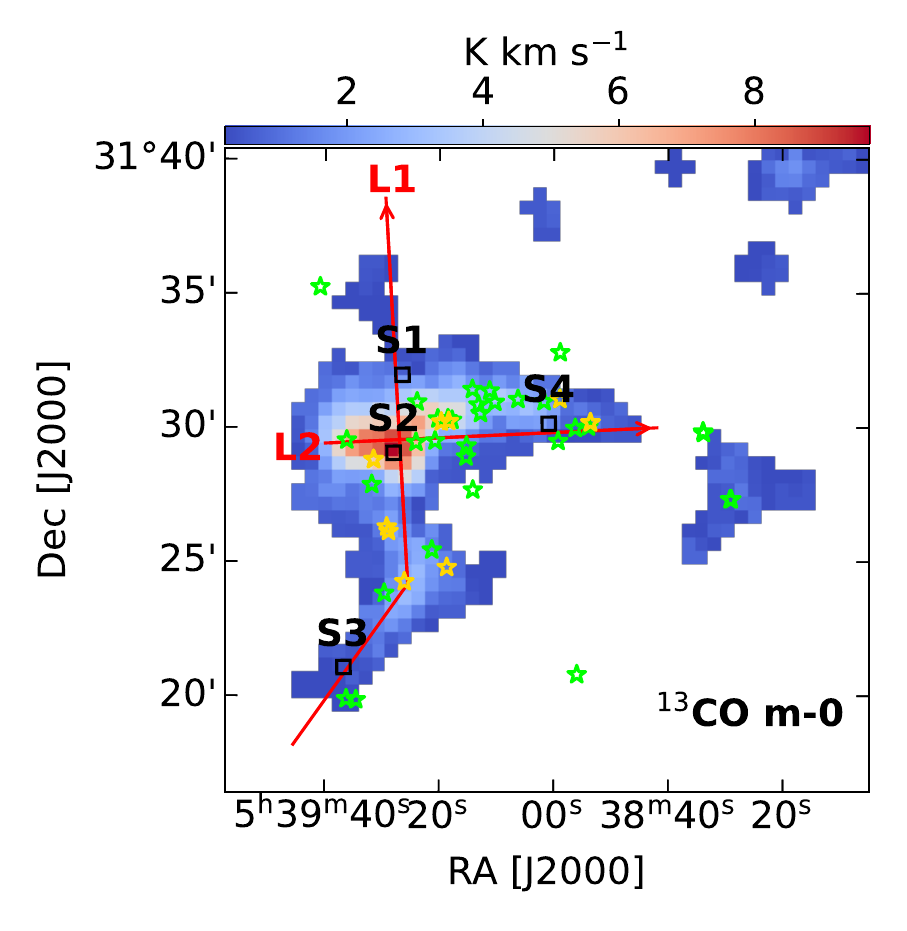}
\end{minipage}
\begin{minipage}{0.65\textwidth}
\includegraphics[width=\textwidth]{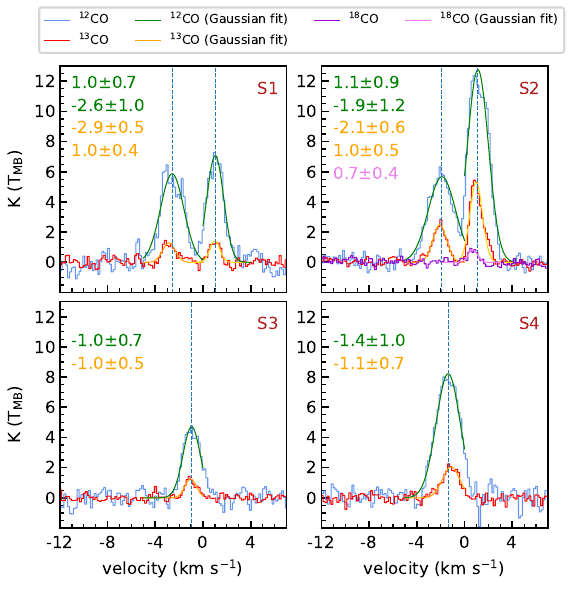}
\end{minipage}
\caption{ Upper left panel: $^{12}$CO$(J=1-0)$ m-0, and
lower left panel: $^{13}$CO$(J=1-0)$ m-0 maps overlaid with elected locations (S1 - S4) to extract spectra and red colored path to extract the PV diagram in the velocity range [$-4$, 6] \kms\, for KPS0620 along with YSOs (Class I (yellow asterisk) and Class II (green asterisk)). Right panel: $^{12}$CO$(J=1-0)$ (blue color) and $^{13}$CO$(J=1-0)$ (red color) spectra at the selected locations marked in the left panel. The green and yellow curves represent the Gaussian fit/s to these spectra. The vertical dashed (blue) line represents the $^{12}$CO Gaussian fit/s peak.}
     \label{fig:co_spectra_k620}
\end{figure*}

To scrutinize the gas flow/kinematics in KPS0620, we show the PV diagram in Figure \ref{fig:pv_spectra_k620}. The path L1 and L2, chosen to map the PV diagram, are shown in the upper and lower left panels of Figure \ref{fig:co_spectra_k620} depicting the $^{12/13}$CO m-0 maps in the velocity range [$-4$, 6] \kms. The PV diagram (generated using $^{13}$CO emission) reveals an appreciable gradient and the presence of two velocity peaks at the hub, and this existence is also confirmed by the spectra (shown in the right panel of Figure \ref{fig:co_spectra_k620}) at location S2 in Figure \ref{fig:co_spectra_k620}.
The velocity gradient in the PV slice points towards the gas, spiraling into the central region.

\begin{figure}[!ht]
    \includegraphics[width=0.48\textwidth]{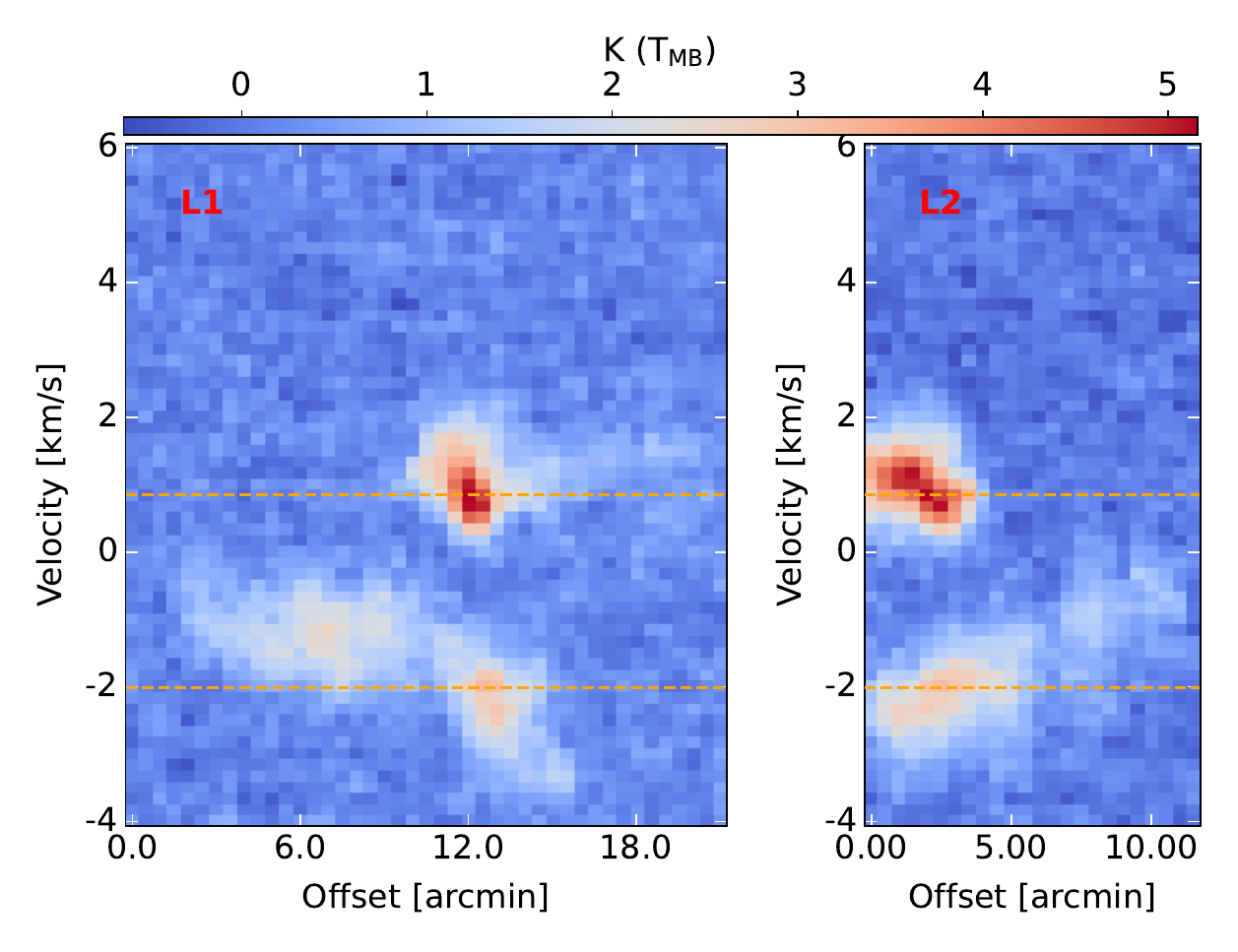}
    \caption{PV diagram generated using $^{13}$CO emission, along the path (red vectors) overlaid in Figure \ref{fig:co_spectra_k620} for the southern molecular structure. In the direction of location S2 (see Figure \ref{fig:co_spectra_k620}), at least two velocity peaks are seen, which are highlighted by straight orange lines.}
    \label{fig:pv_spectra_k620}
\end{figure}

Figure \ref{fig:moment_maps_-4_6_iras} depicts the $^{12}$CO and $^{13}$CO m-0, m-1, and linewidth maps \citep{2023ApJ...944..228M,2023JApA...44...34M} for the IRAS 05331$+$3115 in the velocity range [$-4$, 6] \kms. The inspection of $^{12}$CO and $^{13}$CO m-0 maps (cf. panels (a) and (d) of Figure \ref{fig:moment_maps_-4_6_iras}, respectively) reveals
    it has a dense core with a peak velocity $\sim4.3$ \kms. This peak velocity is determined from the $^{12}$CO and $^{13}$CO m-0 maps (cf. panels (b) and (e) of Figure \ref{fig:moment_maps_-4_6_iras}, respectively). This dense core is also associated with larger linewidth dispersion (cf. panels (c) and (f) of Figure \ref{fig:moment_maps_-4_6_iras}).

\begin{figure*}[!ht]
    \centering
    \includegraphics[width=0.99\textwidth]{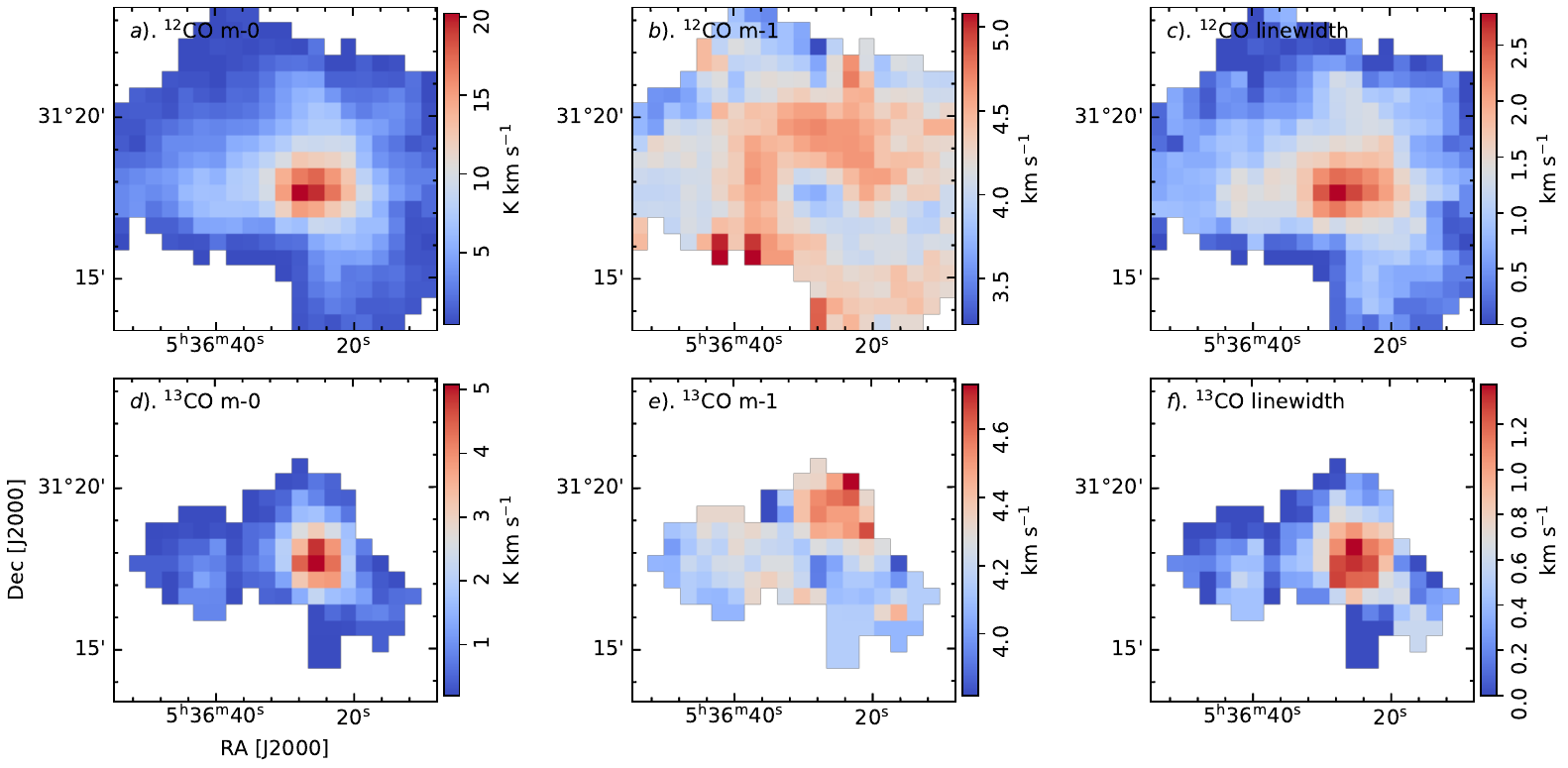}
    \caption{m-0, m-1 and linewidth maps (column-wise) for $^{12}$CO$(J=1-0)$ and $^{13}$CO$(J=1-0)$ emission, respectively in the velocity range [$-4$, 6] \kms\, for IRAS 05331$+$3115. The emission is depicted above 5$\sigma$ value ($\sigma$ being the rms noise for respective spectral cubes).}
    \label{fig:moment_maps_-4_6_iras}
\end{figure*}

The left, middle, and right panels of Figure \ref{fig:rgb_images_k620} depict the Herschel column density, temperature, and Spitzer ratio (4.5 $\mu$m/3.6 $\mu$m) maps, for KPS0620 (upper panels) and IRAS 05331$+$3115 (bottom panels). These maps indicate the existence of cold filamentary structures and a ``hub'' with higher molecular content in KP0620. These structures also host YSOs. The  IRAS 05331$+$3115 also shows a compact, high column density but no filamentary structure. The YSOs are mainly associated with this clump. The Spitzer ratio map does not show the distribution of PDR or outflow activities for both regions. 

\begin{figure*}[!ht]
    \centering
    \includegraphics[width=0.32\textwidth]{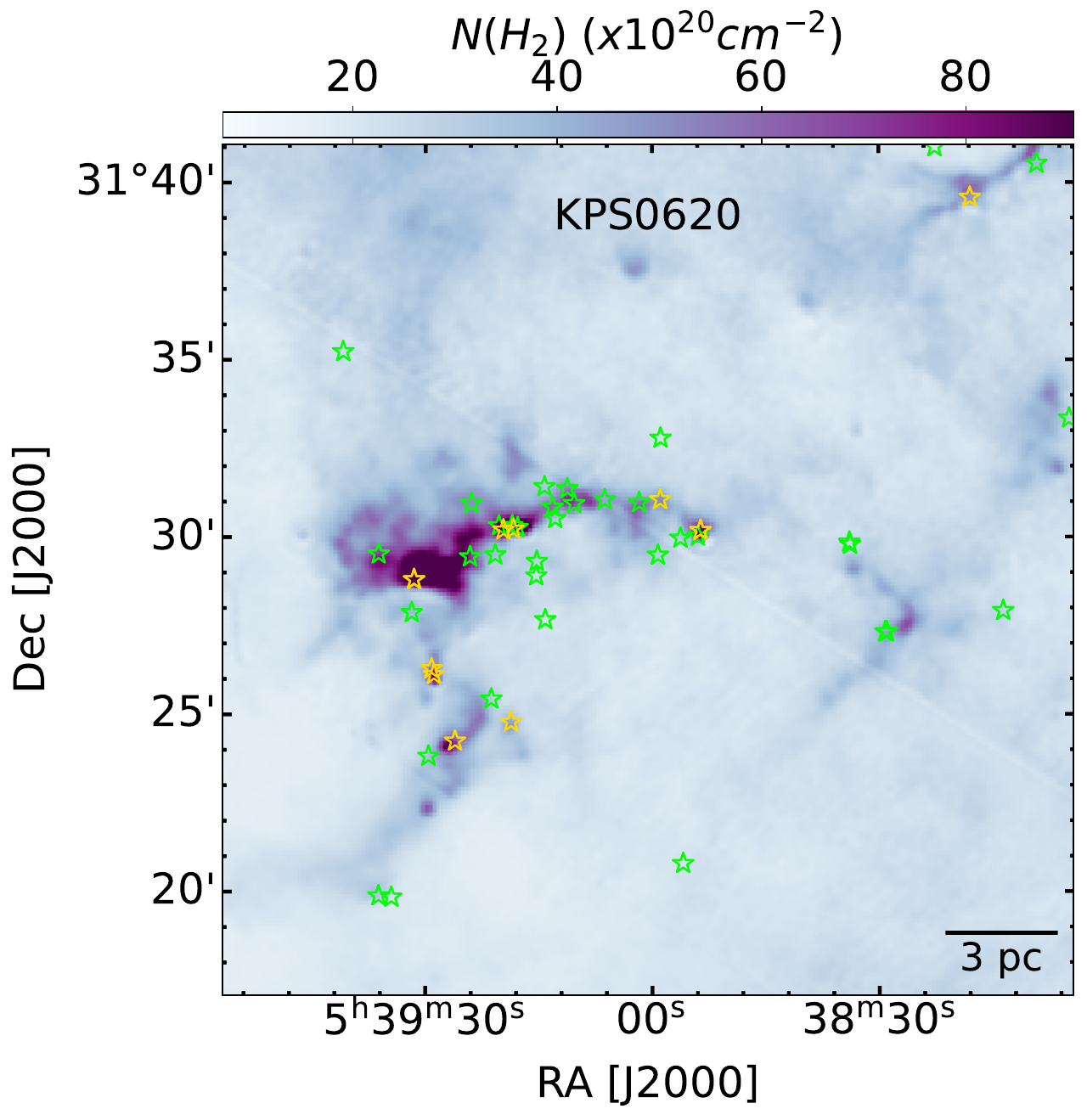}
    \includegraphics[width=0.32\textwidth]{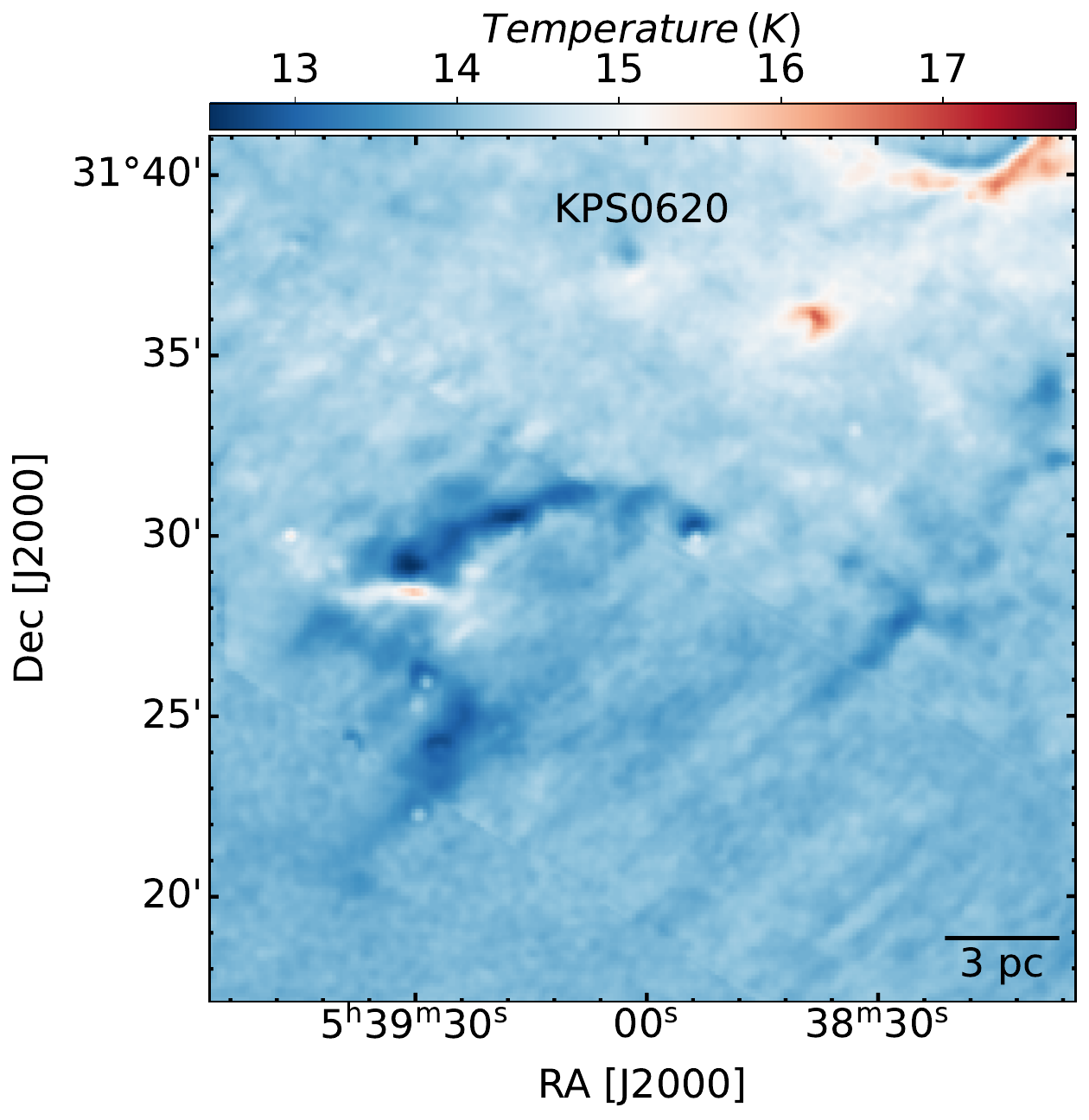}
    \includegraphics[width=0.32\textwidth]{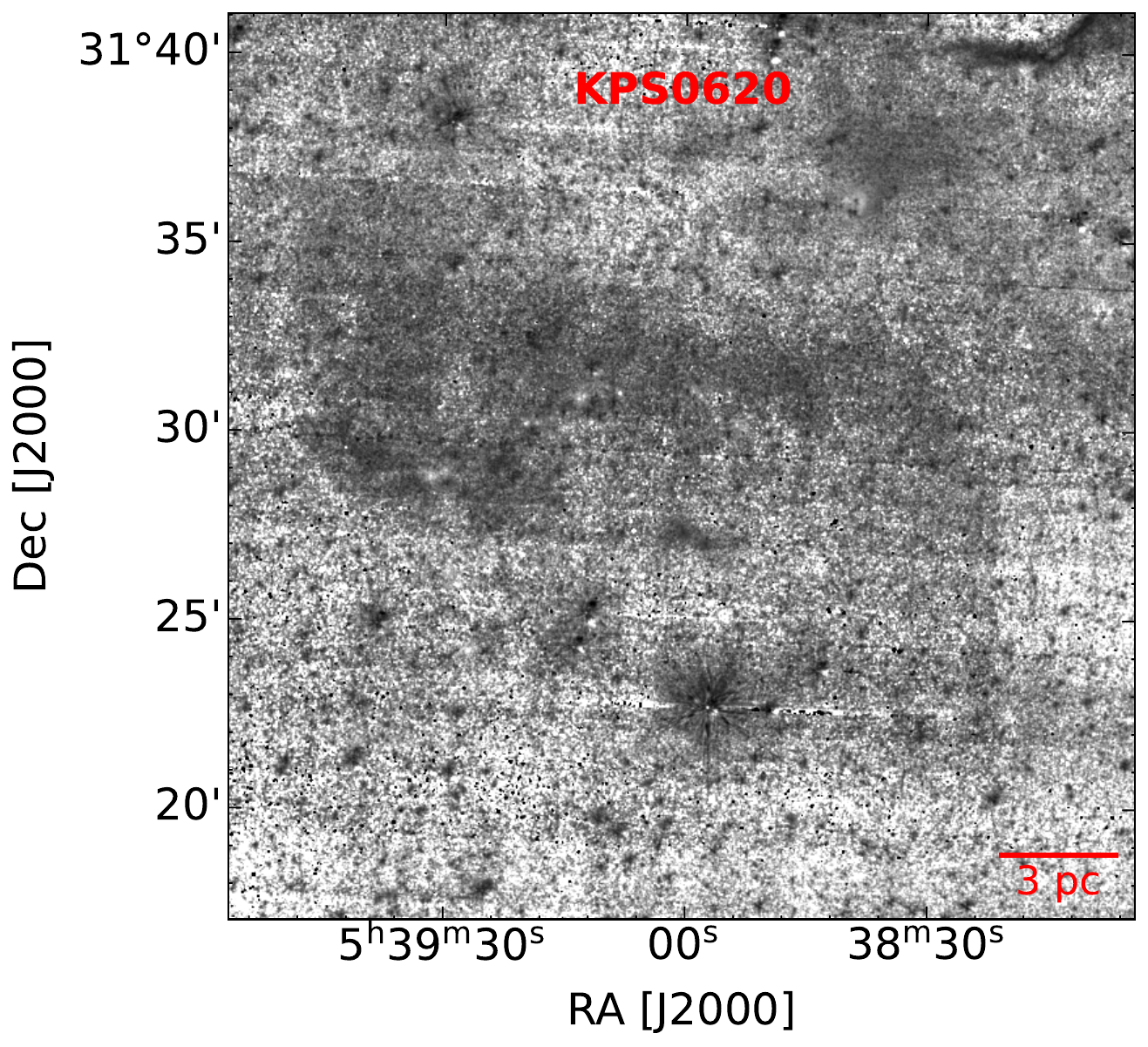}
    \includegraphics[width=0.32\textwidth]{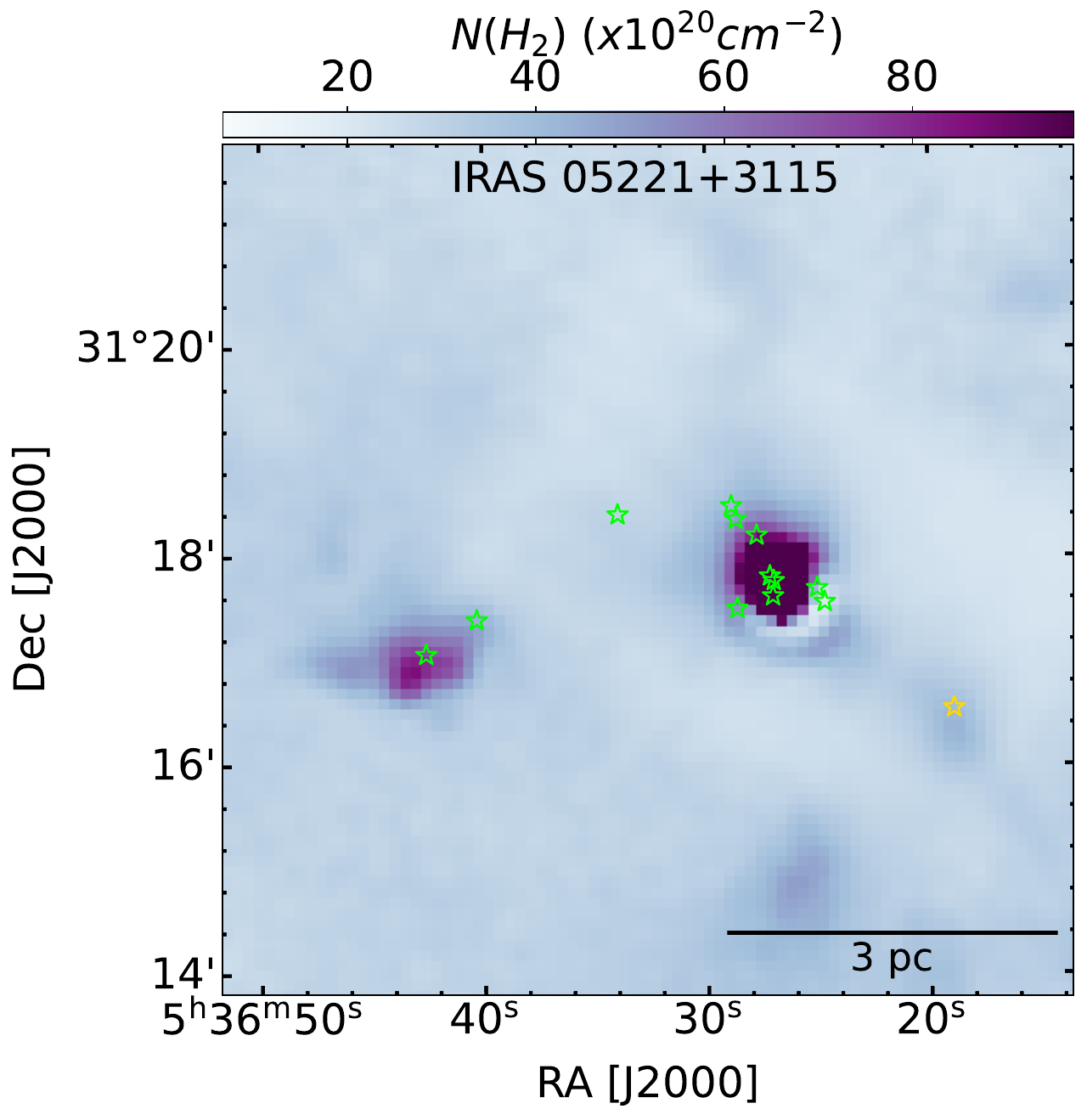}
    \includegraphics[width=0.32\textwidth]{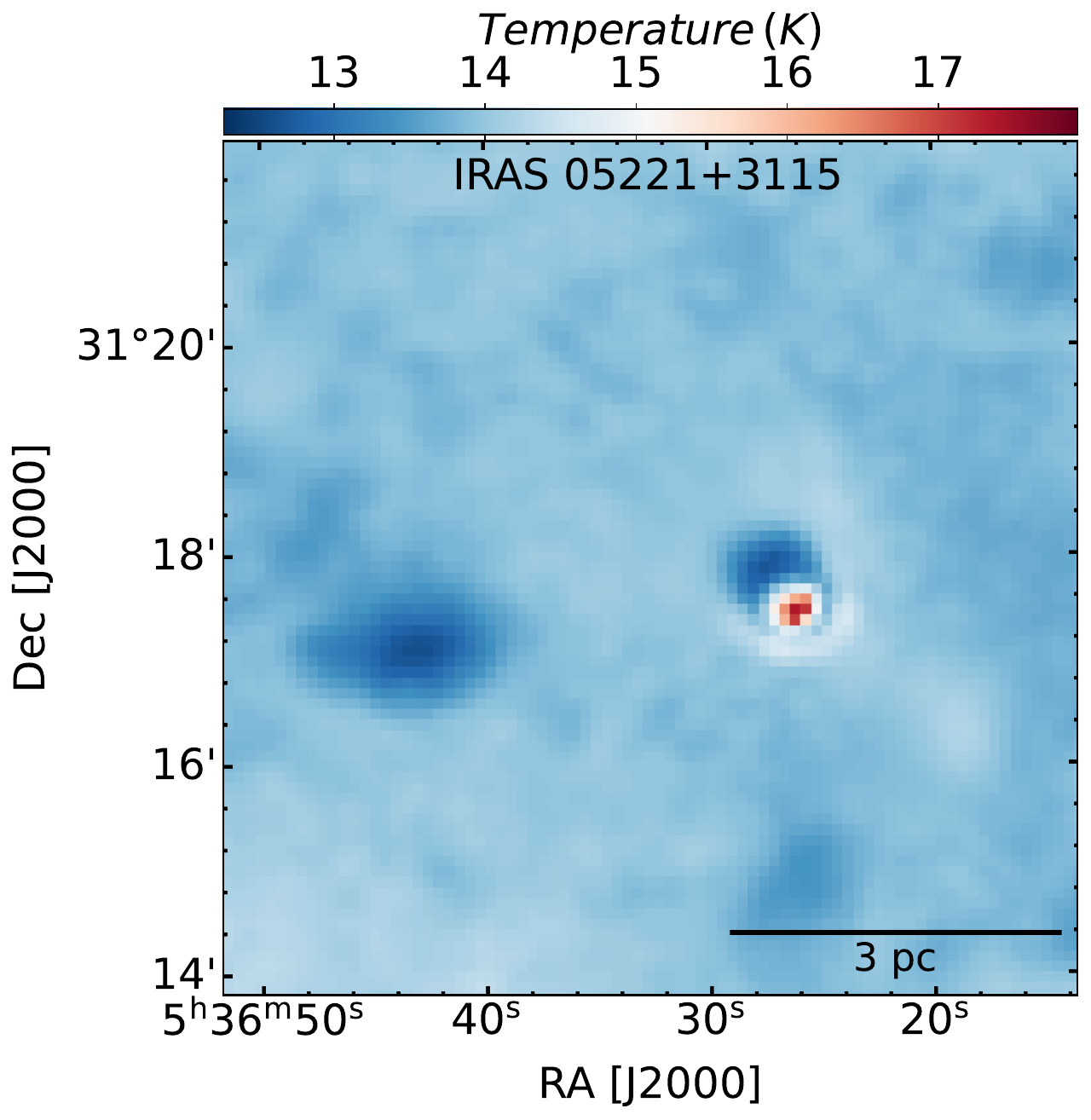}
    \includegraphics[width=0.32\textwidth]{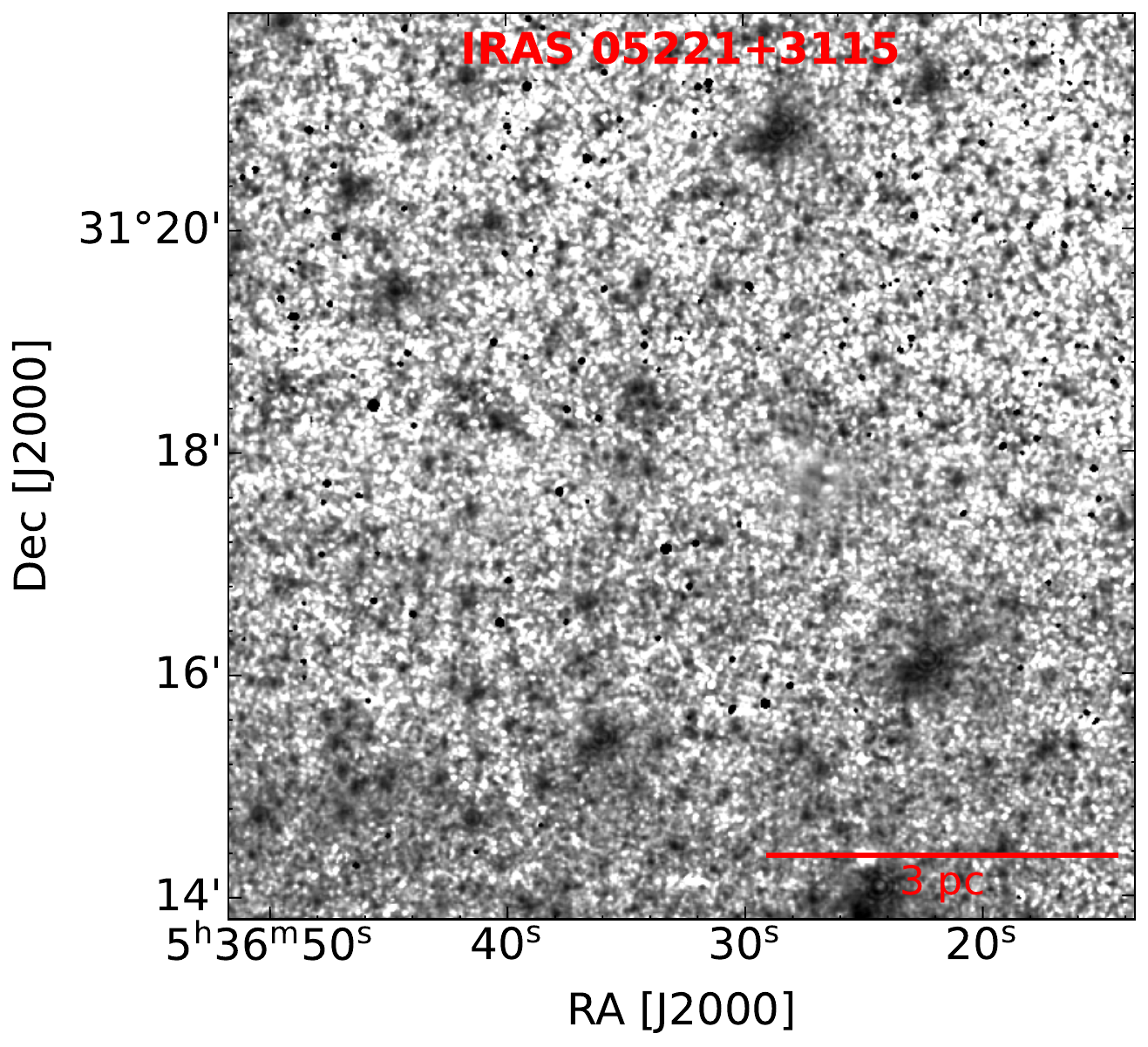}
    \caption{Left panels: Herschel column density map overlaid with the locations of Class I (yellow asterisk) and Class II (green asterisk) YSOs. Middle panels: Herschel temperature map. Right Panels: Spitzer ratio map (4.5 $\mu$m$/$3.6 $\mu$m emission) for KPS0620 (upper panels) and IRAS 05331$+$3115 (bottom panels).}
    \label{fig:rgb_images_k620}
\end{figure*}

\subsection{Physical Parameters of the Stellar Clustering in the Selected Target Area}

\begin{figure*}[!ht]
    \centering
    \includegraphics[width=0.48\textwidth]{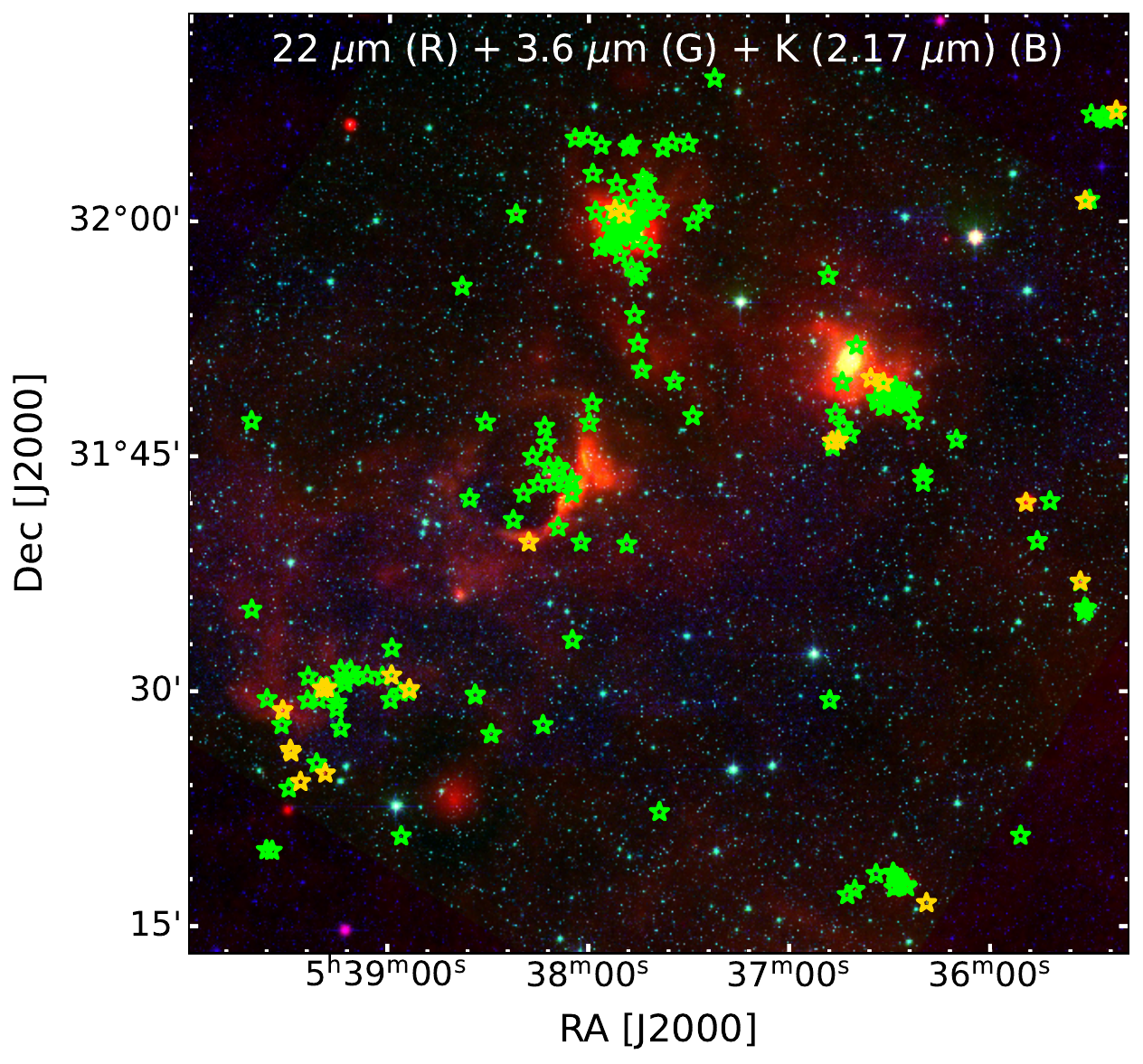}
    \includegraphics[width=0.48\textwidth]{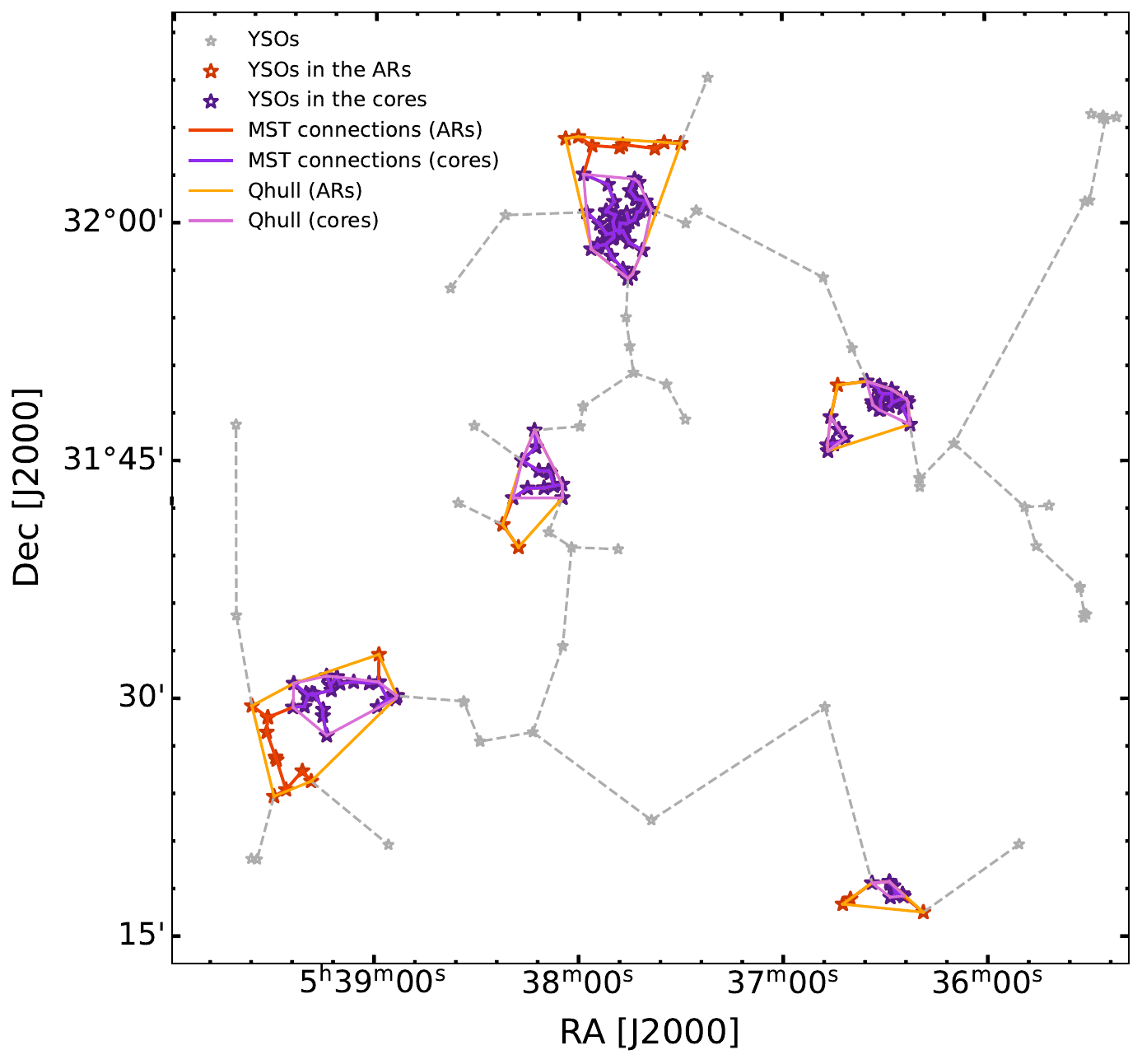}
    \includegraphics[width=0.48\textwidth]{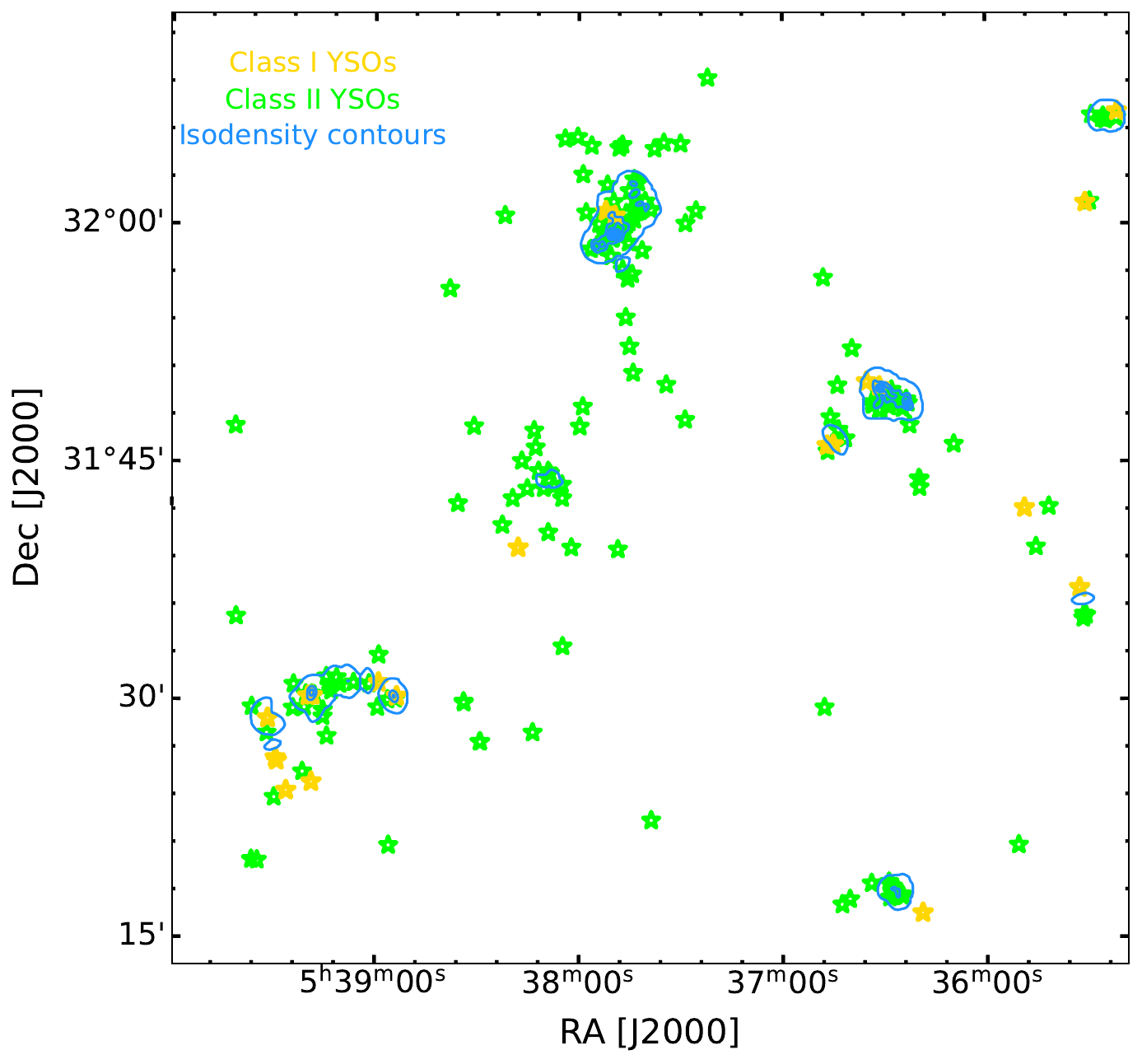}
    \includegraphics[width=0.48\textwidth]{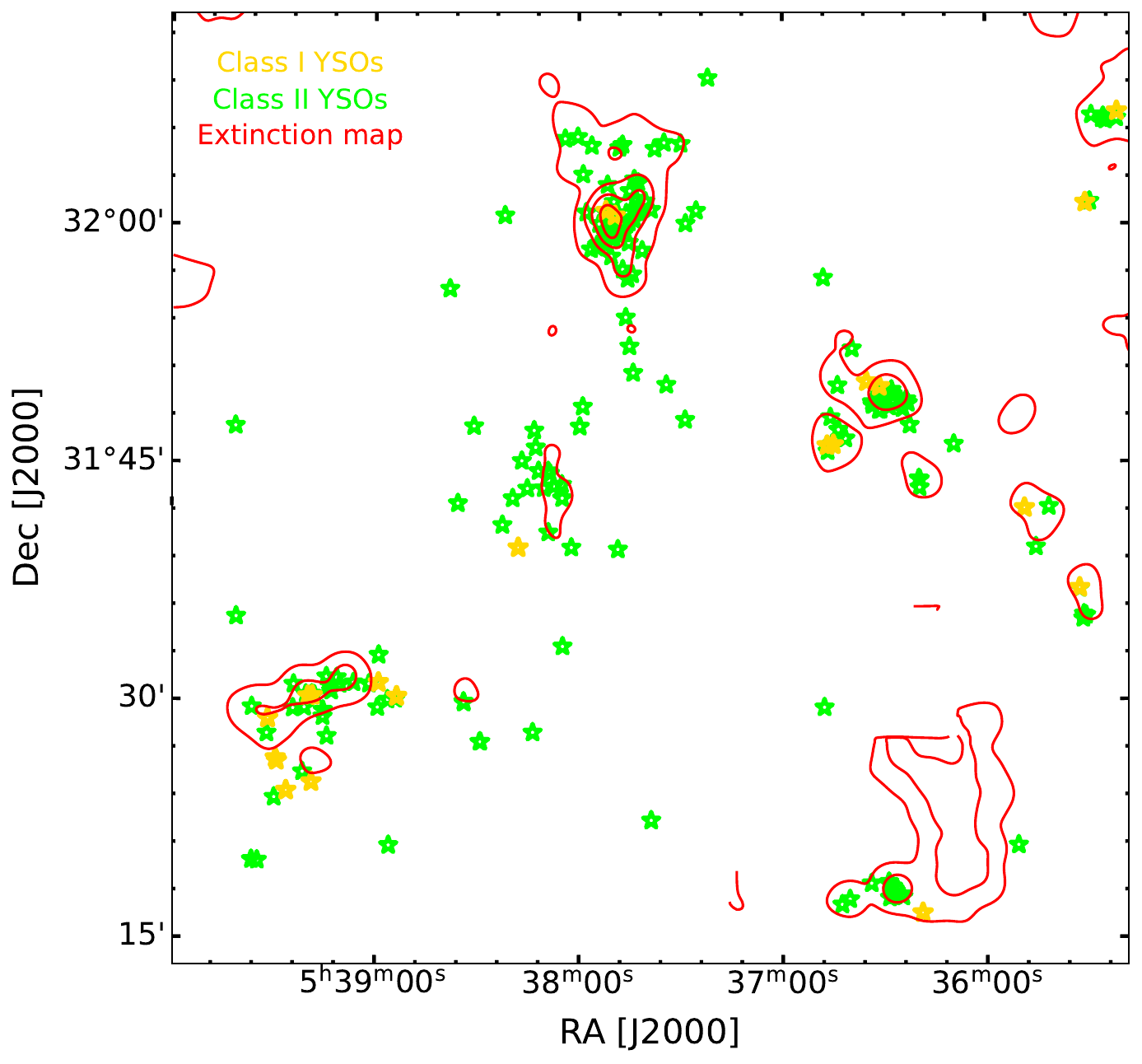}
     \caption{Upper-left panel: Color-composite image (Red: WISE 12 $\mu$m; Green: Spitzer 3.6 $\mu$m, and Blue: 2MASS K-band 2.17 $\mu$m) of the region overlaid with the locations of identified YSOs (Class I (yellow asterisk) and Class II (green asterisk)). Upper-right panel: Representation of MST connections in the cluster (orange) and core (purple) along with the locations of identified YSOs (orange asterisks for the YSOs in the cores and purple asterisks for the locations of YSOs in clusters). The extracted clusters/cores are also enclosed by the \emph{Convex hulls} using yellow and purple line segments, respectively. Black asterisks also mark the locations of all the identified YSOs.
     Lower-left panel: Surface density map of all stellar sources. Lower-right panel: Extinction map overlaid with the locations of identified YSOs (Class I (yellow asterisk) and Class II (green asterisk)). The lowest contour level for the generation of the surface density map is 1 YSO arcmin$^{-2}$ with a step size of 29.5 stars arcmin$^{-2}$ whereas for the extinction map, it is 2.7 mag with a step size of 1.92 mag.}   
    \label{fig:intro_rgb}
\end{figure*}

The upper-left panel of Figure \ref{fig:intro_rgb} shows the color-composite image generated using WISE 22 $\mu$m, Spitzer 3.6 $\mu$m and 2MASS K (2.17 $\mu$m) band images of the region overlaid with the identified YSOs. The WISE 22 $\mu$m indicates the distribution of warm gas, whereas WISE 12 $\mu$m indicates PAH featuring at 11.3 $\mu$m. Furthermore, the MIR emission in 3.6 $\mu$m wavelength depicts the distribution of gas and dust. All these features show solid evidence of star formation activities in these regions.  It is easily visible that most YSOs are associated with the MIR emissions and the target star-forming sites. However, some Class II YSOs are randomly distributed. This section will briefly study the physical properties of the young star clusters embedded in this region.


\subsubsection{Embedded Young Population}\label{sec:mst}

To isolate and quantify young and embedded stellar clustering in our selected region, we applied an empirical technique \emph{`Minimal Spanning Tree'} (MST; \citealt{2009ApJS..184...18G,2014MNRAS.439.1765K}) to the 204 YSOs identified based on excess infrared (IR) emission using the Spitzer \citep{2009ApJS..184...18G}, 2MASS-UKIDSS \citep{2004ApJ...608..797O}, and WISE \citep{2014ApJ...791..131K}  data \citep[for details, please refer,][]{2023JApA...44...52V,2023ApJ...953..145V,2016AJ....151..126S}. 
It is among the best techniques to isolate the clusterings without any bias or smoothening and prevents underlying geometry \citep{2004MNRAS.348..589C,2009ApJS..184...18G, 2016AJ....151..126S, 2020ApJ...891...81P}. The extracted MST is plotted in the upper-right panel of Figure \ref{fig:intro_rgb}, which indicates five prominent stellar groups/clustering in the region; the distribution of YSOs has different concentrations in different groups/clustering. The asterisks and line segments in different colors represent the YSOs and the MST branches, respectively. We then estimated the critical branch lengths to isolate the YSOs associated with the active regions (or cluster regions) and their cores by using the histogram and cumulative distribution of the MST branch lengths \citep[for more details, please refer,][]{2009ApJS..184...18G,2016AJ....151..126S,2020ApJ...891...81P}. 
The YSOs and MST connections for the clusters and the cores are represented by orange and purple colored asterisks and line segments in the upper-right panel of Figure \ref{fig:intro_rgb}. 
Since the above distribution of YSOs in the clusters/cores is not symmetrical/circular, we used the \emph{convex hull}\footnote{\emph{Convex hull} is a polygon enclosing all the points in a grouping with internal angles between two contiguous sides of less than 180$^{\circ}$.} generated from their position in cluster/core to estimate their area, i.e., $A_{cluster}$, which is normalized by a geometric factor (\citealt{2006A&A...449..151S}, \citealt{2016AJ....151..126S}, \citealt{2020MNRAS.498.2309S}), given as:
\begin{equation}
    \begin{split}
        A_{cluster}=\frac{A_{hull}}{1-\frac{n_{hull}}{n_{total}}}
    \end{split}
\end{equation}
where, $A_{hull}$ is the area of the hull, $n_{hull}$ is the total number of vertices on the hull, and $n_{total}$ is the total number of points (YSOs) inside the hull. The radius of the cluster/core $R_{cluster}$ is defined as the radius of the circle whose area equals $A_{cluster}$. 
The extracted clusters and cores are enclosed by the \emph{Convex hulls} using the yellow and purple line segments in the upper-right panel of Figure \ref{fig:intro_rgb}. 
We also estimated the $R_{circ}$ (half of the farthest distance between two hull objects) and the aspect ratio ($\frac{R^2_{circ}}{R^2_{cluster}}$) of the extracted clusters/cores. Table \ref{tab:mst_parameter} gives the above derived parameters.

\begin{table*}[!ht]
    \centering
    \caption{Properties of the identified clusters and their cores based on MST Analysis. 
    The number of enclosed YSOs is mentioned in column 2; $R_{cluster}$ is mentioned in column 3, $R_{circ}$ in column 4; aspect ratio in column 5; the mean and peak stellar surface densities are in columns 6 and 7, respectively; the mean MST branch length in column 8; Q parameter in column 9; fraction of Class I YSOs column 10; total molecular mass in column 11; $\lambda_J$ in column 12; and finally SFE is mentioned in column 13.}

    \begin{tabular}{@{}r r @{ }c@{ } c@{ } c@{ } c@{ } c@{ } c@{ } c@{ } c@{ } c@{ } c@{ } c@{ } c@{}}
    \hline
    Region & N$^a$ &  $R_{cluster}$ & $R_{circ}$ & Aspect & $\rho_{mean}$ & $\rho_{peak}$ & MST & $Q$ & fraction$^b$ & Molecular & $\lambda_J$ & SFE\\
     &  & (pc) & (pc) & Ratio & (pc$^{-2}$) & (pc$^{-2}$) & (pc) &  & & mass$^c$ & (pc) & (\%)\\
    \hline
    Clusters:\\
    AFGL 5157    & 60  & 1.75 & 2.25 & 1.64 & 3.1 & 355.5 & 0.24 & 0.76 & 0.03 & 5.4 & 1.1 & 7.7\\
    FSR0807        & 28  & 1.26 & 1.38 & 1.20 & 11.2 & 217.0 & 0.27 & 0.68 & 0.14 & 1.8 & 1.2 & 10.6\\
    E70         & 14  & 2.59 & 3.53 & 1.86 & 0.2 &  2.2 & 0.99 & 0.81 & 0.07 & 6.9 & 1.8 & 1.5\\
    KPS0620        & 36  & 3.95 & 5.27 & 1.78 & 1.1 & 22.2 & 0.78 & 0.66 & 0.31 & 19.2 & 2.1 & 1.4\\
    IRAS 05331$+$3115        & 13  & 1.44 & 2.50 & 3.04 & 1.7 & 10.0 & 0.60 & 0.77 & 0.08 & 1.9 & 1.4 & 4.8\\
    \hline\\
    Cores:\\
    AFGL 5157 & 52 & 1.29 & 1.66 & 1.67 & 16.4 & 355.5 & 0.20 & 0.81 & 0.04 & 2.5 & 1.1 & 13.4\\
    FSR0807 (west) & 20 & 0.63 & 0.89 & 2.00 & 35.9 & 217.0 & 0.20 & 0.84 & 0.10 & 0.4 & 0.8 & 27.0\\
    FSR0807 (east) & 7 & 0.44 & 0.50 & 1.31 & 7.3 & 18.5 & 0.27 & 0.68 & 0.29 & 0.2 & 0.8 & 25.8\\
    E70 & 12 & 2.59 & 2.19 & 0.71 & 0.4 & 2.2 & 0.85 & 0.86 & - & 3.8 & 1.8 & 2.3\\
    KPS0620 & 25 & 2.61 & 3.10 & 1.41 & 2.0 & 22.2 & 0.61 & 0.64 & 0.20 & 6.9 & 2.1 & 2.7\\
    IRAS 05331$+$3115 & 10 & 0.75 & 2.00 & 7.17 & 4.5 & 10.0 & 0.36 & 0.83 & - & 0.4 & 1.4 & 16.3\\
    \hline\\
    \end{tabular}
    
    a: Number of enclosed YSOs\\
    b: Ratio of the enclosed Class I YSOs to total (= Class I + Class II) YSOs\\
    c: $\times 10^{2}$ $M_\odot$
    \label{tab:mst_parameter}
\end{table*}

\subsubsection{Isodensity and Extinction Maps}

We also used the stellar surface density map generated by the nearest neighbor (NN) method to understand the stellar clustering of all point sources in our target area (refer \citealt{2005ApJ...632..397G,2009ApJS..184...18G}). The stellar surface density $\sigma$ at any grid position [i, j] is evaluated by:
\begin{equation}
    \sigma(i,j) = \frac{N}{\pi r_N^2(i,j)}
\end{equation}

Here $r_N(i,j)$ is the projected radial distance to the \emph{N$^{th}$} nearest star.
The lower-left panel of Figure \ref{fig:intro_rgb} represents the stellar surface density map (blue contours) with a grid size 6$\arcsec$, for the twentieth nearest neighbor star. The lowest contour level is one stars arcmin$^{-2}$ with a step size of 29.5 stars arcmin$^{-2}$. This map is generated for the stellar sources detected in the NIR point source catalog generated using the UKIDSS and 2MASS archives (refer \citealt{2011ApJ...739...84G,2016AJ....151..126S}). The mean and peak value of stellar surface densities ($\rho_{mean}$ and $\rho_{peak}$, respectively) for each of the cluster/core are tabulated in Table \ref{tab:mst_parameter}.

We also used the above NIR catalog to generate the extinction map from the $(H-K)$ colors of the main-sequence (MS) stellar sources. This map was also generated using the NN method after removing the YSOs population (cf. \citealt{2005ApJ...632..397G,2009ApJS..184...18G}). As both isodensity and extinction maps have been generated from the same catalog, they will represent the same depth in sensitivity. They can be used to see the association of stars with the gas and dust distribution.
We determined the $(H-K)$ colors at each grid position (i, j) with a step size 6$\arcsec$ for the twentieth nearest stars. The sources having a value above 3$\sigma$ were removed to determine the final mean color at each grid position. $A_K$ values are determined by the relation $A_K=1.82\times((H-K)_{obs}-(H-K)_{int})$ \citep{2007ApJ...663.1069F}. We assumed $(H-K)_{int}=$ 0.2 mag as the average intrinsic color of the MS stars \citep{2008ApJ...675..491A,2009ApJS..184...18G}. This assumption might introduce errors in estimating $A_K$ values, but these effects are generally small \citep{2016AJ....151..126S}. Then we converted these $A_K$ values into $A_V$ by using the relation given by \citet{1985ApJ...288..618R}. To dispose of the contributions of foreground stars in the extinction measurements, we considered the stars having $A_V > 0.85 \times D$, where $D$ is the distance of the clusters in kpc (refer Section \ref{sec:distance}). Then, we plotted the $A_V$ extinction map for our target area as shown in the lower-right panel of Figure \ref{fig:intro_rgb}. 

\subsubsection{Distance}\label{sec:distance}

From the isodensity contours, we isolated the cluster region in the five star-forming sites in our target area. The distance of these clusterings can be derived using the Gaia proper motion data \citep[for details, please refer][]{2020MNRAS.498.2309S}, provided they have enough optically detected member stars.  The clusters AFGL 5157 and FSR0807, and E70 have a good number of optically (Gaia) detected stars; thus, we can estimate their distance using the procedure outlines in \citet{2020MNRAS.498.2309S} as  1.60 $\pm$ 0.20 kpc and 1.60 $\pm$ 0.20 kpc, 3.26 $\pm$ 0.45 kpc, respectively. The distance estimation of the E70 bubble has already been discussed in \citet{2023ApJ...953..145V}.

\begin{figure*}[!ht]
    \includegraphics[width=0.245\textwidth]{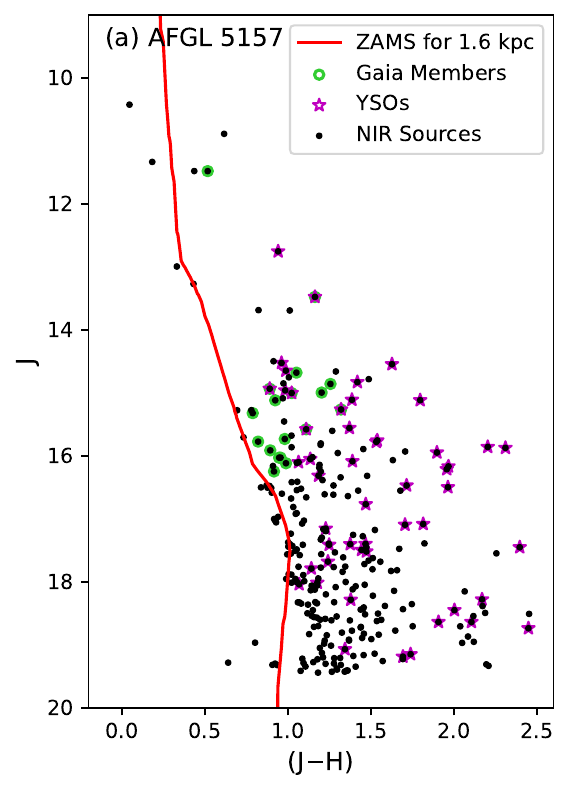}
    \includegraphics[width=0.245\textwidth]{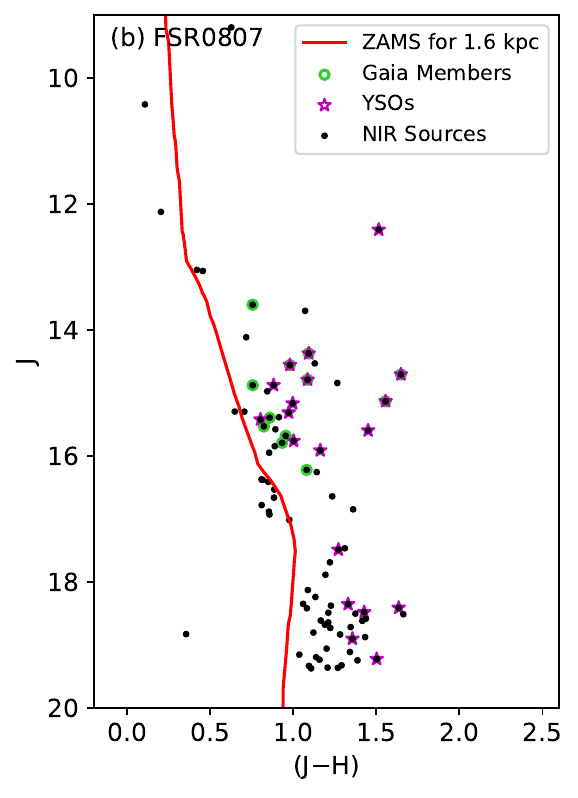}
    \includegraphics[width=0.245\textwidth]{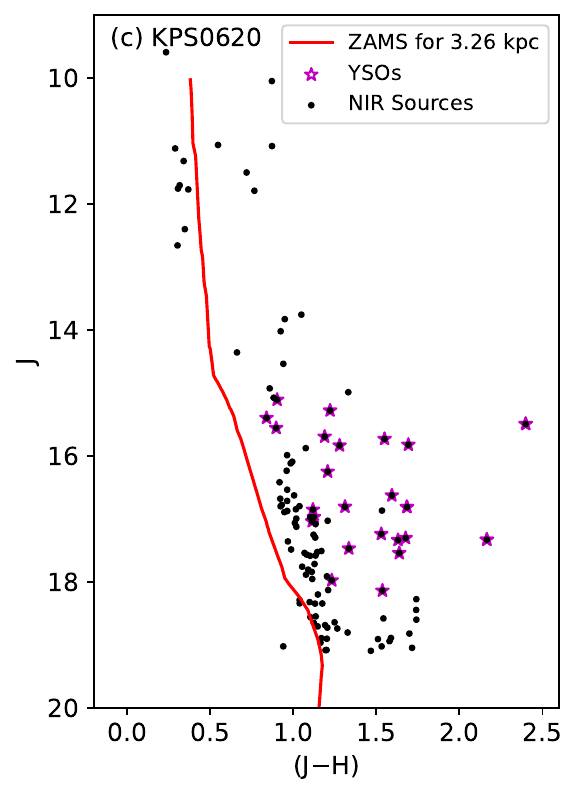}
    \includegraphics[width=0.245\textwidth]{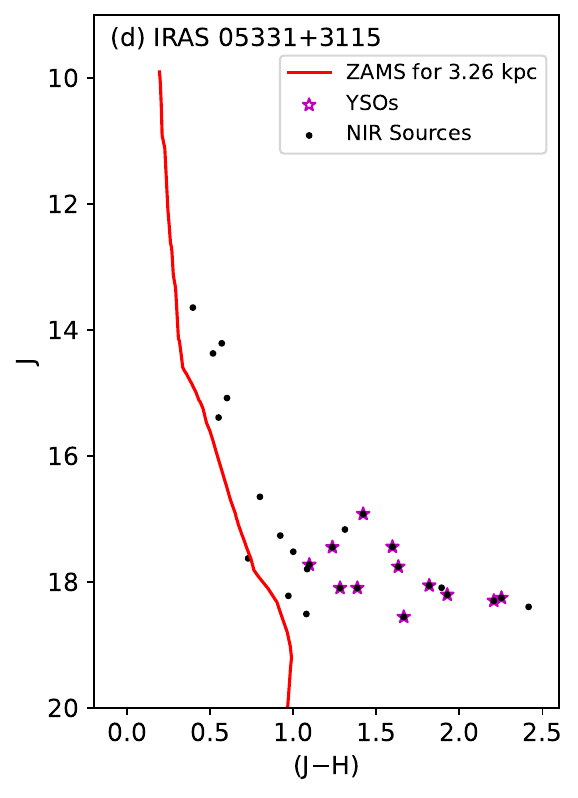}
    \caption{Panel (a): $J$ vs. $(J-H)$ CMD generated using NIR catalog for the stars within the AFGL 5157 region. The red solid curve defines the ZAMS isochrone by \citet{2013ApJS..208....9P}, corrected for the distance 1.6 kpc and $A_V=2.24$ mag. Panel (b): $J$ vs. $(J-H)$ CMD for the stars within the FSR0807 cluster. The red solid curve defines the ZAMS isochrone by \citet{2013ApJS..208....9P}, corrected for the distance 1.6 kpc and zero-reddening. Panel (c): $J$ vs. $(J-H)$ CMD for the stars within the KPS0620 cluster. The red solid curve defines the ZAMS isochrone by \citet{2013ApJS..208....9P}, corrected for the distance of 3.26 kpc and $A_V=2.56$ mag. Panel (d): $J$ vs. $(J-H)$ CMD for the stars within the IRAS 05331$+$3115 cluster. The red solid curve defines the ZAMS isochrone by \citet{2013ApJS..208....9P}, corrected for the distance of 3.26 kpc and $A_V=3.20$ mag. 
    The YSOs and Gaia DR3 members are marked in all the panels with magenta asterisks and green circles, respectively. Due to less detection of optical clustering, the Gaia members are not marked for KPS0620 and  IRAS 05331$+$3115.}
    \label{fig:nir_tcd}
\end{figure*}

Furthermore, we utilized $J$ vs. $(J-H)$ color-magnitude diagram (CMD) generated using the deep NIR catalog (see Figure \ref{fig:nir_tcd}),  to confirm the estimated distances of AFGL 5157 and FSR0807 (\citealt{2020MNRAS.492.2446P, 2020ApJ...891...81P, 2020ApJ...896...29K,2023JApA...44...66K,2023JApA...44...52V}). These CMDs are statistically subtracted from the field star population using the CMD of a nearby reference field (for details, please refer to \citealt{2017MNRAS.467.2943S}). Probable Gaia members and the identified young PMS stars are also marked with green circles and magenta asterisks. The intrinsic ZAMS (red solid curve) taken from \citet{2013ApJS..208....9P} is also plotted for the estimated Gaia distances and the extinction. The extinction value ($A_V$) is determined using the reddening value (E(g$-$r); \citealt{2019ApJ...877..116W}) from Bayestar19 3D dust map \citep{2019ApJ...887...93G}. The ZAMS seems to be very well-fitted to the blue envelope of the distribution of stars in AFGL 5157 and FSR0807, thus validating our distance estimates (for a detailed description of CMD isochrones fitting, see \citealt{1994ApJS...90...31P}). 



As the clusters KPS0620 and IRAS 05331$+$3115 have very few optically detected (Gaia) sources, we tried the isochrone fitting method to constrain their distance (see right panel of Figure \ref{fig:nir_tcd}). However, the ZAMS is not well defined for both clusters, even though we can see a prominent population of low-mass young stars in IRAS 05331$+$3115. Thus, it isn't easy to ascertain the distance of these clusters from the CMD or even with the Gaia PM data. We took the distance and reddening of KPS0620 cluster similar to the E70 cluster (3.26 kpc and $A_V=2.56$ mag), as the E70 bubble seems to be associated with the southern molecular structure associated with KPS0620 and IRAS 05331$+$3115 (refer to Figure 14 of \citet[][]{2023ApJ...953..145V}, and Figure 2 of this paper). However, for IRAS 05331$+$3115, we considered a distance of 3.26 kpc (same as E70) and $A_V=3.20$ mag as it is more fitted for this extinction value. 



\section{Discussion}\label{sec:discussion}

We did a comprehensive study of the $1^\circ\times1^\circ$ region showing signatures of recent star formation. 
In the subsequent section, we explored the physical properties of the identified stellar clustering in our target area, studied their physical environment, and then explored molecular gas kinematics/morphology to conclude the overall star formation scenario.

\subsection{Physical Properties of the Stellar Clustering in the Region}

We identified five young star clusters and their cores in our target area and reported their various physical parameters as listed in Table \ref{tab:mst_parameter}  \citep[for details, please refer,][]{2009ApJS..184...18G,2011ApJ...739...84G,2014MNRAS.439.3719C, 2016AJ....151..126S}.
Briefly, the $R_{cluster}$ for the clusters varies between 1.26 - 3.95 pc, whereas the aspect ratio varies between 1.20 to 3.04, indicating that all the clusters have elongated morphology. The cores are a bit less elongated in comparison to clusters. The number of YSOs enclosed by each cluster varies between 13 - 60 such that the cluster with the maximum number of YSOs has maximum peak stellar surface density and vice versa. The mean MST branch length for the clusters varies between 0.24 - 0.99 pc, indicating that the YSOs in the E70 have comparatively larger separation than the other clusters.

To assess the strength of hierarchical versus radial distributions of YSOs in these clustering, we estimated the $Q$ parameter. It is defined as the ratio of the normalized mean MST branch ($\bar{l}_{MST}$) to the normalized mean separation ($\bar{s}$) between objects (refer \citealt{2006A&A...449..151S,2014MNRAS.439.3719C,Ascenso2018}),

\begin{equation}
    Q=\frac{\bar{l}_{MST}}{\bar{s}}
\end{equation}

Here mean MST branch (${l}_{MST}$) is normalized by a factor $\sqrt{A_{cluster}/n_{total}}$ and mean separation (${s}$) is normalized by $R_{cluster}$.

\citet{2004MNRAS.348..589C} reported that if $Q>$ 0.8, the points are distributed radially, whereas if $Q<$ 0.8, the points possess a more fractal distribution.
The estimated $Q$ parameters  for our sample are listed in
 Table \ref{tab:mst_parameter}. These values indicate that fractal distribution exists in all the clusters except E70.

We evaluated the molecular mass of the clusters and cores using the Herschel $H_2$ column density maps 
using the following relation given in \citet{2017ApJ...845...34D}:

\begin{equation}
    M =  \mu_{H_2}\,m_H\,a_{pixel}\,\Sigma N(H_2)
\end{equation}
Here, $\mu_{H_2}$ denotes the mean molecular weight per hydrogen molecule (i.e., 2.8), 
$a_{pixel}$ denotes the area subtended by 1 pixel, and $N(H_2)$ denotes the total column density of that particular region. The calculated values are given in Table \ref{tab:mst_parameter}. The cores and the clusters have molecular masses ranging from 20 - 690 $M_\odot$ and 180 - 1920 $M_\odot$ with a mean value of 236 $M_\odot$ and 704 $M_\odot$, respectively.

\citet{2009MNRAS.392..868B} reported that Jeans fragmentation is an initial point to understand the primordial structure in star-forming regions. The Jeans length $\lambda_J$ is defined as the minimum radius needed for the gravitational collapse of a homogeneous isothermal sphere. We calculated $\lambda_J$ using the formula \citep{2014MNRAS.439.3719C},

\begin{equation}
    \lambda_J = \left(\frac{15 k T}{4 \pi G m_H \rho_0}\right)^{1/2}, 
\end{equation}

where T is the temperature of the cloud, $m_H$ is the atomic mass of the hydrogen, and $\rho_0$ is the mean density defined as,

\begin{equation}
    \rho_0 = \frac{3 M_T}{4 \pi R_H^3}, 
\end{equation}

where $M_T$ is the total mass of the cluster (gas and stars). 

The $\lambda_J$ values for the clusters and cores range between 0.75 - 2.06 pc with a mean value of 1.15 pc (cf. Table \ref{tab:mst_parameter}). We also compared $\lambda_J$ to the mean separation between the YSOs, i.e., the mean branch length for the cores. We found that current results support the existence of non-thermal driven fragmentation due to smaller separation between the YSOs than the $\lambda_J$ \citep{2014MNRAS.439.3719C,2020ApJ...891...81P}.

We obtained a wide range of surface density distribution of identified YSOs, which gives a scope to relate these values with the SFE and properties of associated MC \citep{2011ApJ...739...84G}. SFE is described as the percentage of gas and dust mass inside the convex hull area of the cluster converted into stars (cf. \citealt{2008ApJ...688.1142K}). The calculated values of the SFE for the present study are given in Table \ref{tab:mst_parameter}. The SFE value for the current sample of clusters/cores ranges from $\sim$1.4\% to 16.3\%, typical of star formation sites \citep{2016AJ....151..126S}. Several studies have shown that increased SFE is associated with increased stellar density. \citet{2009ApJS..181..321E} reported the existence of higher SFE (30\%) at the locations with higher stellar density than the surroundings with lower stellar density having lower SFE (3\% - 6\%). Similarly, \citet{2008ApJ...688.1142K} reported higher SFE ($>$ 3\% - 6\%) for the higher stellar density region than the lower stellar density region with 3\% SFE. \citet{2014MNRAS.439.3719C} reported a range of SFE (3\% - 45\% with an average of 20\%) for embedded clusters. We also found that the clusters having maximum stellar density (AFGL 5157 and FSR0807) have maximum SFE.

Overall, from the value of the physical parameters of the clusters/cores given in Table \ref{tab:mst_parameter}, we found that E70, having an MIR bubble, is an evolved cluster with a very low fraction of younger YSOs, low SFE, and low stellar density. The AFGL 5157 and FSR0807 have higher SFE and density-packed stars and seem younger than E70.  FSR0807 seems younger than AFGL 5157, showing a higher fraction of younger YSOs.
The KPS0620, with its largest fraction of younger YSOs, shows a bit low SFE. This is because of the massive clump of gas at the hub of this system. This suggests that KPS0620 is even younger than the above cluster. The IRAS 05331$+$3115 shows 5\% SFE with a Class I fraction of 8\%.
All young clusters except E70 (oldest) show a hierarchical distribution of stars with Q $<$ 0.8. The higher Q value hints towards the radial distribution of stars in E70, hinting at the dynamical evolution of stars in the cluster.
The highest stellar density of the clusters lies in their cores, which generally show radial density distribution and larger SFE.   

We also found an inverse linear relation between SFE and Jeans' fragmentation scale, i.e., the larger the $\lambda_J$ value, the lesser the SFE of the cores. A similar trend was also observed in the case of Sh 2-305 by \citet{2020ApJ...891...81P}. This suggests that star formation is more efficient at smaller Jeans lengths.

\subsection{Star Formation Scenario in the Region}

The star formation activities generally occur inside the dense cores of MCs, and the YSOs follow the clumpy structures of their natal MCs (e.g., \citealt{1996AJ....111.1964L,1998A&A...336..150M,2005ApJ...632..397G,2007ApJ...669..493W}). From the distribution of column density and extinction maps, we found that almost all younger YSOs (Class I) are clustered at the regions of denser molecular material. All five sites in our selected target area show prominent star formation activities, although they seem to have different evolutionary statuses and/or are going through different star-forming scenarios.

Broadly, we found two different molecular structures with two different velocity components in our target area from the CO emission maps.
One is blue-shifted (northern, ranging between $-22$ to $-12$ \kms\, and comprising AFGL 5157 and FSR0807), and another is red-shifted (southern, ranging between $-4$ to 6 \kms\, and comprising KPS0620 and IRAS 05331$+$3115). The E70 bubble seems to be in between these two structures. 

The AFGL 5157 and FSR0807 are located at the same distance of 1.6 kpc and are connected through a cold filamentary structure in the northern molecular structure. These clusters show strong evidence of recent star formation activities and feedback from the massive stars  \citep[see also,][]{2019ApJ...884...84D}. These clusters also show massive molecular clumps at their center and several filamentary sub-structures. AFGL 5157 has three times more molecular mass than FSR0807. The PV diagrams indicate gas flow towards AFGL 5157. Being a massive core, AFGL 5157 seems to be a potential well for this system and driving gas flow towards itself. It also shows two velocity components and a larger dispersion in the velocities, further confirming the above scenario.
In the literature, \citet{2018ARA&A..56...41M}  presented the `global nonisotropic collapse' (GNIC) scenario for massive star formation. This empirical model displays aspects of global hierarchical collapse and clump-feed accretion scenario and suggests that the formation of massive protostellar cores can occur from the low-mass protostellar cores, which subsequently accrete material from the parent massive dense core. Such a scenario requires evidence of a hub-filamentary system comprising the highest column density region with multiple accreting filaments. AFGL 5157 and FSR0807 seem to show the same GNIC scenario.

From our cluster analysis, the AFGL 5157 seems to be older than FSR0807. We also found evidence of longitudinal flow of material towards AFGL 5157. It seems that star formation started at AFGL 5157 and then later proceeded to FSR0807. A similar scenario was also recently observed in the NGC 2316 region \citep{2024AJ....167..106S}. They found that the NGC 2316  cluster is part of an HFS, in which the NGC 2316 cluster is probably the hub, and there are other filamentary structures. Being the gravity well of this HFS, star formation started first in the NGC 2316 region and went on to the other filamentary nodes  \citep{2024AJ....167..106S}. 
Thus, star formation occurs in the northern clouds due to filamentary structures and is an ideal site to investigate further. 

We found no connection between E70, KPS0620, and IRAS 05331$+$3115 in the southern molecular structure. They seem to be forming stars independently through different star formation processes. E70 has a MIR bubble, which seems to be an evolved cluster, and is forming stars due to the positive feedback from a massive star (for details, please refer \citealt{2023ApJ...953..145V}). 
The KPS0620 shows strong evidence of an HFS, having a massive hub and prominent cold filamentary structures. Star formation activities are ongoing in these structures, as evidenced by the location of YSOs. However, this system has no signature of massive star feedback. 
The velocity gradient in the PV slices hints that the gas is spiraling into the cluster's central region. A similar nature of the molecular gas was presented by \citep{2023ApJ...944..228M}  in Sh2-138. 
The IRAS 05331$+$3115 has neither a filamentary nor a bubble structure associated with it. It is a single molecular gas clump where low-mass star formation occurs. The CMD shown in panel (d) of Figure \ref{fig:nir_tcd} shows a striking population of only low-mass stars in IRAS 05331$+$3115. The center of this clump has a higher molecular column density and larger dispersion on the velocity map. This clump seems to have collapsed under its own gravity, and the low-mass star formation is going on in the innermost region \citep{2007prpl.conf...17D}.

Briefly, we have studied the star formation activities in a wide FOV comprising active star-forming sites/clusters. We have found two molecular structures at two different velocities (distances).  The clusters associated with the northern molecular cloud (i.e., AFGL 5157 and FSR0807) are located closer ($\sim$1.6 kpc) and show inter-connected hub-filamentary structures and the presence of feedback from the massive star/s and star-forming activities. Whereas the clusters associated with the southern molecular cloud (i.e., KPS0620 and IRAS 05331$+$3115) are located at a farther distance ($\sim$3.2 kpc). They have a `miniature spiral galaxy' like hub-filamentary system (the similar feature was shown by the star-forming complex Monoceros R2; \citealt{2019A&A...629A..81T,2022A&A...658A.114K}) and a single molecular clump, respectively, showing signatures of active star formation but lacks feedback signatures from massive stars. 
E70 is a bit evolved region with a MIR bubble and no signatures of filamentary structures, and star formation is happening because of the feedback from the massive stars.
All these clusters in our target area are found to be in different evolutionary stages of star formation. The KPS0620 and IRAS 05331$+$3115 seem to be the youngest sites,  AFGL 5157 and FSR0807 formed earlier, and E70 is the oldest. Our target area is a menagerie of star-forming sites where the formation of the stars happens through different processes.

\section{Conclusion}\label{sec:conclusion}

We used a multiwavelength approach to examine the global star formation scenario concerning MC morphology at AFGL 5157, FSR0807, KPS0620, and  IRAS 05331$+$3115 located towards $l=176^{\circ}$ to $177^{\circ}$, $b=-0^{\circ}.44$ to $0.^{\circ}20$ in the outer galactic arm. We used quantitative techniques to analyze the spatial structure of YSOs. The following are the main conclusions from the present study:

\begin{enumerate}
   \item With the help of sensitive CO molecular data analysis, we found that there are two molecular structures in our target area: `northern molecular structure/cloud' that has blue-shifted velocity components ranging from $-22$ to $-12$ \kms\, and `southern molecular structure/cloud' that has a red-shifted velocity component ranging between $-4$ to 6 \kms. The `northern molecular cloud' comprises AFGL 5157 and FSR0807, whereas the `southern molecular cloud' comprises KPS0620 and IRAS 05331$+$3115.
    \item To understand the stellar clustering in the region, we used deep NIR data. Five major clusters were found in the region, consisting of 204 YSOs. The YSOs were classified based on their excess IR emission. The presence of these YSOs is an indication of recent star formation activities.
    \item We compared the spatial distribution of YSOs with the associated MC, deduced by the isodensity map and extinction map. We found that most of the YSOs are populated at higher extinction regions. Most Class I objects belong to these regions compared to the Class II objects distributed randomly. This result agrees with the notion that star formation often occurs inside the dense cores of the MC.
    \item We found that the E70 bubble is an evolved cluster with a low fraction of younger YSOs, low SFE, and low stellar density. The AFGL 5157 and FSR0807 have higher SFE and density-packed stars and seem younger than E70.  FSR0807 seems younger than AFGL 5157, showing a higher fraction of younger YSOs. The KPS0620, with its most significant fraction of younger YSOs, suggests this region is even younger than the other clusters. 
    \item Longer Jeans length `$\lambda_J$' compared to the separation between the YSOs reveals the non-thermally driven fragmentation in the clusters. We also found an inverse linear relation between the SFE and Jeans $\lambda_J$ value, suggesting that star formation is more efficient at smaller Jeans lengths.    
    \item All young clusters except E70 (oldest) show a hierarchical distribution of stars with Q $<$ 0.8. The higher Q value hints towards the radial distribution of stars in E70, hinting at the dynamical evolution of stars in the cluster.
    \item In the northern molecular cloud, we found a large spread in velocity at AFGL 5157 and FSR0807, which are connected by a structure of non-uniform velocity in velocity space.
    \item  In the southern molecular cloud, KPS0620 seems to have a filamentary structure, with the central hub having two velocity peaks. In contrast, IRAS 05331$+$3115 is a single clump of molecular gas where low-mass star formation occurs. 
    \item AFGL 5157, FSR0807, and KPS0620 show velocity gradients that indicate the existence of the longitudinal flow of gas converging onto the central dense clump.
    \item Our target area is a menagerie of star-forming sites where the formation of the stars happens through different processes.
\end{enumerate}


\begin{acknowledgments}
\section*{Acknowledgements}
We thank the anonymous referee for constructive and valuable comments that greatly improved the overall quality of the paper. This research used data from the Milky Way Imaging Scroll Painting (MWISP) project, a multi-line survey in 12CO/13CO/C18O along the northern galactic plane with a PMO-13.7m telescope. We are grateful to all the members of the MWISP working group, particularly the staff members at the PMO-13.7m telescope, for their long-term support. MWISP was sponsored by National Key R\&D Program of China with grant 2017YFA0402701 and CAS Key Research Program of Frontier Sciences with grant QYZDJ-SSW-SLH047. This publication uses data from the Two Micron All Sky Survey, a joint project of the University of Massachusetts, and the Infrared Processing and Analysis Center/California Institute of Technology, funded by the National Aeronautics and Space Administration and the National Science Foundation. This work is based on observations made with the Spitzer Space Telescope, operated by the Jet Propulsion Laboratory, California Institute of Technology, under a contract with the National Aeronautics and Space Administration. This publication uses data products from the Wide-field Infrared Survey Explorer, a joint project of the University of California, Los Angeles, and the Jet Propulsion Laboratory/California Institute of Technology, funded by the National Aeronautics and Space Administration.
A.V. acknowledges the financial support of DST-INSPIRE (No.$\colon$ DST/INSPIRE Fellowship/2019/IF190550). D.K.O. acknowledges the support of the Department of Atomic Energy, Government of India, under Project Identification No. RTI 4002. L.K.D. acknowledges the support at the Physical Research Laboratory by the Department of Space, Government of India.
\end{acknowledgments}






\bibliography{outer_arm}{}

\begin{thebibliography}{}
\expandafter\ifx\csname natexlab\endcsname\relax\def\natexlab#1{#1}\fi
\providecommand{\url}[1]{\href{#1}{#1}}
\providecommand{\dodoi}[1]{doi:~\href{http://doi.org/#1}{\nolinkurl{#1}}}
\providecommand{\doeprint}[1]{\href{http://ascl.net/#1}{\nolinkurl{http://ascl.net/#1}}}
\providecommand{\doarXiv}[1]{\href{https://arxiv.org/abs/#1}{\nolinkurl{https://arxiv.org/abs/#1}}}

\bibitem[{{Aarseth} \& {Hills}(1972)}]{1972A&A....21..255A}
{Aarseth}, S.~J., \& {Hills}, J.~G. 1972, \aap, 21, 255

\bibitem[{{Allen} {et~al.}(2008){Allen}, {Pipher}, {Gutermuth}, {Megeath},
  {Adams}, {Herter}, {Williams}, {Goetz-Bixby}, {Allen}, \&
  {Myers}}]{2008ApJ...675..491A}
{Allen}, T.~S., {Pipher}, J.~L., {Gutermuth}, R.~A., {et~al.} 2008, \apj, 675,
  491, \dodoi{10.1086/525241}

\bibitem[{Ascenso(2018)}]{Ascenso2018}
Ascenso, J. 2018, Embedded Clusters, ed. S.~Stahler (Cham: Springer
  International Publishing), 1--37, \dodoi{10.1007/978-3-319-22801-3_1}

\bibitem[{{Bastian} {et~al.}(2009){Bastian}, {Gieles}, {Ercolano}, \&
  {Gutermuth}}]{2009MNRAS.392..868B}
{Bastian}, N., {Gieles}, M., {Ercolano}, B., \& {Gutermuth}, R. 2009, \mnras,
  392, 868, \dodoi{10.1111/j.1365-2966.2008.14107.x}

\bibitem[{Beichman {et~al.}(1988)Beichman, Neugebauer, Habing, Clegg, \&
  Chester}]{beichman1988infrared}
Beichman, C., Neugebauer, G., Habing, H., Clegg, P., \& Chester, T.~J. 1988,
  Infrared astronomical satellite (IRAS) catalogs and atlases. Volume 1:
  Explanatory supplement, Tech. rep.

\bibitem[{{Benjamin} {et~al.}(2005){Benjamin}, {Churchwell}, {Babler},
  {Indebetouw}, {Meade}, {Whitney}, {Watson}, {Wolfire}, {Wolff}, {Ignace},
  {Bania}, {Bracker}, {Clemens}, {Chomiuk}, {Cohen}, {Dickey}, {Jackson},
  {Kobulnicky}, {Mercer}, {Mathis}, {Stolovy}, \&
  {Uzpen}}]{2005ApJ...630L.149B}
{Benjamin}, R.~A., {Churchwell}, E., {Babler}, B.~L., {et~al.} 2005, \apjl,
  630, L149, \dodoi{10.1086/491785}

\bibitem[{{Bronfman} {et~al.}(1996){Bronfman}, {Nyman}, \&
  {May}}]{1996A&AS..115...81B}
{Bronfman}, L., {Nyman}, L.~A., \& {May}, J. 1996, \aaps, 115, 81

\bibitem[{{Buckner} \& {Froebrich}(2013)}]{2013MNRAS.436.1465B}
{Buckner}, A. S.~M., \& {Froebrich}, D. 2013, \mnras, 436, 1465,
  \dodoi{10.1093/mnras/stt1665}

\bibitem[{{Cartwright} \& {Whitworth}(2004)}]{2004MNRAS.348..589C}
{Cartwright}, A., \& {Whitworth}, A.~P. 2004, \mnras, 348, 589,
  \dodoi{10.1111/j.1365-2966.2004.07360.x}

\bibitem[{{Chavarr{\'\i}a} {et~al.}(2014){Chavarr{\'\i}a}, {Allen}, {Brunt},
  {Hora}, {Muench}, \& {Fazio}}]{2014MNRAS.439.3719C}
{Chavarr{\'\i}a}, L., {Allen}, L., {Brunt}, C., {et~al.} 2014, \mnras, 439,
  3719, \dodoi{10.1093/mnras/stu224}

\bibitem[{{Chen} {et~al.}(1999){Chen}, {Yao}, {Yang}, {Hirao}, {Ishii},
  {Nagata}, \& {Sato}}]{1999AJ....117..446C}
{Chen}, Y., {Yao}, Y., {Yang}, J., {et~al.} 1999, \aj, 117, 446,
  \dodoi{10.1086/300706}

\bibitem[{{Chen} {et~al.}(2003){Chen}, {Yao}, {Yang}, {Zeng}, \&
  {Sato}}]{2003A&A...405..655C}
{Chen}, Y., {Yao}, Y., {Yang}, J., {Zeng}, Q., \& {Sato}, S. 2003, \aap, 405,
  655, \dodoi{10.1051/0004-6361:20030666}

\bibitem[{DengRong {et~al.}(2018)DengRong, Jixian, Zherui, Liang, \&
  Ji}]{633694461037117445}
DengRong, L., Jixian, S., Zherui, Y., Liang, L., \& Ji, Y. 2018, {Data of the
  MWISP Sky Survey (2011 - 2017)}, V1,  Science Data Bank,
  \dodoi{10.11922/sciencedb.570}

\bibitem[{{Dewangan}(2019)}]{2019ApJ...884...84D}
{Dewangan}, L.~K. 2019, \apj, 884, 84, \dodoi{10.3847/1538-4357/ab4189}

\bibitem[{{Dewangan} {et~al.}(2017){Dewangan}, {Baug}, {Ojha}, {Janardhan},
  {Devaraj}, \& {Luna}}]{2017ApJ...845...34D}
{Dewangan}, L.~K., {Baug}, T., {Ojha}, D.~K., {et~al.} 2017, \apj, 845, 34,
  \dodoi{10.3847/1538-4357/aa7da2}

\bibitem[{{di Francesco} {et~al.}(2007){di Francesco}, {Evans}, {Caselli},
  {Myers}, {Shirley}, {Aikawa}, \& {Tafalla}}]{2007prpl.conf...17D}
{di Francesco}, J., {Evans}, N.~J., I., {Caselli}, P., {et~al.} 2007, in
  Protostars and Planets V, ed. B.~{Reipurth}, D.~{Jewitt}, \& K.~{Keil}, 17,
  \dodoi{10.48550/arXiv.astro-ph/0602379}

\bibitem[{{Dib} {et~al.}(2018){Dib}, {Schmeja}, \&
  {Parker}}]{2018MNRAS.473..849D}
{Dib}, S., {Schmeja}, S., \& {Parker}, R.~J. 2018, \mnras, 473, 849,
  \dodoi{10.1093/mnras/stx2413}

\bibitem[{{Efremov}(1978)}]{1978PAZh....4..125E}
{Efremov}, Y.~N. 1978, Pisma v Astronomicheskii Zhurnal, 4, 125

\bibitem[{{Elmegreen}(1993)}]{1993prpl.conf...97E}
{Elmegreen}, B.~G. 1993, in Protostars and Planets III, ed. E.~H. {Levy} \&
  J.~I. {Lunine}, 97

\bibitem[{{Elmegreen} \& {Falgarone}(1996)}]{1996ApJ...471..816E}
{Elmegreen}, B.~G., \& {Falgarone}, E. 1996, \apj, 471, 816,
  \dodoi{10.1086/178009}

\bibitem[{{Evans} \& {Lada}(1991)}]{1991IAUS..147..293E}
{Evans}, N.~J., I., \& {Lada}, E.~A. 1991, in Fragmentation of Molecular Clouds
  and Star Formation, ed. E.~{Falgarone}, F.~{Boulanger}, \& G.~{Duvert}, Vol.
  147, 293

\bibitem[{{Evans} {et~al.}(2009){Evans}, {Dunham}, {J{\o}rgensen}, {Enoch},
  {Mer{\'\i}n}, {van Dishoeck}, {Alcal{\'a}}, {Myers}, {Stapelfeldt}, {Huard},
  {Allen}, {Harvey}, {van Kempen}, {Blake}, {Koerner}, {Mundy}, {Padgett}, \&
  {Sargent}}]{2009ApJS..181..321E}
{Evans}, Neal~J., I., {Dunham}, M.~M., {J{\o}rgensen}, J.~K., {et~al.} 2009,
  \apjs, 181, 321, \dodoi{10.1088/0067-0049/181/2/321}

\bibitem[{{Flaherty} {et~al.}(2007){Flaherty}, {Pipher}, {Megeath}, {Winston},
  {Gutermuth}, {Muzerolle}, {Allen}, \& {Fazio}}]{2007ApJ...663.1069F}
{Flaherty}, K.~M., {Pipher}, J.~L., {Megeath}, S.~T., {et~al.} 2007, \apj, 663,
  1069, \dodoi{10.1086/518411}

\bibitem[{{Gaia Collaboration} {et~al.}(2016){Gaia Collaboration}, {Prusti},
  {de Bruijne}, {Brown}, {Vallenari}, {Babusiaux}, {Bailer-Jones}, {Bastian},
  {Biermann}, {Evans}, {Eyer}, {Jansen}, {Jordi}, {Klioner}, {Lammers},
  {Lindegren}, {Luri}, {Mignard}, {Milligan}, {Panem}, {Poinsignon},
  {Pourbaix}, {Randich}, {Sarri}, {Sartoretti}, {Siddiqui}, {Soubiran},
  {Valette}, {van Leeuwen}, {Walton}, {Aerts}, {Arenou}, {Cropper}, {Drimmel},
  {H{\o}g}, {Katz}, {Lattanzi}, {O'Mullane}, {Grebel}, {Holland}, {Huc},
  {Passot}, {Bramante}, {Cacciari}, {Casta{\~n}eda}, {Chaoul}, {Cheek}, {De
  Angeli}, {Fabricius}, {Guerra}, {Hern{\'a}ndez}, {Jean-Antoine-Piccolo},
  {Masana}, {Messineo}, {Mowlavi}, {Nienartowicz}, {Ord{\'o}{\~n}ez-Blanco},
  {Panuzzo}, {Portell}, {Richards}, {Riello}, {Seabroke}, {Tanga},
  {Th{\'e}venin}, {Torra}, {Els}, {Gracia-Abril}, {Comoretto},
  {Garcia-Reinaldos}, {Lock}, {Mercier}, {Altmann}, {Andrae}, {Astraatmadja},
  {Bellas-Velidis}, {Benson}, {Berthier}, {Blomme}, {Busso}, {Carry},
  {Cellino}, {Clementini}, {Cowell}, {Creevey}, {Cuypers}, {Davidson}, {De
  Ridder}, {de Torres}, {Delchambre}, {Dell'Oro}, {Ducourant}, {Fr{\'e}mat},
  {Garc{\'\i}a-Torres}, {Gosset}, {Halbwachs}, {Hambly}, {Harrison}, {Hauser},
  {Hestroffer}, {Hodgkin}, {Huckle}, {Hutton}, {Jasniewicz}, {Jordan},
  {Kontizas}, {Korn}, {Lanzafame}, {Manteiga}, {Moitinho}, {Muinonen},
  {Osinde}, {Pancino}, {Pauwels}, {Petit}, {Recio-Blanco}, {Robin}, {Sarro},
  {Siopis}, {Smith}, {Smith}, {Sozzetti}, {Thuillot}, {van Reeven}, {Viala},
  {Abbas}, {Abreu Aramburu}, {Accart}, {Aguado}, {Allan}, {Allasia},
  {Altavilla}, {{\'A}lvarez}, {Alves}, {Anderson}, {Andrei}, {Anglada Varela},
  {Antiche}, {Antoja}, {Ant{\'o}n}, {Arcay}, {Atzei}, {Ayache}, {Bach},
  {Baker}, {Balaguer-N{\'u}{\~n}ez}, {Barache}, {Barata}, {Barbier}, {Barblan},
  {Baroni}, {Barrado y Navascu{\'e}s}, {Barros}, {Barstow}, {Becciani},
  {Bellazzini}, {Bellei}, {Bello Garc{\'\i}a}, {Belokurov}, {Bendjoya},
  {Berihuete}, {Bianchi}, {Bienaym{\'e}}, {Billebaud}, {Blagorodnova},
  {Blanco-Cuaresma}, {Boch}, {Bombrun}, {Borrachero}, {Bouquillon}, {Bourda},
  {Bouy}, {Bragaglia}, {Breddels}, {Brouillet}, {Br{\"u}semeister},
  {Bucciarelli}, {Budnik}, {Burgess}, {Burgon}, {Burlacu}, {Busonero}, {Buzzi},
  {Caffau}, {Cambras}, {Campbell}, {Cancelliere}, {Cantat-Gaudin}, {Carlucci},
  {Carrasco}, {Castellani}, {Charlot}, {Charnas}, {Charvet}, {Chassat},
  {Chiavassa}, {Clotet}, {Cocozza}, {Collins}, {Collins}, {Costigan}, {Crifo},
  {Cross}, {Crosta}, {Crowley}, {Dafonte}, {Damerdji}, {Dapergolas}, {David},
  {David}, {De Cat}, {de Felice}, {de Laverny}, {De Luise}, {De March}, {de
  Martino}, {de Souza}, {Debosscher}, {del Pozo}, {Delbo}, {Delgado},
  {Delgado}, {di Marco}, {Di Matteo}, {Diakite}, {Distefano}, {Dolding}, {Dos
  Anjos}, {Drazinos}, {Dur{\'a}n}, {Dzigan}, {Ecale}, {Edvardsson}, {Enke},
  {Erdmann}, {Escolar}, {Espina}, {Evans}, {Eynard Bontemps}, {Fabre},
  {Fabrizio}, {Faigler}, {Falc{\~a}o}, {Farr{\`a}s Casas}, {Faye}, {Federici},
  {Fedorets}, {Fern{\'a}ndez-Hern{\'a}ndez}, {Fernique}, {Fienga}, {Figueras},
  {Filippi}, {Findeisen}, {Fonti}, {Fouesneau}, {Fraile}, {Fraser}, {Fuchs},
  {Furnell}, {Gai}, {Galleti}, {Galluccio}, {Garabato}, {Garc{\'\i}a-Sedano},
  {Gar{\'e}}, {Garofalo}, {Garralda}, {Gavras}, {Gerssen}, {Geyer}, {Gilmore},
  {Girona}, {Giuffrida}, {Gomes}, {Gonz{\'a}lez-Marcos},
  {Gonz{\'a}lez-N{\'u}{\~n}ez}, {Gonz{\'a}lez-Vidal}, {Granvik}, {Guerrier},
  {Guillout}, {Guiraud}, {G{\'u}rpide}, {Guti{\'e}rrez-S{\'a}nchez}, {Guy},
  {Haigron}, {Hatzidimitriou}, {Haywood}, {Heiter}, {Helmi}, {Hobbs},
  {Hofmann}, {Holl}, {Holland}, {Hunt}, {Hypki}, {Icardi}, {Irwin}, {Jevardat
  de Fombelle}, {Jofr{\'e}}, {Jonker}, {Jorissen}, {Julbe}, {Karampelas},
  {Kochoska}, {Kohley}, {Kolenberg}, {Kontizas}, {Koposov}, {Kordopatis},
  {Koubsky}, {Kowalczyk}, {Krone-Martins}, {Kudryashova}, {Kull}, {Bachchan},
  {Lacoste-Seris}, {Lanza}, {Lavigne}, {Le Poncin-Lafitte}, {Lebreton},
  {Lebzelter}, {Leccia}, {Leclerc}, {Lecoeur-Taibi}, {Lemaitre}, {Lenhardt},
  {Leroux}, {Liao}, {Licata}, {Lindstr{\o}m}, {Lister}, {Livanou}, {Lobel},
  {L{\"o}ffler}, {L{\'o}pez}, {Lopez-Lozano}, {Lorenz}, {Loureiro},
  {MacDonald}, {Magalh{\~a}es Fernandes}, {Managau}, {Mann}, {Mantelet},
  {Marchal}, {Marchant}, {Marconi}, {Marie}, {Marinoni}, {Marrese},
  {Marschalk{\'o}}, {Marshall}, {Mart{\'\i}n-Fleitas}, {Martino}, {Mary},
  {Matijevi{\v{c}}}, {Mazeh}, {McMillan}, {Messina}, {Mestre}, {Michalik},
  {Millar}, {Miranda}, {Molina}, {Molinaro}, {Molinaro}, {Moln{\'a}r},
  {Moniez}, {Montegriffo}, {Monteiro}, {Mor}, {Mora}, {Morbidelli}, {Morel},
  {Morgenthaler}, {Morley}, {Morris}, {Mulone}, {Muraveva}, {Musella},
  {Narbonne}, {Nelemans}, {Nicastro}, {Noval}, {Ord{\'e}novic},
  {Ordieres-Mer{\'e}}, {Osborne}, {Pagani}, {Pagano}, {Pailler}, {Palacin},
  {Palaversa}, {Parsons}, {Paulsen}, {Pecoraro}, {Pedrosa}, {Pentik{\"a}inen},
  {Pereira}, {Pichon}, {Piersimoni}, {Pineau}, {Plachy}, {Plum}, {Poujoulet},
  {Pr{\v{s}}a}, {Pulone}, {Ragaini}, {Rago}, {Rambaux}, {Ramos-Lerate},
  {Ranalli}, {Rauw}, {Read}, {Regibo}, {Renk}, {Reyl{\'e}}, {Ribeiro},
  {Rimoldini}, {Ripepi}, {Riva}, {Rixon}, {Roelens}, {Romero-G{\'o}mez},
  {Rowell}, {Royer}, {Rudolph}, {Ruiz-Dern}, {Sadowski}, {Sagrist{\`a}
  Sell{\'e}s}, {Sahlmann}, {Salgado}, {Salguero}, {Sarasso}, {Savietto},
  {Schnorhk}, {Schultheis}, {Sciacca}, {Segol}, {Segovia}, {Segransan},
  {Serpell}, {Shih}, {Smareglia}, {Smart}, {Smith}, {Solano}, {Solitro},
  {Sordo}, {Soria Nieto}, {Souchay}, {Spagna}, {Spoto}, {Stampa}, {Steele},
  {Steidelm{\"u}ller}, {Stephenson}, {Stoev}, {Suess}, {S{\"u}veges}, {Surdej},
  {Szabados}, {Szegedi-Elek}, {Tapiador}, {Taris}, {Tauran}, {Taylor},
  {Teixeira}, {Terrett}, {Tingley}, {Trager}, {Turon}, {Ulla}, {Utrilla},
  {Valentini}, {van Elteren}, {Van Hemelryck}, {van Leeuwen}, {Varadi},
  {Vecchiato}, {Veljanoski}, {Via}, {Vicente}, {Vogt}, {Voss}, {Votruba},
  {Voutsinas}, {Walmsley}, {Weiler}, {Weingrill}, {Werner}, {Wevers},
  {Whitehead}, {Wyrzykowski}, {Yoldas}, {{\v{Z}}erjal}, {Zucker}, {Zurbach},
  {Zwitter}, {Alecu}, {Allen}, {Allende Prieto}, {Amorim},
  {Anglada-Escud{\'e}}, {Arsenijevic}, {Azaz}, {Balm}, {Beck}, {Bernstein},
  {Bigot}, {Bijaoui}, {Blasco}, {Bonfigli}, {Bono}, {Boudreault}, {Bressan},
  {Brown}, {Brunet}, {Bunclark}, {Buonanno}, {Butkevich}, {Carret}, {Carrion},
  {Chemin}, {Ch{\'e}reau}, {Corcione}, {Darmigny}, {de Boer}, {de Teodoro}, {de
  Zeeuw}, {Delle Luche}, {Domingues}, {Dubath}, {Fodor}, {Fr{\'e}zouls},
  {Fries}, {Fustes}, {Fyfe}, {Gallardo}, {Gallegos}, {Gardiol}, {Gebran},
  {Gomboc}, {G{\'o}mez}, {Grux}, {Gueguen}, {Heyrovsky}, {Hoar}, {Iannicola},
  {Isasi Parache}, {Janotto}, {Joliet}, {Jonckheere}, {Keil}, {Kim},
  {Klagyivik}, {Klar}, {Knude}, {Kochukhov}, {Kolka}, {Kos}, {Kutka}, {Lainey},
  {LeBouquin}, {Liu}, {Loreggia}, {Makarov}, {Marseille}, {Martayan},
  {Martinez-Rubi}, {Massart}, {Meynadier}, {Mignot}, {Munari}, {Nguyen},
  {Nordlander}, {Ocvirk}, {O'Flaherty}, {Olias Sanz}, {Ortiz}, {Osorio},
  {Oszkiewicz}, {Ouzounis}, {Palmer}, {Park}, {Pasquato}, {Peltzer}, {Peralta},
  {P{\'e}turaud}, {Pieniluoma}, {Pigozzi}, {Poels}, {Prat}, {Prod'homme},
  {Raison}, {Rebordao}, {Risquez}, {Rocca-Volmerange}, {Rosen}, {Ruiz-Fuertes},
  {Russo}, {Sembay}, {Serraller Vizcaino}, {Short}, {Siebert}, {Silva},
  {Sinachopoulos}, {Slezak}, {Soffel}, {Sosnowska}, {Strai{\v{z}}ys}, {ter
  Linden}, {Terrell}, {Theil}, {Tiede}, {Troisi}, {Tsalmantza}, {Tur},
  {Vaccari}, {Vachier}, {Valles}, {Van Hamme}, {Veltz}, {Virtanen}, {Wallut},
  {Wichmann}, {Wilkinson}, {Ziaeepour}, \& {Zschocke}}]{2016gaia}
{Gaia Collaboration}, {Prusti}, T., {de Bruijne}, J.~H.~J., {et~al.} 2016,
  \aap, 595, A1, \dodoi{10.1051/0004-6361/201629272}

\bibitem[{{Gaia Collaboration} {et~al.}(2023){Gaia Collaboration}, {Vallenari},
  {Brown}, {Prusti}, {de Bruijne}, {Arenou}, {Babusiaux}, {Biermann},
  {Creevey}, {Ducourant}, {Evans}, {Eyer}, {Guerra}, {Hutton}, {Jordi},
  {Klioner}, {Lammers}, {Lindegren}, {Luri}, {Mignard}, {Panem}, {Pourbaix},
  {Randich}, {Sartoretti}, {Soubiran}, {Tanga}, {Walton}, {Bailer-Jones},
  {Bastian}, {Drimmel}, {Jansen}, {Katz}, {Lattanzi}, {van Leeuwen}, {Bakker},
  {Cacciari}, {Casta{\~n}eda}, {De Angeli}, {Fabricius}, {Fouesneau},
  {Fr{\'e}mat}, {Galluccio}, {Guerrier}, {Heiter}, {Masana}, {Messineo},
  {Mowlavi}, {Nicolas}, {Nienartowicz}, {Pailler}, {Panuzzo}, {Riclet}, {Roux},
  {Seabroke}, {Sordo}, {Th{\'e}venin}, {Gracia-Abril}, {Portell}, {Teyssier},
  {Altmann}, {Andrae}, {Audard}, {Bellas-Velidis}, {Benson}, {Berthier},
  {Blomme}, {Burgess}, {Busonero}, {Busso}, {C{\'a}novas}, {Carry}, {Cellino},
  {Cheek}, {Clementini}, {Damerdji}, {Davidson}, {de Teodoro}, {Nu{\~n}ez
  Campos}, {Delchambre}, {Dell'Oro}, {Esquej}, {Fern{\'a}ndez-Hern{\'a}ndez},
  {Fraile}, {Garabato}, {Garc{\'\i}a-Lario}, {Gosset}, {Haigron}, {Halbwachs},
  {Hambly}, {Harrison}, {Hern{\'a}ndez}, {Hestroffer}, {Hodgkin}, {Holl},
  {Jan{\ss}en}, {Jevardat de Fombelle}, {Jordan}, {Krone-Martins}, {Lanzafame},
  {L{\"o}ffler}, {Marchal}, {Marrese}, {Moitinho}, {Muinonen}, {Osborne},
  {Pancino}, {Pauwels}, {Recio-Blanco}, {Reyl{\'e}}, {Riello}, {Rimoldini},
  {Roegiers}, {Rybizki}, {Sarro}, {Siopis}, {Smith}, {Sozzetti}, {Utrilla},
  {van Leeuwen}, {Abbas}, {{\'A}brah{\'a}m}, {Abreu Aramburu}, {Aerts},
  {Aguado}, {Ajaj}, {Aldea-Montero}, {Altavilla}, {{\'A}lvarez}, {Alves},
  {Anders}, {Anderson}, {Anglada Varela}, {Antoja}, {Baines}, {Baker},
  {Balaguer-N{\'u}{\~n}ez}, {Balbinot}, {Balog}, {Barache}, {Barbato},
  {Barros}, {Barstow}, {Bartolom{\'e}}, {Bassilana}, {Bauchet}, {Becciani},
  {Bellazzini}, {Berihuete}, {Bernet}, {Bertone}, {Bianchi}, {Binnenfeld},
  {Blanco-Cuaresma}, {Blazere}, {Boch}, {Bombrun}, {Bossini}, {Bouquillon},
  {Bragaglia}, {Bramante}, {Breedt}, {Bressan}, {Brouillet}, {Brugaletta},
  {Bucciarelli}, {Burlacu}, {Butkevich}, {Buzzi}, {Caffau}, {Cancelliere},
  {Cantat-Gaudin}, {Carballo}, {Carlucci}, {Carnerero}, {Carrasco},
  {Casamiquela}, {Castellani}, {Castro-Ginard}, {Chaoul}, {Charlot}, {Chemin},
  {Chiaramida}, {Chiavassa}, {Chornay}, {Comoretto}, {Contursi}, {Cooper},
  {Cornez}, {Cowell}, {Crifo}, {Cropper}, {Crosta}, {Crowley}, {Dafonte},
  {Dapergolas}, {David}, {David}, {de Laverny}, {De Luise}, {De March}, {De
  Ridder}, {de Souza}, {de Torres}, {del Peloso}, {del Pozo}, {Delbo},
  {Delgado}, {Delisle}, {Demouchy}, {Dharmawardena}, {Di Matteo}, {Diakite},
  {Diener}, {Distefano}, {Dolding}, {Edvardsson}, {Enke}, {Fabre}, {Fabrizio},
  {Faigler}, {Fedorets}, {Fernique}, {Fienga}, {Figueras}, {Fournier},
  {Fouron}, {Fragkoudi}, {Gai}, {Garcia-Gutierrez}, {Garcia-Reinaldos},
  {Garc{\'\i}a-Torres}, {Garofalo}, {Gavel}, {Gavras}, {Gerlach}, {Geyer},
  {Giacobbe}, {Gilmore}, {Girona}, {Giuffrida}, {Gomel}, {Gomez},
  {Gonz{\'a}lez-N{\'u}{\~n}ez}, {Gonz{\'a}lez-Santamar{\'\i}a},
  {Gonz{\'a}lez-Vidal}, {Granvik}, {Guillout}, {Guiraud},
  {Guti{\'e}rrez-S{\'a}nchez}, {Guy}, {Hatzidimitriou}, {Hauser}, {Haywood},
  {Helmer}, {Helmi}, {Sarmiento}, {Hidalgo}, {Hilger}, {H{\l}adczuk}, {Hobbs},
  {Holland}, {Huckle}, {Jardine}, {Jasniewicz}, {Jean-Antoine Piccolo},
  {Jim{\'e}nez-Arranz}, {Jorissen}, {Juaristi Campillo}, {Julbe}, {Karbevska},
  {Kervella}, {Khanna}, {Kontizas}, {Kordopatis}, {Korn}, {K{\'o}sp{\'a}l},
  {Kostrzewa-Rutkowska}, {Kruszy{\'n}ska}, {Kun}, {Laizeau}, {Lambert},
  {Lanza}, {Lasne}, {Le Campion}, {Lebreton}, {Lebzelter}, {Leccia}, {Leclerc},
  {Lecoeur-Taibi}, {Liao}, {Licata}, {Lindstr{\o}m}, {Lister}, {Livanou},
  {Lobel}, {Lorca}, {Loup}, {Madrero Pardo}, {Magdaleno Romeo}, {Managau},
  {Mann}, {Manteiga}, {Marchant}, {Marconi}, {Marcos}, {Marcos Santos},
  {Mar{\'\i}n Pina}, {Marinoni}, {Marocco}, {Marshall}, {Martin Polo},
  {Mart{\'\i}n-Fleitas}, {Marton}, {Mary}, {Masip}, {Massari},
  {Mastrobuono-Battisti}, {Mazeh}, {McMillan}, {Messina}, {Michalik}, {Millar},
  {Mints}, {Molina}, {Molinaro}, {Moln{\'a}r}, {Monari}, {Mongui{\'o}},
  {Montegriffo}, {Montero}, {Mor}, {Mora}, {Morbidelli}, {Morel}, {Morris},
  {Muraveva}, {Murphy}, {Musella}, {Nagy}, {Noval}, {Oca{\~n}a}, {Ogden},
  {Ordenovic}, {Osinde}, {Pagani}, {Pagano}, {Palaversa}, {Palicio},
  {Pallas-Quintela}, {Panahi}, {Payne-Wardenaar}, {Pe{\~n}alosa Esteller},
  {Penttil{\"a}}, {Pichon}, {Piersimoni}, {Pineau}, {Plachy}, {Plum}, {Poggio},
  {Pr{\v{s}}a}, {Pulone}, {Racero}, {Ragaini}, {Rainer}, {Raiteri}, {Rambaux},
  {Ramos}, {Ramos-Lerate}, {Re Fiorentin}, {Regibo}, {Richards}, {Rios Diaz},
  {Ripepi}, {Riva}, {Rix}, {Rixon}, {Robichon}, {Robin}, {Robin}, {Roelens},
  {Rogues}, {Rohrbasser}, {Romero-G{\'o}mez}, {Rowell}, {Royer}, {Ruz Mieres},
  {Rybicki}, {Sadowski}, {S{\'a}ez N{\'u}{\~n}ez}, {Sagrist{\`a} Sell{\'e}s},
  {Sahlmann}, {Salguero}, {Samaras}, {Sanchez Gimenez}, {Sanna},
  {Santove{\~n}a}, {Sarasso}, {Schultheis}, {Sciacca}, {Segol}, {Segovia},
  {S{\'e}gransan}, {Semeux}, {Shahaf}, {Siddiqui}, {Siebert}, {Siltala},
  {Silvelo}, {Slezak}, {Slezak}, {Smart}, {Snaith}, {Solano}, {Solitro},
  {Souami}, {Souchay}, {Spagna}, {Spina}, {Spoto}, {Steele},
  {Steidelm{\"u}ller}, {Stephenson}, {S{\"u}veges}, {Surdej}, {Szabados},
  {Szegedi-Elek}, {Taris}, {Taylor}, {Teixeira}, {Tolomei}, {Tonello}, {Torra},
  {Torra}, {Torralba Elipe}, {Trabucchi}, {Tsounis}, {Turon}, {Ulla}, {Unger},
  {Vaillant}, {van Dillen}, {van Reeven}, {Vanel}, {Vecchiato}, {Viala},
  {Vicente}, {Voutsinas}, {Weiler}, {Wevers}, {Wyrzykowski}, {Yoldas}, {Yvard},
  {Zhao}, {Zorec}, {Zucker}, \& {Zwitter}}]{2023gaia}
{Gaia Collaboration}, {Vallenari}, A., {Brown}, A.~G.~A., {et~al.} 2023, \aap,
  674, A1, \dodoi{10.1051/0004-6361/202243940}

\bibitem[{{GLIMPSE Team}(2020)}]{https://doi.org/10.26131/irsa214}
{GLIMPSE Team}. 2020, GLIMPSE 360 Catalog,  IPAC, \dodoi{10.26131/IRSA214}

\bibitem[{{Grasha} {et~al.}(2017){Grasha}, {Elmegreen}, {Calzetti}, {Adamo},
  {Aloisi}, {Bright}, {Cook}, {Dale}, {Fumagalli}, {Gallagher}, {Gouliermis},
  {Grebel}, {Kahre}, {Kim}, {Krumholz}, {Lee}, {Messa}, {Ryon}, \&
  {Ubeda}}]{2017ApJ...842...25G}
{Grasha}, K., {Elmegreen}, B.~G., {Calzetti}, D., {et~al.} 2017, \apj, 842, 25,
  \dodoi{10.3847/1538-4357/aa740b}

\bibitem[{{Grasha} {et~al.}(2018){Grasha}, {Calzetti}, {Bittle}, {Johnson},
  {Donovan Meyer}, {Kennicutt}, {Elmegreen}, {Adamo}, {Krumholz}, {Fumagalli},
  {Grebel}, {Gouliermis}, {Cook}, {Gallagher}, {Aloisi}, {Dale}, {Linden},
  {Sacchi}, {Thilker}, {Walterbos}, {Messa}, {Wofford}, \&
  {Smith}}]{2018MNRAS.481.1016G}
{Grasha}, K., {Calzetti}, D., {Bittle}, L., {et~al.} 2018, \mnras, 481, 1016,
  \dodoi{10.1093/mnras/sty2154}

\bibitem[{{Green} {et~al.}(2019){Green}, {Schlafly}, {Zucker}, {Speagle}, \&
  {Finkbeiner}}]{2019ApJ...887...93G}
{Green}, G.~M., {Schlafly}, E., {Zucker}, C., {Speagle}, J.~S., \&
  {Finkbeiner}, D. 2019, \apj, 887, 93, \dodoi{10.3847/1538-4357/ab5362}

\bibitem[{{Gutermuth} {et~al.}(2009){Gutermuth}, {Megeath}, {Myers}, {Allen},
  {Pipher}, \& {Fazio}}]{2009ApJS..184...18G}
{Gutermuth}, R.~A., {Megeath}, S.~T., {Myers}, P.~C., {et~al.} 2009, \apjs,
  184, 18, \dodoi{10.1088/0067-0049/184/1/18}

\bibitem[{{Gutermuth} {et~al.}(2005){Gutermuth}, {Megeath}, {Pipher},
  {Williams}, {Allen}, {Myers}, \& {Raines}}]{2005ApJ...632..397G}
{Gutermuth}, R.~A., {Megeath}, S.~T., {Pipher}, J.~L., {et~al.} 2005, \apj,
  632, 397, \dodoi{10.1086/432460}

\bibitem[{{Gutermuth} {et~al.}(2011){Gutermuth}, {Pipher}, {Megeath}, {Myers},
  {Allen}, \& {Allen}}]{2011ApJ...739...84G}
{Gutermuth}, R.~A., {Pipher}, J.~L., {Megeath}, S.~T., {et~al.} 2011, \apj,
  739, 84, \dodoi{10.1088/0004-637X/739/2/84}

\bibitem[{{Hanaoka} {et~al.}(2019){Hanaoka}, {Kaneda}, {Suzuki}, {Kokusho},
  {Oyabu}, {Ishihara}, {Kohno}, {Furuta}, {Tsuchikawa}, \&
  {Saito}}]{2019PASJ...71....6H}
{Hanaoka}, M., {Kaneda}, H., {Suzuki}, T., {et~al.} 2019, \pasj, 71, 6,
  \dodoi{10.1093/pasj/psy126}

\bibitem[{{Kaur} {et~al.}(2020){Kaur}, {Sharma}, {Dewangan}, {Ojha},
  {Durgapal}, \& {Panwar}}]{2020ApJ...896...29K}
{Kaur}, H., {Sharma}, S., {Dewangan}, L.~K., {et~al.} 2020, \apj, 896, 29,
  \dodoi{10.3847/1538-4357/ab9122}

\bibitem[{{Kaur} {et~al.}(2023){Kaur}, {Sharma}, {Durgapal}, {Dewangan},
  {Verma}, {Panwar}, {Pandey}, \& {Ghosh}}]{2023JApA...44...66K}
{Kaur}, H., {Sharma}, S., {Durgapal}, A., {et~al.} 2023, Journal of
  Astrophysics and Astronomy, 44, 66, \dodoi{10.1007/s12036-023-09953-9}

\bibitem[{{Kharchenko} {et~al.}(2013){Kharchenko}, {Piskunov}, {Roeser},
  {Schilbach}, \& {Scholz}}]{2013yCat..35580053K}
{Kharchenko}, N.~V., {Piskunov}, A.~E., {Roeser}, S., {Schilbach}, E., \&
  {Scholz}, R.~D. 2013, VizieR Online Data Catalog, J/A+A/558/A53

\bibitem[{{Kirk} {et~al.}(2014){Kirk}, {Offner}, \&
  {Redmond}}]{2014MNRAS.439.1765K}
{Kirk}, H., {Offner}, S. S.~R., \& {Redmond}, K.~J. 2014, \mnras, 439, 1765,
  \dodoi{10.1093/mnras/stu052}

\bibitem[{{Koenig} {et~al.}(2008){Koenig}, {Allen}, {Gutermuth}, {Hora},
  {Brunt}, \& {Muzerolle}}]{2008ApJ...688.1142K}
{Koenig}, X.~P., {Allen}, L.~E., {Gutermuth}, R.~A., {et~al.} 2008, \apj, 688,
  1142, \dodoi{10.1086/592322}

\bibitem[{{Koenig} \& {Leisawitz}(2014)}]{2014ApJ...791..131K}
{Koenig}, X.~P., \& {Leisawitz}, D.~T. 2014, \apj, 791, 131,
  \dodoi{10.1088/0004-637X/791/2/131}

\bibitem[{{Kuhn} {et~al.}(2014){Kuhn}, {Feigelson}, {Getman}, {Baddeley},
  {Broos}, {Sills}, {Bate}, {Povich}, {Luhman}, {Busk}, {Naylor}, \&
  {King}}]{2014ApJ...787..107K}
{Kuhn}, M.~A., {Feigelson}, E.~D., {Getman}, K.~V., {et~al.} 2014, \apj, 787,
  107, \dodoi{10.1088/0004-637X/787/2/107}

\bibitem[{{Kumar} {et~al.}(2022){Kumar}, {Arzoumanian}, {Men'shchikov},
  {Palmeirim}, {Matsumura}, \& {Inutsuka}}]{2022A&A...658A.114K}
{Kumar}, M.~S.~N., {Arzoumanian}, D., {Men'shchikov}, A., {et~al.} 2022, \aap,
  658, A114, \dodoi{10.1051/0004-6361/202140363}

\bibitem[{{Lada} {et~al.}(1996){Lada}, {Alves}, \&
  {Lada}}]{1996AJ....111.1964L}
{Lada}, C.~J., {Alves}, J., \& {Lada}, E.~A. 1996, \aj, 111, 1964,
  \dodoi{10.1086/117933}

\bibitem[{{Lada} \& {Lada}(2003)}]{2003ARA&A..41...57L}
{Lada}, C.~J., \& {Lada}, E.~A. 2003, \araa, 41, 57,
  \dodoi{10.1146/annurev.astro.41.011802.094844}

\bibitem[{{Liu} {et~al.}(2011){Liu}, {Wu}, {Liu}, {Qin}, {Su}, {Chen}, \&
  {Ren}}]{2011ApJ...730..102L}
{Liu}, T., {Wu}, Y., {Liu}, S.-Y., {et~al.} 2011, \apj, 730, 102,
  \dodoi{10.1088/0004-637X/730/2/102}

\bibitem[{{Lucas} {et~al.}(2008){Lucas}, {Hoare}, {Longmore}, {Schr{\"o}der},
  {Davis}, {Adamson}, {Bandyopadhyay}, {de Grijs}, {Smith}, {Gosling},
  {Mitchison}, {G{\'a}sp{\'a}r}, {Coe}, {Tamura}, {Parker}, {Irwin}, {Hambly},
  {Bryant}, {Collins}, {Cross}, {Evans}, {Gonzalez-Solares}, {Hodgkin},
  {Lewis}, {Read}, {Riello}, {Sutorius}, {Lawrence}, {Drew}, {Dye}, \&
  {Thompson}}]{2008MNRAS.391..136L}
{Lucas}, P.~W., {Hoare}, M.~G., {Longmore}, A., {et~al.} 2008, \mnras, 391,
  136, \dodoi{10.1111/j.1365-2966.2008.13924.x}

\bibitem[{{Mallick} {et~al.}(2023{\natexlab{a}}){Mallick}, {Dewangan}, {Ojha},
  {Baug}, \& {Zinchenko}}]{2023ApJ...944..228M}
{Mallick}, K.~K., {Dewangan}, L.~K., {Ojha}, D.~K., {Baug}, T., \& {Zinchenko},
  I.~I. 2023{\natexlab{a}}, \apj, 944, 228, \dodoi{10.3847/1538-4357/acb8bc}

\bibitem[{{Mallick} {et~al.}(2023{\natexlab{b}}){Mallick}, {Sharma},
  {Dewangan}, {Ojha}, {Panwar}, \& {Baug}}]{2023JApA...44...34M}
{Mallick}, K.~K., {Sharma}, S., {Dewangan}, L.~K., {et~al.} 2023{\natexlab{b}},
  Journal of Astrophysics and Astronomy, 44, 34,
  \dodoi{10.1007/s12036-023-09930-2}

\bibitem[{{Marsh} {et~al.}(2015){Marsh}, {Whitworth}, \&
  {Lomax}}]{2015MNRAS.454.4282M}
{Marsh}, K.~A., {Whitworth}, A.~P., \& {Lomax}, O. 2015, \mnras, 454, 4282,
  \dodoi{10.1093/mnras/stv2248}

\bibitem[{{Marsh} {et~al.}(2017){Marsh}, {Whitworth}, {Lomax}, {Ragan},
  {Becciani}, {Cambr{\'e}sy}, {Di Giorgio}, {Eden}, {Elia}, {Kacsuk},
  {Molinari}, {Palmeirim}, {Pezzuto}, {Schneider}, {Sciacca}, \&
  {Vitello}}]{2017MNRAS.471.2730M}
{Marsh}, K.~A., {Whitworth}, A.~P., {Lomax}, O., {et~al.} 2017, \mnras, 471,
  2730, \dodoi{10.1093/mnras/stx1723}

\bibitem[{{Molinari} {et~al.}(2010{\natexlab{a}}){Molinari}, {Swinyard},
  {Bally}, {Barlow}, {Bernard}, {Martin}, {Moore}, {Noriega-Crespo}, {Plume},
  {Testi}, {Zavagno}, {Abergel}, {Ali}, {Andr{\'e}}, {Baluteau}, {Benedettini},
  {Bern{\'e}}, {Billot}, {Blommaert}, {Bontemps}, {Boulanger}, {Brand},
  {Brunt}, {Burton}, {Campeggio}, {Carey}, {Caselli}, {Cesaroni}, {Cernicharo},
  {Chakrabarti}, {Chrysostomou}, {Codella}, {Cohen}, {Compiegne}, {Davis}, {de
  Bernardis}, {de Gasperis}, {Di Francesco}, {di Giorgio}, {Elia}, {Faustini},
  {Fischera}, {Fukui}, {Fuller}, {Ganga}, {Garcia-Lario}, {Giard}, {Giardino},
  {Glenn}, {Goldsmith}, {Griffin}, {Hoare}, {Huang}, {Jiang}, {Joblin},
  {Joncas}, {Juvela}, {Kirk}, {Lagache}, {Li}, {Lim}, {Lord}, {Lucas},
  {Maiolo}, {Marengo}, {Marshall}, {Masi}, {Massi}, {Matsuura}, {Meny},
  {Minier}, {Miville-Desch{\^e}nes}, {Montier}, {Motte}, {M{\"u}ller},
  {Natoli}, {Neves}, {Olmi}, {Paladini}, {Paradis}, {Pestalozzi}, {Pezzuto},
  {Piacentini}, {Pomar{\`e}s}, {Popescu}, {Reach}, {Richer}, {Ristorcelli},
  {Roy}, {Royer}, {Russeil}, {Saraceno}, {Sauvage}, {Schilke},
  {Schneider-Bontemps}, {Schuller}, {Schultz}, {Shepherd}, {Sibthorpe},
  {Smith}, {Smith}, {Spinoglio}, {Stamatellos}, {Strafella}, {Stringfellow},
  {Sturm}, {Taylor}, {Thompson}, {Tuffs}, {Umana}, {Valenziano}, {Vavrek},
  {Viti}, {Waelkens}, {Ward-Thompson}, {White}, {Wyrowski}, {Yorke}, \&
  {Zhang}}]{2010PASP..122..314M}
{Molinari}, S., {Swinyard}, B., {Bally}, J., {et~al.} 2010{\natexlab{a}},
  \pasp, 122, 314, \dodoi{10.1086/651314}

\bibitem[{{Molinari} {et~al.}(2010{\natexlab{b}}){Molinari}, {Swinyard},
  {Bally}, {Barlow}, {Bernard}, {Martin}, {Moore}, {Noriega-Crespo}, {Plume},
  {Testi}, {Zavagno}, {Abergel}, {Ali}, {Anderson}, {Andr{\'e}}, {Baluteau},
  {Battersby}, {Beltr{\'a}n}, {Benedettini}, {Billot}, {Blommaert}, {Bontemps},
  {Boulanger}, {Brand}, {Brunt}, {Burton}, {Calzoletti}, {Carey}, {Caselli},
  {Cesaroni}, {Cernicharo}, {Chakrabarti}, {Chrysostomou}, {Cohen},
  {Compiegne}, {de Bernardis}, {de Gasperis}, {di Giorgio}, {Elia}, {Faustini},
  {Flagey}, {Fukui}, {Fuller}, {Ganga}, {Garcia-Lario}, {Glenn}, {Goldsmith},
  {Griffin}, {Hoare}, {Huang}, {Ikhenaode}, {Joblin}, {Joncas}, {Juvela},
  {Kirk}, {Lagache}, {Li}, {Lim}, {Lord}, {Marengo}, {Marshall}, {Masi},
  {Massi}, {Matsuura}, {Minier}, {Miville-Desch{\^e}nes}, {Montier}, {Morgan},
  {Motte}, {Mottram}, {M{\"u}ller}, {Natoli}, {Neves}, {Olmi}, {Paladini},
  {Paradis}, {Parsons}, {Peretto}, {Pestalozzi}, {Pezzuto}, {Piacentini},
  {Piazzo}, {Polychroni}, {Pomar{\`e}s}, {Popescu}, {Reach}, {Ristorcelli},
  {Robitaille}, {Robitaille}, {Rod{\'o}n}, {Roy}, {Royer}, {Russeil},
  {Saraceno}, {Sauvage}, {Schilke}, {Schisano}, {Schneider}, {Schuller},
  {Schulz}, {Sibthorpe}, {Smith}, {Smith}, {Spinoglio}, {Stamatellos},
  {Strafella}, {Stringfellow}, {Sturm}, {Taylor}, {Thompson}, {Traficante},
  {Tuffs}, {Umana}, {Valenziano}, {Vavrek}, {Veneziani}, {Viti}, {Waelkens},
  {Ward-Thompson}, {White}, {Wilcock}, {Wyrowski}, {Yorke}, \&
  {Zhang}}]{2010A&A...518L.100M}
---. 2010{\natexlab{b}}, \aap, 518, L100, \dodoi{10.1051/0004-6361/201014659}

\bibitem[{{Motte} {et~al.}(1998){Motte}, {Andre}, \&
  {Neri}}]{1998A&A...336..150M}
{Motte}, F., {Andre}, P., \& {Neri}, R. 1998, \aap, 336, 150

\bibitem[{{Motte} {et~al.}(2018){Motte}, {Bontemps}, \&
  {Louvet}}]{2018ARA&A..56...41M}
{Motte}, F., {Bontemps}, S., \& {Louvet}, F. 2018, \araa, 56, 41,
  \dodoi{10.1146/annurev-astro-091916-055235}

\bibitem[{{Myers}(1998)}]{1998sao..rept.....M}
{Myers}, P.~C. 1998, {Observational and Theoretical Studies of Low-Mass Star
  Formation}, Technical Report, NASA/CR-1998-207452; NAS 1.26:207452

\bibitem[{{Myers}(2009)}]{2009ApJ...700.1609M}
---. 2009, \apj, 700, 1609, \dodoi{10.1088/0004-637X/700/2/1609}

\bibitem[{{Ojha} {et~al.}(2004){Ojha}, {Tamura}, {Nakajima}, {Fukagawa},
  {Sugitani}, {Nagashima}, {Nagayama}, {Nagata}, {Sato}, {Pickles}, \&
  {Ogura}}]{2004ApJ...608..797O}
{Ojha}, D.~K., {Tamura}, M., {Nakajima}, Y., {et~al.} 2004, \apj, 608, 797,
  \dodoi{10.1086/420876}

\bibitem[{{Pandey} {et~al.}(2020{\natexlab{a}}){Pandey}, {Sharma}, {Kobayashi},
  {Sarugaku}, \& {Ogura}}]{2020MNRAS.492.2446P}
{Pandey}, A.~K., {Sharma}, S., {Kobayashi}, N., {Sarugaku}, Y., \& {Ogura}, K.
  2020{\natexlab{a}}, \mnras, 492, 2446, \dodoi{10.1093/mnras/stz3596}

\bibitem[{{Pandey} {et~al.}(2020{\natexlab{b}}){Pandey}, {Sharma}, {Panwar},
  {Dewangan}, {Ojha}, {Bisen}, {Sinha}, {Ghosh}, \&
  {Pandey}}]{2020ApJ...891...81P}
{Pandey}, R., {Sharma}, S., {Panwar}, N., {et~al.} 2020{\natexlab{b}}, \apj,
  891, 81, \dodoi{10.3847/1538-4357/ab6dc7}

\bibitem[{{Pandey} {et~al.}(2024){Pandey}, {Sharma}, {Dewangan}, {Ojha},
  {Panwar}, {Ghosh}, {Sinha}, {Verma}, \& {Kaur}}]{2024MNRAS.527.9626P}
{Pandey}, R., {Sharma}, S., {Dewangan}, L., {et~al.} 2024, \mnras, 527, 9626,
  \dodoi{10.1093/mnras/stad2944}

\bibitem[{{Paulson} {et~al.}(2024){Paulson}, {Mallick}, \&
  {Ojha}}]{2024MNRAS.530.1516P}
{Paulson}, S.~T., {Mallick}, K.~K., \& {Ojha}, D.~K. 2024, \mnras, 530, 1516,
  \dodoi{10.1093/mnras/stae917}

\bibitem[{{Pecaut} \& {Mamajek}(2013)}]{2013ApJS..208....9P}
{Pecaut}, M.~J., \& {Mamajek}, E.~E. 2013, \apjs, 208, 9,
  \dodoi{10.1088/0067-0049/208/1/9}

\bibitem[{{Phelps} \& {Janes}(1994)}]{1994ApJS...90...31P}
{Phelps}, R.~L., \& {Janes}, K.~A. 1994, \apjs, 90, 31, \dodoi{10.1086/191857}

\bibitem[{{Reipurth} \& {Yan}(2008)}]{2008hsf1.book..869R}
{Reipurth}, B., \& {Yan}, C.~H. 2008, in Handbook of Star Forming Regions,
  Volume I, ed. B.~{Reipurth}, Vol.~4 (The Northern Sky ASP Monograph
  Publications), 869

\bibitem[{{Rieke} \& {Lebofsky}(1985)}]{1985ApJ...288..618R}
{Rieke}, G.~H., \& {Lebofsky}, M.~J. 1985, \apj, 288, 618,
  \dodoi{10.1086/162827}

\bibitem[{{S{\'a}nchez} {et~al.}(2010){S{\'a}nchez}, {A{\~n}ez}, {Alfaro}, \&
  {Crone Odekon}}]{2010ApJ...720..541S}
{S{\'a}nchez}, N., {A{\~n}ez}, N., {Alfaro}, E.~J., \& {Crone Odekon}, M. 2010,
  \apj, 720, 541, \dodoi{10.1088/0004-637X/720/1/541}

\bibitem[{{Schmeja} \& {Klessen}(2006)}]{2006A&A...449..151S}
{Schmeja}, S., \& {Klessen}, R.~S. 2006, \aap, 449, 151,
  \dodoi{10.1051/0004-6361:20054464}

\bibitem[{{Sharma} {et~al.}(2017){Sharma}, {Pandey}, {Ojha}, {Bhatt}, {Ogura},
  {Kobayashi}, {Yadav}, \& {Pandey}}]{2017MNRAS.467.2943S}
{Sharma}, S., {Pandey}, A.~K., {Ojha}, D.~K., {et~al.} 2017, \mnras, 467, 2943,
  \dodoi{10.1093/mnras/stx014}

\bibitem[{{Sharma} {et~al.}(2016){Sharma}, {Pandey}, {Borissova}, {Ojha},
  {Ivanov}, {Ogura}, {Kobayashi}, {Kurtev}, {Gopinathan}, \& {Kesh
  Yadav}}]{2016AJ....151..126S}
{Sharma}, S., {Pandey}, A.~K., {Borissova}, J., {et~al.} 2016, \aj, 151, 126,
  \dodoi{10.3847/0004-6256/151/5/126}

\bibitem[{{Sharma} {et~al.}(2020){Sharma}, {Ghosh}, {Ojha}, {Pandey}, {Sinha},
  {Pandey}, {Ghosh}, {Panwar}, \& {Pandey}}]{2020MNRAS.498.2309S}
{Sharma}, S., {Ghosh}, A., {Ojha}, D.~K., {et~al.} 2020, \mnras, 498, 2309,
  \dodoi{10.1093/mnras/staa2412}

\bibitem[{{Sharma} {et~al.}(2024){Sharma}, {Verma}, {Mallick}, {Dewangan},
  {Kaur}, {Yadav}, {Panwar}, {Ojha}, {Chand}, \&
  {Agarwal}}]{2024AJ....167..106S}
{Sharma}, S., {Verma}, A., {Mallick}, K., {et~al.} 2024, \aj, 167, 106,
  \dodoi{10.3847/1538-3881/ad19cd}

\bibitem[{{Sills} {et~al.}(2018){Sills}, {Rieder}, {Scora}, {McCloskey}, \&
  {Jaffa}}]{2018MNRAS.477.1903S}
{Sills}, A., {Rieder}, S., {Scora}, J., {McCloskey}, J., \& {Jaffa}, S. 2018,
  \mnras, 477, 1903, \dodoi{10.1093/mnras/sty681}

\bibitem[{{Skrutskie} {et~al.}(2003){Skrutskie}, {Cutri}, {Stiening},
  {Weinberg}, {Schneider}, {Carpenter}, {Beichman}, {Capps}, {Chester},
  {Elias}, {Huchra}, {Liebert}, {Lonsdale}, {Monet}, {Price}, {Seitzer},
  {Jarrett}, {Kirkpatrick}, {Gizis}, {Howard}, {Evans}, {Fowler}, {Fullmer},
  {Hurt}, {Light}, {Kopan}, {Marsh}, {McCallon}, {Tam}, {Van Dyk}, \&
  {Wheelock}}]{https://doi.org/10.26131/irsa2}
{Skrutskie}, M.~F., {Cutri}, R.~M., {Stiening}, R., {et~al.} 2003, 2MASS
  All-Sky Point Source Catalog,  IPAC, \dodoi{10.26131/IRSA2}

\bibitem[{{Skrutskie} {et~al.}(2006){Skrutskie}, {Cutri}, {Stiening},
  {Weinberg}, {Schneider}, {Carpenter}, {Beichman}, {Capps}, {Chester},
  {Elias}, {Huchra}, {Liebert}, {Lonsdale}, {Monet}, {Price}, {Seitzer},
  {Jarrett}, {Kirkpatrick}, {Gizis}, {Howard}, {Evans}, {Fowler}, {Fullmer},
  {Hurt}, {Light}, {Kopan}, {Marsh}, {McCallon}, {Tam}, {Van Dyk}, \&
  {Wheelock}}]{2006AJ....131.1163S}
---. 2006, \aj, 131, 1163, \dodoi{10.1086/498708}

\bibitem[{{Su} {et~al.}(2019){Su}, {Yang}, {Zhang}, {Gong}, {Wang}, {Zhou},
  {Wang}, {Chen}, {Sun}, {Chen}, {Xu}, \& {Jiang}}]{Su_MWISP_2019ApJS}
{Su}, Y., {Yang}, J., {Zhang}, S., {et~al.} 2019, \apjs, 240, 9,
  \dodoi{10.3847/1538-4365/aaf1c8}

\bibitem[{{Torrelles} {et~al.}(1992){Torrelles}, {Eiroa}, {Mauersberger},
  {Estalella}, {Miranda}, \& {Anglada}}]{1992ApJ...384..528T}
{Torrelles}, J.~M., {Eiroa}, C., {Mauersberger}, R., {et~al.} 1992, \apj, 384,
  528, \dodoi{10.1086/170895}

\bibitem[{{T{\'o}th} {et~al.}(2014){T{\'o}th}, {Marton}, {Zahorecz},
  {Bal{\'a}zs}, {Ueno}, {Tamura}, {Kawamura}, {Kiss}, \&
  {Kitamura}}]{2014PASJ...66...17T}
{T{\'o}th}, L.~V., {Marton}, G., {Zahorecz}, S., {et~al.} 2014, \pasj, 66, 17,
  \dodoi{10.1093/pasj/pst017}

\bibitem[{{Trevi{\~n}o-Morales} {et~al.}(2019){Trevi{\~n}o-Morales}, {Fuente},
  {S{\'a}nchez-Monge}, {Kainulainen}, {Didelon}, {Suri}, {Schneider},
  {Ballesteros-Paredes}, {Lee}, {Hennebelle}, {Pilleri},
  {Gonz{\'a}lez-Garc{\'\i}a}, {Kramer}, {Garc{\'\i}a-Burillo}, {Luna},
  {Goicoechea}, {Tremblin}, \& {Geen}}]{2019A&A...629A..81T}
{Trevi{\~n}o-Morales}, S.~P., {Fuente}, A., {S{\'a}nchez-Monge}, {\'A}.,
  {et~al.} 2019, \aap, 629, A81, \dodoi{10.1051/0004-6361/201935260}

\bibitem[{{Verma} {et~al.}(2023{\natexlab{a}}){Verma}, {Sharma}, {Dewangan},
  {Pandey}, {Baug}, {Ojha}, {Ghosh}, \& {Kaur}}]{2023JApA...44...52V}
{Verma}, A., {Sharma}, S., {Dewangan}, L., {et~al.} 2023{\natexlab{a}}, Journal
  of Astrophysics and Astronomy, 44, 52, \dodoi{10.1007/s12036-023-09932-0}

\bibitem[{{Verma} {et~al.}(2023{\natexlab{b}}){Verma}, {Sharma}, {Mallick},
  {Dewangan}, {Ojha}, {Yadav}, {Pandey}, {Ghosh}, {Kaur}, {Panwar}, \&
  {Chand}}]{2023ApJ...953..145V}
{Verma}, A., {Sharma}, S., {Mallick}, K.~K., {et~al.} 2023{\natexlab{b}}, \apj,
  953, 145, \dodoi{10.3847/1538-4357/acdeef}

\bibitem[{{Wang} \& {Chen}(2019)}]{2019ApJ...877..116W}
{Wang}, S., \& {Chen}, X. 2019, \apj, 877, 116,
  \dodoi{10.3847/1538-4357/ab1c61}

\bibitem[{{Winston} {et~al.}(2007){Winston}, {Megeath}, {Wolk}, {Muzerolle},
  {Gutermuth}, {Hora}, {Allen}, {Spitzbart}, {Myers}, \&
  {Fazio}}]{2007ApJ...669..493W}
{Winston}, E., {Megeath}, S.~T., {Wolk}, S.~J., {et~al.} 2007, \apj, 669, 493,
  \dodoi{10.1086/521384}

\bibitem[{{Wolf-Chase} {et~al.}(2017){Wolf-Chase}, {Arvidsson}, \&
  {Smutko}}]{2017ApJ...844...38W}
{Wolf-Chase}, G., {Arvidsson}, K., \& {Smutko}, M. 2017, \apj, 844, 38,
  \dodoi{10.3847/1538-4357/aa762a}

\bibitem[{{Wouterloot} \& {Brand}(1989)}]{1989A&AS...80..149W}
{Wouterloot}, J.~G.~A., \& {Brand}, J. 1989, \aaps, 80, 149

\bibitem[{{Wright} {et~al.}(2010){Wright}, {Eisenhardt}, {Mainzer}, {Ressler},
  {Cutri}, {Jarrett}, {Kirkpatrick}, {Padgett}, {McMillan}, {Skrutskie},
  {Stanford}, {Cohen}, {Walker}, {Mather}, {Leisawitz}, {Gautier}, {McLean},
  {Benford}, {Lonsdale}, {Blain}, {Mendez}, {Irace}, {Duval}, {Liu}, {Royer},
  {Heinrichsen}, {Howard}, {Shannon}, {Kendall}, {Walsh}, {Larsen}, {Cardon},
  {Schick}, {Schwalm}, {Abid}, {Fabinsky}, {Naes}, \&
  {Tsai}}]{2010AJ....140.1868W}
{Wright}, E.~L., {Eisenhardt}, P. R.~M., {Mainzer}, A.~K., {et~al.} 2010, \aj,
  140, 1868, \dodoi{10.1088/0004-6256/140/6/1868}

\bibitem[{{Wright} {et~al.}(2019){Wright}, {Eisenhardt}, {Mainzer}, {Ressler},
  {Cutri}, {Jarrett}, {Kirkpatrick}, {Padgett}, {McMillan}, {Skrutskie},
  {Stanford}, {Cohen}, {Walker}, {Mather}, {Leisawitz}, {Gautier}, {McLean},
  {Benford}, {Lonsdale}, {Blain}, {Mendez}, {Irace}, {Duval}, {Liu}, {Royer},
  {Heinrichsen}, {Howard}, {Shannon}, {Kendall}, {Walsh}, {Larsen}, {Cardon},
  {Schick}, {Schwalm}, {Abid}, {Fabinsky}, {Naes}, \&
  {Tsai}}]{https://doi.org/10.26131/irsa1}
---. 2019, AllWISE Source Catalog,  IPAC, \dodoi{10.26131/IRSA1}

\bibitem[{{Yang} {et~al.}(2002){Yang}, {Jiang}, {Wang}, {Ju}, \&
  {Wang}}]{2002ApJS..141..157Y}
{Yang}, J., {Jiang}, Z., {Wang}, M., {Ju}, B., \& {Wang}, H. 2002, \apjs, 141,
  157, \dodoi{10.1086/340038}

\end{thebibliography}
\bibliographystyle{aasjournal}



\end{document}